\documentclass[aps,pra,twocolumn,superscriptaddress]{revtex4}

\usepackage{amssymb}
\usepackage{graphicx}
\usepackage{amsmath}
\usepackage{bbold}
\usepackage[dvipsnames]{xcolor}
\usepackage{verbatim}
\usepackage[utf8]{inputenc}

\usepackage{hyperref}
\usepackage{tikz}
\usepackage{tikz-cd}
\usetikzlibrary{shapes.misc, positioning}
\usetikzlibrary{quantikz}
\usetikzlibrary{calc, arrows, intersections, arrows.meta}

\usetikzlibrary{decorations.markings}
\tikzset{double line with arrow/.style args={#1,#2}{decorate,decoration={markings,%
mark=at position 0 with {\coordinate (ta-base-1) at (0,1pt);
\coordinate (ta-base-2) at (0,-1pt);},
mark=at position 1 with {\draw[#1] (ta-base-1) -- (0,1pt);
\draw[#2] (ta-base-2) -- (0,-1pt);
}}}}

\tikzset{
operator/.append style={fill=black!5}}

\newcommand{\beq}{\begin{equation}}
\newcommand{\eeq}{\end{equation}}
\newcommand{\ba}{\begin{eqnarray}}
\newcommand{\ea}{\end{eqnarray}}

\newcommand\noquad{{}}

\begin{document}

\title{Universal quantum computation with symmetric qubit clusters coupled to an environment}

\author{Christian Boudreault}
\email{christian.boudreault@cmrsj-rmcsj.ca}
\affiliation{D{\'e}partement des sciences de la nature, Coll{\`e}ge militaire royal de Saint-Jean
15 Jacques-Cartier Nord, Saint-Jean-sur-Richelieu, QC, Canada, J3B 8R8}
\affiliation{D{\'e}partement de physique, Universit{\'e} de Montr{\'e}al, Complexe des Sciences,  C.P. 6128, succursale Centre-ville, Montr{\'e}al, QC, Canada, H3C 3J7}
\author{Hichem Eleuch}
\email{heleuch@fulbrightmail.org}
\affiliation{Department of Applied Physics and Astronomy, University of Sharjah, Sharjah, United Arab Emirates}
\affiliation{College of Arts and Sciences, Abu Dhabi University, Abu Dhabi 59911, United Arab Emirates }
\author{Michael Hilke}
\email{hilke@physics.mcgill.ca}
\affiliation{Department of Physics, McGill University, Montr\'eal, QC, Canada, H3A 2T8}
\author{Richard MacKenzie}
\email{richard.mackenzie@umontreal.ca}
\affiliation{D{\'e}partement de physique, Universit{\'e} de Montr{\'e}al, Complexe des Sciences,  C.P. 6128, succursale Centre-ville, Montr{\'e}al, QC, Canada, H3C 3J7}

\begin{abstract}
One of the most challenging problems for the realization of a scalable quantum computer is to design a physical device that keeps the error rate for each quantum processing operation low. These errors can originate from the accuracy of quantum manipulation, such as the sweeping of a gate voltage in solid state qubits or the duration of a laser pulse in optical schemes. Errors also result from decoherence, which is often regarded as more crucial in the sense that it is inherent to the quantum system, being fundamentally a consequence of the coupling to the external environment.

Grouping small collections of qubits into clusters with symmetries may serve to protect parts of the calculation from decoherence. In this work, we use 4-level cores with a straightforward generalization of discrete rotational symmetry, called $\omega$-rotation invariance, to encode pairs of coupled qubits and universal 2-qubit logical gates. We include quantum errors as a main source of decoherence, and show that symmetry makes logical operations particularly resilient to untimely anisotropic qubit rotations. We propose a scalable scheme for universal quantum computation where cores play the role of quantum-computational transistors, or \textit{quansistors} for short.

Initialization and readout are achieved by tunnel-coupling the quansistor to leads. The external leads are explicitly considered and are assumed to be the other main source of decoherence. We show that quansistors can be dynamically decoupled from the leads by tuning their internal parameters, giving them the versatility required to act as controllable quantum memory units. With this dynamical decoupling, logical operations within quansistors are also symmetry-protected from unbiased noise in their parameters. We identify technologies that could implement $\omega$-rotation invariance. Many of our results can be generalized to higher-level $\omega$-rotation-invariant systems, or adapted to clusters with other symmetries.
\end{abstract}

\maketitle


\section{Introduction}

Quantum information theory has become a mature field of research over the last three decades, equipped with its own objectives towards quantum computation and communication~\cite{nielsen2010quantum}, as well as quantum simulation~\cite{altman2021quantum}, while at the same time allowing entirely novel perspectives on other established fields, in particular an algorithmic approach to quantum systems, a structure-of-entanglement characterization of large classes of many-body quantum states (matrix product states, tensor networks)~\cite{zeng2015quantum}, and quantum-enhanced measurements reaching the Heisenberg precision limit (quantum metrology)~\cite{giovannetti2004quantum-enchanced}. 

Quantum information processing departed from its classical counterpart with the proof that two-qubit gates~\cite{reck1994experimental, divincenzo1995twobit,Barenco1995universal} can simulate arbitrary unitary matrices, followed by the identification of `simple' quantum universal sets like single-qubit gates with $\mathsf{CNOT}$~\cite{barenco1995elementary}, and \textit{finite} quantum universal sets like Toffoli with Hadamard and $\frac{\pi}{4}$-gate~\cite{kitaev1997quantum}, or $\mathsf{SWAP}$ with \textit{almost any} two-qubit gate~\cite{deutsch1995universality,lloyd1995almost}. The deep theorem of Solovay and Kitaev showed that it is possible to translate between strictly universal sets with at most polylogarithmic overhead~\cite{kitaev1997quantum}. Alongside strict universality, encoded universality~\cite{bacon2001encoded, kempe2001encoded} and computational universality~\cite{aharonov2003simple} allow even more systems to qualify as universal quantum computers.

Circumstantial evidence suggests that quantum computers might achieve superpolynomial speedups over probabilistic classical ones. Lloyd's universal quantum simulator and Shor's algorithms for integer factorization and for discrete logarithms are prominent examples of efficient quantum solutions for problems suspected to be not computable in polynomial time classically~\cite{shor1997polynomial, lloyd1996universal}. Quantum communication protocols are provably exponentially faster than classical-probabilistic ones for specific communication complexity problems~\cite{raz1999exponential, gavinsky2007exponential}, and there exist problems that space-bounded quantum algorithms can solve using exponentially less work space than any classical algorithm~\cite{legall2006exponential}. Nonetheless, large classes of quantum tasks involving highly entangled states are efficiently simulatable classically. Quantum teleportation, superdense coding and computation using only Hadamard,  $\mathsf{CNOT}$, and measurements fall into this category according to the Gottesman-Knill theorem~\cite{gottesman1997stabilizer, nielsen2010quantum}. Fermionic linear optics with measurements, and more generally matchgate computation, are also known to be classically simulatable in polynomial time~\cite{valiant2001quantum, terhal2001classical, knill2001fermionic}. (This is in contrast to universal bosonic linear optics with measurements~\cite{knill2001scheme} and universal fermionic \textit{nonlinear} optics with measurements~\cite{bravyi2000fermionic}. The computational difference between particle-number-preserving fermions and bosons arises as a result of the easy task of computing a (Slater) determinant in the case of fermions versus the hard task ($\sharp \mathsf{P}$-complete) of computing a permanent \cite{valiant1979complexity} in the case of bosons.)

The physical realization of quantum computers and quantum communication channels is a major endeavor. Most building blocks of quantum computers are based on qubits, which are quantum two-level systems. They form the unit cells that allow us to exploit the potential of quantum information processing, when many of these qubits are coherently coupled and manipulated so as to perform various coherent quantum operations. While many different types of qubits have been developed, such as semiconductor technologies in quantum dots~\cite{loss1998quantum}, including silicon~\cite{zwanenburg2013silicon, watson2018programmable}, or  GaAs~\cite{li2018controlled}, in superconducting technologies~\cite{nakamura1999coherent,devoret2013superconducting}, in all-optical technologies~\cite{o2003demonstration}, and in hybrid technologies such as ion traps~\cite{blatt2008entangled}, cold atoms and nitrogen-vacancy centers in diamond~\cite{yao2012scalable} that require quantum systems and a laser for control. Topological technologies can also form the basis of qubits~\cite{nayak2008non}, though their experimental realization is much harder. These technologies all share the same basic principle of operating as a quantum two-level system. 

In this work we explore a quantum processing unit based on a four-level system. While there have been some earlier works on such higher-level systems, including multilevel superconducting circuits as single qu\textit{d}its and two-qubit gates~\cite{kiktenko2015multilevel,kiktenko2015single}, here we consider a special four-level system with $\omega$-rotation invariance, defined and discussed below, that we will compare to a pair of qubits in order to address one of the major challenges in quantum information processing, namely the fidelity of two-qubit operations against environment-induced effects~\cite{yan2018tunable}.

Indeed, one of the biggest obstructions for a competitive quantum computation is to keep the error rate low for each quantum operation~\cite{gottesman1998theory}. These errors can stem from the precision of the quantum manipulation, like the sweeping of a gate voltage in solid state qubits or the duration of a laser pulse in optical schemes~\cite{sanders2015bounding}. In addition, there are errors due to decoherence~\cite{shor1995scheme}. These are often considered more fundamental in the sense that they don't depend on the precision of the instrumentation but are intrinsic to the quantum system considered. They are a reflection of the coupling to the outside environment. Sources of decoherence can be leads, nuclear spins, optical absorption, phonons, and non-linearities. Most of these environments fall into the category of fermionic or bosonic baths~\cite{palma1996quantum,ischi2005decoherence}.

In our basic quantum information unit, based on a four-level system, untimely single-qubit and double-qubit unitaries will correspond to environment-induced logical errors. We will also consider the effect of external leads as the other main source of decoherence. Indeed, in solid-state-based qubits electric leads are often the main source of decoherence, particularly in superconducting qubits and semiconductor quantum dots \cite{martinis2003decoherence}. While our model is not limited to a particular implementation, we will use the coupled quantum dot geometry as an illustration of our quantum processor unit.

We will restrict ourselves to examining the fidelity and robustness of two double-qubit gates in the presence of a selected error set, and observe what appears to be an improvement in the results arising due to symmetry. {Our results are of immediate relevance to the study of noisy intermediate-scale quantum (NISQ) devices, which could realize useful versions of quantum supremacy in the very-near futur, long before fully operational fault-tolerant architectures become available~\cite{Preskill2018quantum}.} Here, we do not discuss logical error decay under the consumption of a resource, nor do we discuss fault tolerance beyond a few remarks in the Outlook.

\section{Preliminaries}
Let us consider first a physical system formed by a core of four coupled quantum dots with on-site energies $\epsilon_i$. Each quantum dot interacts with all the other dots via complex couplings (we will discuss in section~\ref{S:implementation} how it is possible to realize complex couplings physically). The corresponding isolated Hamiltonian is
\beq
H_{\text{core}}=\frac{1}{2}\sum_{i=1}^4  \epsilon_i a_i^{\dagger}a_i +\frac{1}{2}\sum_{i,j=1}^4 h_{ij} a_i^{\dagger}a_j+ \text{h.c.} 
\eeq
Each dot is now made to interact with a semi-infinite chain consisting of a semi-infinite hopping Hamiltonian with hopping parameter set to unity (thus setting the scale for all energies). The leads have scattering eigenstates with energies $-2<E<2$. Tunnel couplings between dot and chain are initially all identical and are chosen real, positive and small ($0< t_c \ll 1$). As in the case of double- and triple-dots, the Feshbach projector method shows that the effect of each lead is to modify the self-energies of the dots. In this work, we will study a similar core system formed by 4 sites (though not necessarily quantum dots) tunnel-coupled to semi-infinite leads, but \textit{with a crucial additional core symmetry}.

\begin{figure}
\centering
\includegraphics[width=8cm]{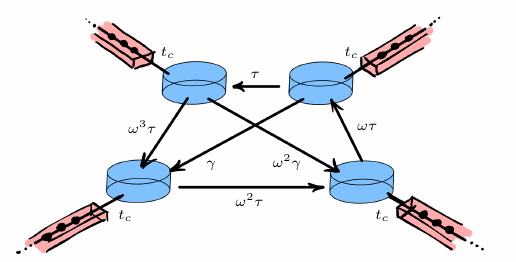}
\caption{(Color on line.) Depiction of the model, consisting of a 4-site core system (disks and dark arrows) tunnel-coupled to four identical semi-infinite leads (rectangular boxes). The core system is $\omega$-rotation-invariant (see Section \ref{S:isolated}) for any $\omega^4=1$. The tunable parameters are $\epsilon, \gamma \in \mathbb{R}$ and $\tau\in\mathbb{C}$. The hopping parameters linking the sites are generally complex. The coupling constants between the core's sites and the leads are equal to $t_c$. All hopping amplitudes along the leads are set to unity.}\label{F:model}
\end{figure}

Specifically, we will consider the single-particle sector of a class of tunable systems possessing a simple geometric symmetry, dubbed $\omega$-\textit{rotation invariance}, to be defined in the next section. A diagram of the model used throughout the paper is displayed in Fig.~\ref{F:model}.
It consists of a completely connected 4-site core system tunnel-coupled to four identical semi-infinite leads (a simple physical example being four quantum dots tunnel-coupled to semi-infinite leads). The Hamiltonian is
\beq\label{E:H_manyparticles}
\begin{aligned}
&\mathcal{H}=\mathcal{H}_{\text{core}} + \mathcal{H}_{\text{int}} + \mathcal{H}_{\text{lead}}\\
 &=\frac{1}{2}\sum_{i,j = 1}^4 h_{ij}a_i^{\dagger}a_j  +t_c  \sum_{i=1}^4 a_i^{\dagger}b_{i,1}^{\phantom{\dagger}}  + \sum_{i=1}^4 \sum_{j=1}^{\infty} b_{i,j}^{\dagger}b_{i,j+1}^{\phantom{\dagger}} +\text{ h.c.}
\end{aligned}
\eeq
restricted to the single-particle sector of Hilbert space. The core couplings $h_{ij}$ are chosen to satisfy relationships ensuring $\omega$-rotation invariance (see Section \ref{S:isolated}). The coupling between site $i$ and lead $i$ is $t_c$, which can be taken real and positive without loss of generality. The Hamiltonian has been normalized such that the hopping parameter within the semi-infinite chains is unity.
The operators $a_i$ and $b_{i,j}$ are annihilation operators, acting respectively on site $i$ of the core system, and on site $j$ of the $i$-th lead. Since we work in the single-particle sector, these operators could be fermionic or bosonic. (An example of each would be a single electron and a Cooper pair, respectively. Cold atoms can realize either choice.) Our choice of a 4-level core is motivated by our desire to describe two coupled qubits. The four semi-infinite leads simulate individual contact with the environment and enable us to reveal selective protection from decoherence. Most of our results can be generalized to an arbitrary number of sites in the core system with corresponding identical leads. The required modifications will be discussed briefly in the Outlook and in the Appendices.

\subsection{Outline of the paper}
In Section~\ref{S:isolated} we focus on the core system. We define $\omega$-rotation invariance as an obvious generalization of discrete rotation invariance, and show that the tunable parameters of an $\omega$-rotation-invariant system give full control over its eigenenergies while the energy eigenstates remain fixed. Independent control over the energy levels will be used frequently and is the main motivation for implementing $\omega$-rotation invariance. Systems with this symmetry could be realized by applying the technique of synthetic gauge fields on a tight-binding Hamiltonian~\cite{aidelsburger2018artificial}. Selecting a representative from two distinct $\omega$-classes and following the scheme of Deutsch et al.~\cite{deutsch1995universality}, we show that our 4-level core system is strictly universal for quantum computation. We then consider one possible two-qubit logical basis and discuss single-pulse logical gates as well as symmetry protection against errors, and qubit initialization and readout.

In Section~\ref{S:envcoupling} we consider the effect of the four identical leads on the core. The effective Hamiltonian of the core will in general be non-Hermitian but will remain $\omega$-rotation-invariant, and as a consequence will still allow independent energy tuning. Our ability to fully control the (potentially complex) eigenenergies will result in the possibility of transmitting an eigenstate through the leads or else of protecting it from decoherence, independently of the other eigenstates. In that sense, the 4-level core may be used as a two-qubit quantum memory unit.

Finally, in Section \ref{S:scalability} we propose a scalable scheme for universal quantum computation based on 4-level cores as the elementary computational units. The number of cores required scales linearly in the number of qubits. Because cores play a role similar to that of transistors in classical computation, we propose to call them quantum-computational transistors, or more succinctly \textit{quansistors}.

Rotation-invariant (circulant) $4\times 4$ Hamiltonians have recently been advocated~\cite{Ivanov2020two-qubit} as a way to implement the adiabatic Fourier transform on two qubits, with gate fidelities and entanglement benefitting from a symmetry that protects against decoherence. The proposal includes a possible physical implementation of circulant symmetry by tuning spin-spin interactions in ion traps. Although our work also utilizes (generalized) circulant symmetry for protection against decoherence, the aim and scope of the present article are somewhat different. We put forward a blueprint for scalable universal quantum computation based on symmetry-protected qubit clusters, with $\omega$-rotation invariance standing out as the prototype of a symmetry which is provably universal, and realistically implementable physically on a variety of platforms.

\section{Core system}\label{S:isolated}
For a $4\times 4$ matrix, we make a slight generalization of the notion of discrete rotational invariance (which can also be viewed as cyclic permutation of the sites) to $\omega$-\textit{rotation invariance}: $M$ is $\omega$-rotation-invariant if
\beq
J_{\omega}^{\dagger} M J_{\omega} = M\hspace{0.5cm},\hspace{0.5cm}\omega^4=1,
\eeq
with a modified shift matrix
\beq
J_{\omega}=\begin{pmatrix}
0 & 1 & 0 & 0\\
0 & 0 & \omega & 0\\
0 & 0 & 0 & \omega^2\\
\omega^3 & 0 & 0 & 0\\
\end{pmatrix}\hspace{0.5cm},\hspace{0.5cm}J_{\omega}^4=\omega^2 \mathbb{1}.
\eeq
Rotational invariance obviously corresponds to the case $\omega=1$. The matrices $J_1$ and $J_{e^{i\pi/2}}$, and their higher dimensional versions, have been discussed in discrete quantum mechanics under the name of Weyl's $X$ and $Y$ matrices, and in quantum information under the name of generalized Pauli $X$ and $Y$ matrices (see Appendix~\ref{S:math}). In the $4\times 4$ case we have
\beq
\begin{aligned}
X&=\begin{pmatrix}
0 & 1 & 0 & 0\\
0 & 0 & 1 & 0\\
0 & 0 & 0 & 1\\
1 & 0 & 0 & 0\\
\end{pmatrix}=J_1 \quad ,\quad  
Z=\begin{pmatrix}
1 & 0 & 0 & 0\\
0 & i & 0 & 0\\
0 & 0 & i^2 & 0\\
0 & 0 & 0 & i^3\\
\end{pmatrix}\\
Y&=ZX=J_{e^{i\pi/2}}.
\end{aligned}
\eeq
(Note that the matrix $X$ is sometimes called $X^{\dagger}$ in the literature.) Just as rotation-invariant matrices are precisely circulant matrices, $\omega$-rotation-invariant matrices correspond to $\omega$-\textit{circulant} matrices, which we will write:
\beq\label{E:omega_circ}
\text{Circ}_\omega(z_0 , \dots , z_3)=\sum_{s=0}^3 z_{s} J_{\omega}^{s},
\eeq
with $z_{s} \in \mathbb{C}$. The terms `$\omega$-rotation-invariant' and `$\omega$-circulant' will be used interchangeably. For Hermitian $\omega$-circulant $4\times 4$ matrices the number of independent \textit{real} parameters is reduced to four. Such matrices constitute what we propose to call a \textit{flat class} : mutually commuting matrices $\{H(\mathbf{g}) \mid \mathbf{g}\in\mathbb{R}^4\}$ with common eigenbasis independent of $\mathbf{g}$, and real eigenvalues $\lambda_1 (\mathbf{g}),	\dots , \lambda_4 (\mathbf{g})$ in one-to-one correspondence with the values of the parameters $\mathbf{g}\in\mathbb{R}^4$.  Each fourth root of unity $\omega$ corresponds to a flat class. (See Appendix~\ref{S:math} for details.)

We consider 4-level cores with the ability to take a nonsymmetric form (the {\em off mode\/}), and a symmetric form (the {\em computational mode\/}). In the off mode, the Hamiltonian is almost diagonal in the single-particle position eigenbasis:
\beq\label{E:H_off_mode_intro}
H_{\text{off}}=-\sum_{\langle ij \rangle} K_{ij}a_i^{\dagger}a_j + \sum_i \epsilon_i a_i^{\dagger}a_i \noquad ,
\eeq
where large energy offsets $|\epsilon_i - \epsilon_j|\gg K_{ij}>0$ effectively suppress spontaneous transitions. The logical states $\{|00\rangle , |01\rangle , |10\rangle , |11\rangle\}$ are naturally chosen to coincide with the position basis eigenstates $\{|m\rangle\}_{m=1,\dots ,4}$. We will come back to this mode later.

In the computational mode, the core system is $\omega$-rotation-invariant in the position basis for all values of its parameters. The matrix of core couplings $\{h_{ij}\}$ in \eqref{E:H_manyparticles} will thus form a class of Hermitian $\omega$-circulant Hamiltonian matrices
\beq\label{E:isolated_hamiltonian}
H^{\text{pos}}(\mathbf{g},\omega)=\sum_{s=0}^3 z_{s}(\mathbf{g}) J_{\omega}^{s} \noquad ,
\eeq
with $z_{s}\in\mathbb{C}$, $\mathbf{g}\in\mathbb{R}^4$, and $\omega^4=1$. (Recall that the complex coefficients $z_s$ are constrained by Hermiticity $H^{\text{pos}} = (H^{\text{pos}})^{\dagger}$, leaving only four real independent parameters $\mathbf{g}$.) For each $q=0,\dots ,3$ the class $\omega=e^{iq\pi/2}$ is flat and diagonalized by a modified quantum Fourier transform $\mathcal{FD}^q$, where $\mathcal{F}$ is the regular quantum Fourier transform
\beq\label{E:regularFourier}
\mathcal{F}_{k,m}=\frac{1}{2}e^{-imk(\pi /2)} \noquad ,
\eeq
and
\beq\label{E:Fourier_D}
\mathcal{D}=\text{diag}(e^{-i\pi/4},1,-e^{-i\pi/4},1)\noquad .
\eeq
The eigenstates are
\beq
|\varphi_{k}^q\rangle = \sum_m (\mathcal{FD}^{q})_{m,k}^{\dagger}|m\rangle = \frac{1}{2}\sum_m (\mathcal{D}_{m,m}^{\dagger})^q e^{imk(\pi /2)} |m\rangle
\eeq
for $q,k=0,\dots , 3$. Note that the first index of a matrix corresponds to a dual vector component, whereas the second index corresponds to a vector component :
\beq
\langle m | \varphi_k^q\rangle = (\mathcal{FD}^{q})_{m,k}^{\dagger} \noquad , \quad \langle \varphi_k^q | m\rangle = (\mathcal{FD}^{q})_{k,m}^{\phantom{\dagger}}.
\eeq
For the purpose of universal quantum computation, two classes of $\omega$-circulant Hamiltonians are necessary and sufficient: for instance, the class $X$ of circulant Hamiltonians ($\omega = 1$), and the class $Y$ of $i$-circulant Hamiltonians ($\omega = e^{i\pi/2}$). In Section~\ref{S:universality} we will build a universal set comprising only \textit{one} Hamiltonian from each class. We will now consider each of these classes in turn.

\subsection{Symmetry class $X$ ($\omega=1$)}
Class $X$ is rotation-invariant in the position eigenbasis:
\beq
H^{\text{pos}}(\mathbf{g},1)=\sum_{s=0}^3 z_{s}(\mathbf{g}) X^{s} \quad (z_{s}\in\mathbb{C},\mathbf{g}\in\mathbb{R}^4).
\eeq
The most general form of the Hamiltonian matrix is
\beq\label{E:H_position}
H^{\text{pos}}(\mathbf{g},1)=\begin{bmatrix}
\epsilon & \tau & \gamma & \phantom{*}\tau^{\dagger}\\
\phantom{*}\tau^{\dagger} & \epsilon & \tau & \gamma\\
\gamma & \phantom{*}\tau^{\dagger} & \epsilon & \tau\\
\tau & \gamma & \phantom{*}\tau^{\dagger} & \epsilon\end{bmatrix}
\eeq
with $\epsilon , \gamma \in\mathbb{R}$ and $\tau=|\tau| e^{i\theta}=\alpha +i\beta$ giving the four real parameters embodied in $\mathbf{g}$. For any value of $\mathbf{g}$ the normalized eigenstates of $H^{\text{pos}}(\mathbf{g},1)$ are
\beq\label{E:phi_k}
|\phi_k\rangle = \sum_m \mathcal{F}_{m,k}^{\dagger}|m\rangle = \frac{1}{2}\sum_m e^{imk(\pi /2)}|m\rangle
\eeq
for $k=1,\dots , 4$, with eigenenergies
\beq\label{E:bare_lambda_k}
\lambda_k = \epsilon + 2|\tau| \cos (\theta + \tfrac{k\pi}{2}) +(-1)^k  \gamma
\eeq
or
\beq\label{E:lambda_para}
\begin{array}{rrrrrrrrr}
\lambda_1 & = & \epsilon & & & - & 2\beta & - & \gamma ,\\
\lambda_2 & = & \epsilon &-  & 2\alpha &  &  & + & \gamma ,\\
\lambda_3 & = & \epsilon & & & + & 2\beta & - & \gamma ,\\
\lambda_4 & = & \epsilon &+ & 2\alpha &  &  & + & \gamma .
\end{array}
\eeq
These can be inverted, giving
\beq
\begin{array}{crrrrrrrr}
\epsilon & = & \frac{\lambda_1}{4} &  + &  \frac{\lambda_2}{4} & + &  \frac{\lambda_3}{4} & + &  \frac{\lambda_4}{4}, \\
\alpha & = &  &  - &  \frac{\lambda_2}{4} &  &  & + &  \frac{\lambda_4}{4} ,\\
\beta & = & -\frac{\lambda_1}{4} &   & & + &  \frac{\lambda_3}{4} & & ,\\
\gamma & = & -\frac{\lambda_1}{4} &  + &  \frac{\lambda_2}{4} & - &  \frac{\lambda_3}{4} & + &  \frac{\lambda_4}{4}.
\end{array}
\eeq
Any path in the $\mathbb{R}_4$ manifold of eigenenergies $(\lambda_1, \lambda_2 , \lambda_3, \lambda_4)$ of the class corresponds to a unique path in the $\mathbb{R}_4$ manifold of parameters $(\epsilon , \alpha , \beta , \gamma )$, giving full control over the energy levels of the class.

\subsection{Symmetry class $Y$ ($\omega=e^{i\pi/2}$)}
Class $Y$ is $e^{i\pi/2}$-rotation-invariant in the position eigenbasis:
\beq
H^{\text{pos}}(\mathbf{g},e^{i\pi/2})=\sum_{s=0}^3 z_{s}(\mathbf{g}) Y^{s} \quad (z_{s}\in\mathbb{C},\mathbf{g}\in\mathbb{R}^4).
\eeq
The most general form of the Hamiltonian is
\beq\label{E:H_position_i}
H^{\text{pos}}(\mathbf{g},e^{i\pi/2})=\begin{bmatrix}
\phantom{-}\epsilon & \phantom{-}\tau & -\gamma & \hspace{6pt}i\tau^{\dagger}\\
\hspace{10pt}\tau^{\dagger} & \phantom{-}\epsilon & \hspace{5pt}i\tau & \hspace{6pt}\gamma\\
-\gamma & \hspace{5pt}-i\tau^{\dagger} & \hspace{5pt}\epsilon & -\tau\\
\hspace{4pt}-i\tau & \hspace{6pt}\gamma & -\tau^{\dagger} & \hspace{6pt}\epsilon\end{bmatrix}
\eeq
where the parameters $\epsilon , \gamma \in\mathbb{R}$ and $\tau\in\mathbb{C}$ are chosen to have exactly the same form as those of \eqref{E:H_position}. The entries in class $X$ and $Y$ are seen to differ by at most a prefactor. For any value of $\mathbf{g}$ the normalized eigenstates of $H^{\text{pos}}(\mathbf{g},e^{i\pi/2})$ are
\beq\label{E:chi_k}
|\chi_k\rangle = \sum_m (\mathcal{FD})_{m,k}^{\dagger}|m\rangle = \frac{1}{2}\sum_m \mathcal{D}_{m,m}^{\dagger} e^{imk(\pi /2)} |m\rangle
\eeq
for $k=1,\dots , 4$, where the coefficients $\mathcal{D}_{m,m}^{\dagger}$ are given in Eq.~\eqref{E:Fourier_D}. The eigenenergies are
\beq\label{E:lambda_para_i}
\begin{array}{rrrrrrrrr}
\lambda'_1 & = & \epsilon &+ & \sqrt{2}\alpha& - & \sqrt{2}\beta & - & \gamma ,\\
\lambda'_2 & = & \epsilon &-  & \sqrt{2}\alpha & - & \sqrt{2}\beta & + & \gamma ,\\
\lambda'_3 & = & \epsilon &- &\sqrt{2}\alpha & + & \sqrt{2}\beta & - & \gamma ,\\
\lambda'_4 & = & \epsilon &+ & \sqrt{2}\alpha & + & \sqrt{2}\beta & + & \gamma ,
\end{array}
\eeq
which can be inverted, giving
\beq
\begin{array}{crrrrrrrr}
\epsilon & = & \frac{\lambda'_1}{4} &  + &  \frac{\lambda'_2}{4} & + &  \frac{\lambda'_3}{4} & + &  \frac{\lambda'_4}{4}, \\
\alpha & = & \frac{\lambda'_1}{4\sqrt{2}} &  - &  \frac{\lambda'_2}{4\sqrt{2}} & - &  \frac{\lambda'_3}{4\sqrt{2}} & + &  \frac{\lambda'_4}{4\sqrt{2}}, \\
\beta & = & -\frac{\lambda'_1}{4\sqrt{2}} &  - &  \frac{\lambda'_2}{4\sqrt{2}} & + &  \frac{\lambda'_3}{4\sqrt{2}} & + &  \frac{\lambda'_4}{4\sqrt{2}}, \\
\gamma & = & -\frac{\lambda'_1}{4} &  + &  \frac{\lambda'_2}{4} & - &  \frac{\lambda'_3}{4} & + &  \frac{\lambda'_4}{4}.
\end{array}
\eeq
Again, any path in the $\mathbb{R}_4$ manifold of eigenenergies $(\lambda'_1, \lambda'_2 , \lambda'_3, \lambda'_4)$ of the class corresponds to a unique path in parameter space $(\epsilon , \alpha , \beta , \gamma )$, giving full control over the energy levels of the class.

\begin{figure}
\includegraphics[]{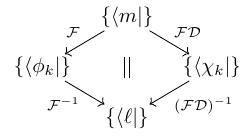}
\caption{Commutative diagram of the four working dual bases and their relationships. {The quantum Fourier transform $\mathcal{F}$ and the matrix $\mathcal{D}$ are defined in Eqns.~\eqref{E:regularFourier} and \eqref{E:Fourier_D}, respectively.} \textbf{Top:} Position basis $\{\langle m|\}$. \textbf{Middle left:} Class $X$ energy basis $\{\langle \phi_k|\}$. \textbf{Middle right:} Class $Y$  energy basis $\{\langle \chi_k|\}$. \textbf{Bottom:} Logical basis $\{\langle \ell|\}$.}\label{F:bases_diagram}
\end{figure}

Independent control over the energy levels will be used later and is a prime motivation for using $\omega$-rotation invariance, but we stress that this choice of symmetry is not unique. (See Section \ref{S:math} for details.) We can now distinguish four `natural' bases for the system, namely the position basis $\{|m\rangle\}$, the energy bases $\{|\phi_k\rangle\}$ and $\{|\chi_k\rangle\}$, and the logical basis $|\ell \rangle = \{|00\rangle,|01\rangle,|10\rangle,|11\rangle\}$, defined by identification with the position eigenstates $|m\rangle$. {Throughout, we will mostly consider the binary-code-ordered logical basis, but at times it will be convenient to consider also the Gray-code-ordered logical basis. Both bases are illustrated in Table~\ref{T:logical_bases}. Our choice of these bases is motivated by simplicity, and also by the proposal for quansistor interaction, to be discussed later.}
\begin{table}[h!]
\centering
 \begin{tabular}{c c c} 
 \hline
 Position & Binary-coded & Gray-coded \\ [0.5ex] 
 \hline\hline
 $|1\rangle$ & $|00\rangle$ & $|00\rangle$\\ 
 $|2\rangle$ & $|01\rangle$ & $|01\rangle$ \\
 $|3\rangle$ & $|10\rangle$ & $|11\rangle$ \\
 $|4\rangle$ & $|11\rangle$ & $|10\rangle$ \\ [1ex] 
 \hline
 \end{tabular}
 \caption{Two possible logical bases. Binary code will be used throughout, unless explicitly stated otherwise.}
\label{T:logical_bases}
\end{table}
The universality result discussed in the next section is independent of this choice. {However, we should point out that the choice of basis is not immaterial. Indeed, the Solovay-Kitaev theorem teaches us that simulating one universal set with another will produce, at worst, polylogarithmic overhead. And while this is considered an acceptable cost in a fault-tolerant setting, such is not the case in a NISQ setting, where the number of qubits is a severe constraint and even constant factor overheads are typically important.} The four working bases $\{|m\rangle\}, \{|\phi_k\rangle\}, \{|\chi_k\rangle\}, \{|\ell\rangle\}$, and their relationships defined in \eqref{E:phi_k},\eqref{E:chi_k}, Tab.~\ref{T:logical_bases}, are summarized in the commutative diagram of Fig.~\ref{F:bases_diagram}. {The diagram actually uses \emph{dual} bases, where an arrow like $\{\langle m |\}\overset{\mathcal{F}}{\to} \{\langle \phi_k |\}$ means $\langle \phi_k | = \sum_m \mathcal{F}_{k,m}\langle m|$. Composition of arrows agrees with conventional matrix composition~\footnote{To agree with conventional matrix composition, from right to left, the arrow corresponding to $| \phi_k \rangle = \sum_m (\mathcal{F}^{\dagger})_{m,k}| m\rangle$ should be $\mathcal{F}^{\dagger}:\{| \phi_k \rangle\}\to\{|m\rangle\}$, which is somewhat counterintuitive.}.}

\subsection{Strict universality on two qubits}\label{S:universality}
The system $H^{\text{pos}}(\mathbf{g},\omega)$ of \eqref{E:isolated_hamiltonian}, with $\omega$ equal to either 1 or $e^{i\pi/2}$, generates a strictly universal set of 2-qubit gates. In fact we prove the stronger result that the finite set $\{\mathsf{V} , \mathsf{W}\}\subset \mathbf{U}(4)$ is strictly universal, where the unitaries $\mathsf{V}$ and $\mathsf{W}$, defined below, belong to classes $X$ and $Y$, respectively. (A word about notation: Sans-serif symbols, like $\mathsf{V}$ and $\mathsf{W}$, will always denote logical gates, other more common examples being the $\pi$-phase shift $\mathsf{Z}$, the qubit flip $\mathsf{X}$ or $\mathsf{NOT}$, the Hadamard gate $\mathsf{H}$, the swapping gate $\mathsf{SWAP}$, and the controlled-not $\mathsf{CNOT}$.) We use the scheme of Ref.~\cite{deutsch1995universality} to prove our claim. We construct sixteen Hermitian $4\times 4$ matrices $H_1 , \dots , H_{16}$ whose evolution unitaries are all within our \textit{repertoire}, meaning that those unitaries can be approximated with arbitrary accuracy by repeatedly applying the gates $\mathsf{V}$ and $\mathsf{W}$. The set $\{H_1 , \dots , H_{16}\}$ is linearly independent over $\mathbb{R}$ so it spans the 16-dimensional $\mathbb{R}$-space of Hermitian $4\times 4$ matrices, which are evolved to generate all $4\times 4$ unitaries. Our repertoire therefore coincides with $\mathbf{U}(4)$, or in other words, is strictly universal on two qubits.  

We first define
\beq\label{E:H_1_universality}
\begin{aligned}
H_1	&=\tfrac{1}{2}\mathbb{1}+(\pi+i)X + X^2 +\text{h.c.}\\
	&=\begin{pmatrix}1& \pi+i& 2& \pi-i\\ \pi-i & 1 & \pi+i& 2\\ 2 & \pi-i& 1& \pi+i \\ \pi+i & 2 & \pi-i & 1\end{pmatrix},
\end{aligned}
\eeq
and the unitary $\mathsf{V}=e^{-iH_1}$, both of class $X$. We also define
\beq\label{E:Q_universality}
\tilde H=\tfrac{1}{2}\mathbb{1} + \pi(1 +\tfrac{i}{4})Y+\text{h.c.}
\eeq
and the unitary $\mathsf{W}=e^{-i\tilde H /\sqrt{2}}$, both of class $Y$. All unitaries of the form $\mathsf{V}^s=e^{-isH_1}$ for $s\in [0,2\pi)$ are in our repertoire, because integers mod $2\pi$ can be found arbitrarily close to $s$. The repertoire also comprises $\mathsf{W V W}^{\dagger}$, and more generally $\mathsf{W V}^s \mathsf{W}^{\dagger}$ for $s\in [0,2\pi)$, which are generated by the Hamiltonian
\beq
H_2=\mathsf{W} H_1 \mathsf{W}^{\dagger}.
\eeq 
(Note that whether or not $H_2$ can be obtained from the system's Hamiltonian is irrelevant. It is sufficient that the unitary $\mathsf{W V}^s \mathsf{W}^{\dagger}$  be in the repertoire for any $s$.) We finally define
\beq
\begin{aligned}
H_j &= i[H_1 , H_{j-1}] \hspace{10pt}, \hspace{10pt} j=3,\dots, 14\\
H_{15}&=i[H_2 , H_3]\\
H_{16}&=i[H_2 , H_5].
\end{aligned}
\eeq
Any unitary generated by $H_j$, for $j\in \{1,\dots ,16\}$, is in the repertoire because of the identity
\beq\label{E:commutator_in_repertoire}
e^{[P,Q]}=\lim_{n\to\infty}\left(e^{-iP/ \sqrt{n}}e^{iQ/ \sqrt{n}}e^{iP/ \sqrt{n}}e^{-iQ/ \sqrt{n}}\right)^n,
\eeq
which ultimately boils down to a sequence of $\mathsf{V}$'s and $\mathsf{W}$'s. Unitaries generated by real linear combinations of the $H_j$'s are in the repertoire as well because of the following identity for non-commuting matrices:
\beq\label{E:linear_in_repertoire}
e^{i(xP+yQ)}=\lim_{n\to\infty}\left(e^{ixP/ n}e^{iyQ/ n}\right)^n.
\eeq
To show that $\{H_1 , \dots , H_{16}\}$ is linearly independent over $\mathbb{R}$ we consider the $4\times4$ matrices $H_j$ as 16-component vectors (obtained by stacking the columns of the matrix one on top of the next from left to right), and compute the determinant of the $16\times 16$ matrix $[H_1 | \cdots | H_{16}]$ whose columns are made of these 16-component vectors. We find $\det[H_1 | \cdots | H_{16}] = P(\pi)$, where $P$ is a polynomial of high order with nontranscendental coefficients (specifically, coefficients in $\mathbb{Q}[\sqrt{2},i]$, that is to say, linear combinations of $\sqrt{2}$ and $i$ with rational coefficients). Since $\pi$ is transcendental we conclude that $P(\pi)\neq 0$ --- actually $|P(\pi)|\sim 10^{73}$ --- so $\{H_1 , \dots , H_{16}\}$ spans the space of $4\times 4$ Hermitian matrices, as required. We have thus proven that the repertoire of $\{\mathsf{V} , \mathsf{W}\}$ is all of $\mathbf{U}(4)$. 

We conclude with a few comments about the Hamiltonians $H_1, \tilde H$ chosen to generate the gates $\mathsf{V} , \mathsf{W}$ in the above construction. Firstly, these Hamiltonians were chosen to produce a sequence of matrices  $H_1 , \dots , H_{16}$ with coefficients in $\mathbb{Q}[\sqrt{2},\pi , i]$, a property used in the proof of linear independence of the $H_j$'s.  This condition is by no means \textit{necessary} for linear independence, and many sets of gates other than $\{\mathsf{V} , \mathsf{W}\}$ would qualify as universal. Secondly, the Hamiltonians $H_1, \tilde H$ were also chosen to be nondegenerate, with spectra $\{3\pm 2\pi, -3,1\}$ and $\{1\pm \tfrac{5}{2\sqrt{2}}\pi, 1\pm \tfrac{3}{2\sqrt{2}}\pi\}$, respectively, a property also shared with the off mode, Eq.~\eqref{E:H_off_mode_intro}. Nondegeneracy plays no role in the above argument, but is desirable in any physical implementation in order to avoid spurious transitions due to coupling with external degrees of freedom.

\subsection{Symmetry-protected logical operations}\label{S:bit_flip}
Now let us consider how robust our proposal is against errors. We first mention a somewhat obvious fact about \textit{parameter} noise. When a quansistor is in its symmetric form, performing a logical operation in either $\omega$-class, its eigenstates are independent of the (real) parameters $\mathbf{g}=(\epsilon, \alpha , \beta , \gamma)$ in the Hamiltonian. As a consequence, when these parameters evolve,
\beq
(\epsilon, \alpha , \beta , \gamma) \to (\epsilon (t), \alpha (t), \beta (t), \gamma (t))\noquad , \qquad t\in [0,T],
\eeq
the corresponding logical gate unitary $U(T)$ is a function of the parameters' time averages only
\beq
U(T)= U(\langle \epsilon \rangle , \langle \alpha \rangle , \langle \beta \rangle , \langle \gamma \rangle ),
\eeq
with $\langle \cdot \rangle = \frac{1}{T}\int_0^T dt (\cdot)$. This is easily seen by recognizing that $H(\epsilon, \alpha , \beta , \gamma)$ is diagonalized by a common unitary $V$ for all values of the parameters:
\beq
V^{\dagger}H(\mathbf{g})V = \text{diag}(\lambda_1 (\mathbf{g}) , \dots , \lambda_4 (\mathbf{g}) ) .
\eeq
Thus
\beq
\begin{aligned}
U(T)&=V \exp \left(-i\int dt\, V^{\dagger}H(\mathbf{g})V\right)V^{\dagger}\\
 &= V \text{diag}\left(e^{-iT\langle\lambda_1 \rangle }, \dots , e^{-iT\langle\lambda_4 \rangle}\right)V^{\dagger}.
\end{aligned}
\eeq
From~\eqref{E:lambda_para} and~\eqref{E:lambda_para_i} we get immediately that $U(T)= U(\langle \epsilon \rangle , \langle \alpha \rangle , \langle \beta \rangle , \langle \gamma \rangle )$. Accordingly, any parameter noise $\mathbf{h} (t)$ without bias, $\langle \mathbf{h}\rangle = 0$, will leave the unitary evolution operator $U(t)$ unaffected:
\beq
U(\mathbf{g}(t)+\mathbf{h}(t)) = U(\langle \mathbf{g}+\mathbf{h}\rangle)=U(\langle \mathbf{g}\rangle).
\eeq
If the quansistor interacts with the environment in such a way that the dominant effect of the latter on the quansistor is unbiased noise in the parameters, then logical operations internal to the quansistor are protected from those influences by symmetry. And if the bias has a nonzero but known value, it is easily compensated for. The argument is valid for any flat class (see Appendix~\ref{S:math}), i.e., generalizing from four states to $N$, any class of Hamiltonians of the form $V\text{diag}(\lambda_1 (\mathbf{g}) , \dots , \lambda_N (\mathbf{g}))V^{\dagger}$ for some unitary $V$, and functions $\lambda_r (\mathbf{g})=\sum_s g_s \lambda_{sr}$ with $\text{det} [\lambda_{sr}]\neq 0$. Of course, the symmetry itself, being the key ingredient here, must be enforced.

On a more interesting level, we now consider the robustness of our logical gates against genuine \textit{quantum} errors. We empirically find that the universal set $\{\mathsf{V} , \mathsf{W}\}$, defined in the previous section, is particularly resilient to small single-qubit $x$- and $z$-rotations, i.e. errors of the form
\beq\label{E:errors_X}
\mathcal{E}^{(1)}_x (\tau)=e^{-i\tau(\sigma_x\otimes\mathbb{1})}\quad , \quad \mathcal{E}^{(2)}_x (\tau)=e^{-i\tau(\mathbb{1}\otimes\sigma_x)},
\eeq
and
\beq\label{E:errors_Z}
\mathcal{E}^{(1)}_z (\tau)=e^{-i\tau(\sigma_z\otimes\mathbb{1})}\quad , \quad \mathcal{E}^{(2)}_z (\tau)=e^{-i\tau(\mathbb{1}\otimes\sigma_z)},
\eeq
for small $\tau$. For particular values $\tau_k=\frac{\pi}{2}+k\pi$, $k\in\mathbb{N}$, the unitaries $\mathcal{E}^{(1,2)}_x$ produce single-qubit flips, while $\mathcal{E}^{(1,2)}_z$ generate phase shifts, 
\beq
\begin{aligned}
&\mathcal{E}^{(1)}_x(\tau_k) \propto \mathsf{X}\otimes\mathbb{1},\quad
&&\mathcal{E}^{(2)}_x(\tau_k) \propto \mathbb{1}\otimes \mathsf{X},\\
&\mathcal{E}^{(1)}_z(\tau_k) \propto \mathsf{Z}\otimes\mathbb{1},\quad
&&\mathcal{E}^{(2)}_z(\tau_k) \propto \mathbb{1}\otimes \mathsf{Z}.
\end{aligned}
\eeq
As a first figure of merit, we have numerically evaluated the average fidelity of computational sequences belonging to the set $\{\mathsf{V} , \mathsf{W}\}$, when affected at each computational step by an error randomly chosen among $\{\mathcal{E}^{(1,2)}_{x}(\tau), \mathcal{E}^{(1,2)}_{z}(\tau)\}$, for a small \textit{fixed} value of $\tau$. (In certain situations, static imperfections are known to dominate random fluctuation errors~\cite{Frahm2004quantum}, to be considered next.) For comparison, we have repeated the same steps with two other computational sequences belonging respectively to two other strictly universal sets, namely the Kitaev set $\{\mathsf{H}\otimes \mathbb{1},\mathsf{CP}(i),\mathsf{SWAP}\}$~\cite{kitaev1997quantum}, and the set $\{\mathsf{A},\mathsf{SWAP}\}$. Here, $\mathsf{H}\otimes \mathbb{1}$ is the first-qubit Hadamard gate, and $\mathsf{CP}(i)$ is the controlled $i$-phase gate
\beq
\mathsf{H}\otimes \mathbb{1} = \frac{1}{\sqrt{2}}\begin{pmatrix}1&\phantom{-}1\\1&-1\end{pmatrix}\otimes\mathbb{1} \quad , \quad \mathsf{CP}(i)=\text{diag}(1,1,1,i).
\eeq
The $\mathsf{A}$ gate is a variant of the $\mathsf{CNOT}$ gate, and is part of a class of unitaries $A(\phi,\alpha,\theta)$ known to be strictly universal individually (in combination with the $\mathsf{SWAP}$ gate) for many values of the parameters~\cite{sleator1995realizable, Barenco1995universal, deutsch1995universality}. Specifically, the $\mathsf{A}$ gate is
\beq
\mathsf{A}=\begin{pmatrix}1&0&0&0\\0&1&0&0\\0&0&i\cos(1)&-\sin(1)\\0&0&-\sin(1)&i\cos(1)\end{pmatrix}.
\eeq

\begin{figure}
  \centering
  	\includegraphics[width=8.5cm,height=10cm,keepaspectratio]{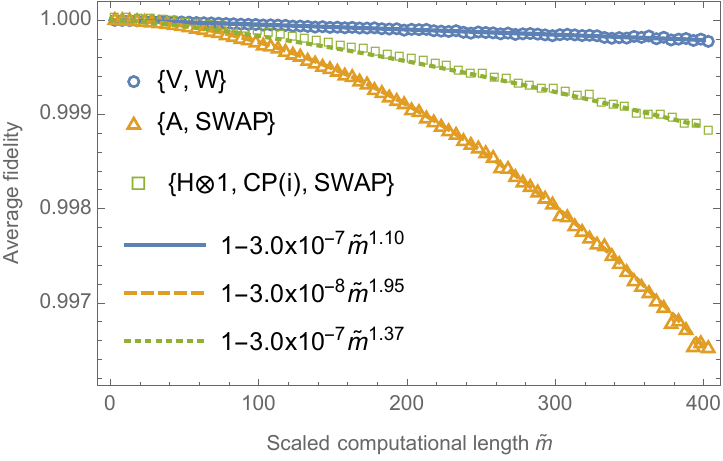}
  \caption{(Color on line.) Average fidelity $F_{\text{avg}}^{xz}$, Eq.~\eqref{E:avg_fid}, against single-qubit $x$- and $z$-rotations, Eqs.~\eqref{E:errors_X},~\eqref{E:errors_Z}, as a function of \emph{scaled} computational length $\tilde m$, for three strictly universal sets: $\{\mathsf{V} , \mathsf{W}\}$ (upper), $\{\mathsf{A},\mathsf{SWAP}\}$ (lower), and $\{\mathsf{H}\otimes \mathbb{1},\mathsf{CP}(i),\mathsf{SWAP}\}$ (middle). The coupling $\tau$ is set to $5\times 10^{-4}$. We find power-law best fits $1-F_{\text{avg}}^{xz}=1-\alpha \tilde m^{\beta}$ with respective powers $\beta_1=1.10$,  $\beta_2=1.95$,  $\beta_3=1.37$. The scaled computational length takes account of scaling effects, as discussed in the text. The maximum value $\tilde m = 403$ corresponds to sequences of length $m=403$ for $\{\mathsf{V} , \mathsf{W}\}$ and $\{\mathsf{A},\mathsf{SWAP}\}$, and $m=253$ for Kitaev.}\label{F:benchmarking_fixedXZ}
\end{figure}

Let us describe the method more precisely. For each universal set $S$ considered, and for integers $m\in [1,400]$, $r\in[1,100]$, we generate the sequence $G_{1,r}^S,\dots ,G_{m,r}^S$ of gates picked randomly from $S$ (with equal probabilities). We call $m$ the \textit{computational length}. We also generate sequences $E_{1,r},\dots ,E_{m-1,r}$ of errors picked randomly (with equal probabilities) from $\{\mathcal{E}^{(1,2)}_{x}(\tau), \mathcal{E}^{(1,2)}_{z}(\tau)\}$, with $\tau=5\times 10^{-4}$. The ideal computations are then
\beq
\mathcal{G}_{m,r}^S=\bigcirc_{i=1}^{m} G_{i,r}^S,
\eeq
while the noisy computations are 
\beq\label{E:faulty_sequence}
\widetilde{\mathcal{G}}_{m,r}^S=G_{m,r}^S\bigcirc_{i=1}^{m-1} (E_{i,r} G_{i,r}^S).
\eeq 
(The symbol ``$\bigcirc$" indicates composition from right to left.) As an illustration, a noisy computation of length $m=5$ from our universal set $\{\mathsf{V} , \mathsf{W}\}$ could be, say, $\widetilde{\mathcal{G}}=\mathsf{V}E_4\mathsf{V}E_3\mathsf{W}E_2\mathsf{V}E_1\mathsf{W}$.
The average fidelity of the $r$th computation $\widetilde{\mathcal{G}}_{m,r}^S$ being
\beq
\tilde F_{\text{avg}}^{xz}(S,m,r)=\left|\frac{1}{4}\text{Tr}(\mathcal{G}_{m,r}^{S,\dagger}\widetilde{\mathcal{G}}_{m,r}^S)\right|^2 ,
\eeq
we finally average over all computations
\beq\label{E:avg_fid}
F_{\text{avg}}^{xz}(S,m)=\frac{1}{100}\sum_{r=1}^{100}\tilde F_{\text{avg}}^{xz}(S,m,r).
\eeq
{While the simplest comparison of the three gate sets would be to plot average fidelity as a function of computational length $m$, this is not necessarily a fair comparison since the Kitaev set, consisting of three rather than two gates, can presumably approximate a given unitary operator to the desired precision in fewer gates by a factor $\log_3 2 \approx 0.63$. Accordingly, we define a \emph{scaled} computational length $\tilde m$ which is $m$ for $\{\mathsf{V} , \mathsf{W}\}$ and $\{\mathsf{A},\mathsf{SWAP}\}$, and $m\log_2 3 \approx 1.59m$ for Kitaev.} A plot of $F_{\text{avg}}^{xz}$ as a function of the scaled computational length $\tilde m$ is given in Fig.~\ref{F:benchmarking_fixedXZ} for each universal set considered, $\{\mathsf{V} , \mathsf{W}\}$, $\{\mathsf{A},\mathsf{SWAP}\}$, and $\{\mathsf{H}\otimes \mathbb{1},\mathsf{CP}(i),\mathsf{SWAP}\}$. The region shown lies within the stage of polynomial decay, and does not show the decaying exponential behavior of the saturation stage, less relevant from the point of view of quantum-coherent computation. The power-law best fit $F_{\text{avg}}^{xz}=1-\alpha \tilde m^{\beta}$ gives
\beq
F_{\text{avg}}^{xz}\approx\begin{cases}
	1-3.0\times 10^{-7} \tilde m^{1.10} & \text{ for }\{\mathsf{V} , \mathsf{W}\}\\
	1-3.0\times 10^{-8} \tilde m^{1.95} & \text{ for }\{\mathsf{A},\mathsf{SWAP}\}\\
	1-3.0\times 10^{-7} \tilde m^{1.37} & \text{ for }\{\mathsf{H}\otimes \mathbb{1},\mathsf{CP},\mathsf{SWAP}\}.
	\end{cases}
\eeq
Manifestly, the set $\{\mathsf{V} , \mathsf{W}\}$ fares much better than the other two against this type of error, with an \textit{almost-linear} decay of fidelity (depending on the degree of error anisotropy).  {When equiprobable $x$-, $y$-, and $z$-rotation errors are considered, with $x,z$ coupling $\tau = 5\times 10^{-4}$ and variable $y$ coupling $\tau_y$, we find that the advantage of $\{\mathsf{V} , \mathsf{W}\}$ over the Kitaev set narrows down with increasing $\tau_y$, vanishing at around $\tau_y / \tau = 0.55$ (not shown). And $\{\mathsf{V} , \mathsf{W}\}$ still outperforms the set $\{\mathsf{A},\mathsf{SWAP}\}$ when $\tau_y /\tau = 1$ (not shown).} This hard-$y$-axis, easy-$xz$-axes anisotropy is a non-trivial property of the set $\{\mathsf{V} , \mathsf{W}\}$. (Additional numerical results point to the special role of the $\mathsf{W}$ gate.) It is worth emphasizing that, while being more sensitive to $y$-rotations, the set $\{\mathsf{V} , \mathsf{W}\}$ outperforms the other two universal sets with respect to both $x$- and $z$-rotations.

\begin{figure}
\includegraphics[]{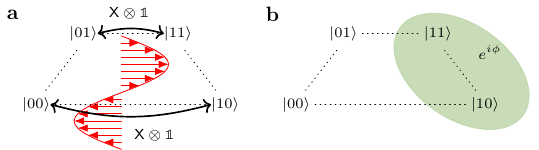}
\caption{(Color on line.) (Fractional) $\mathsf{X}\otimes\mathbb{1}$ and $\mathsf{Z}\otimes\mathbb{1}$ errors in quantum dots with Gray-code logical basis. \textbf{a} Fractional $\mathsf{X}\otimes\mathbb{1}$ induced by an environmental thermal photon. \textbf{b} Fractional $\mathsf{Z}\otimes\mathbb{1}$ induced by the presence of a resonator near the right-hand side of the square. States $|11\rangle$ and $|10\rangle$ acquire an extra phase factor $e^{i\phi}$.}\label{F:gray_error}
\end{figure}

{Let us also mention that the selected type of noise is by no means exhaustive, and was primarily chosen for ease of comparison with more common, qubit-based universal sets. It \emph{could} nevertheless be realistic in certain implementations. For example, in a square of charge quantum dots with Gray-code logical basis (see Fig.~\ref{F:gray_error}), a thermal photon could stimulate tunnelling events along the sides of the square parallel to polarization, increasing the likeliness of the corresponding $x$-rotation, of which $\mathsf{X}\otimes\mathbb{1}$ and $\mathbb{1}\otimes\mathsf{X}$ are particular instances. In the same setup, the presence of a resonator near one side of the square (as discussed in Sec.\ref{S:architecure}) could modify the effective self-energies of the two closest dots. In a first-order treatment, the corresponding logical states would be affected by an identical phase factor, resulting in the dephasing of parallel sides of the square, i.e. a $z$-rotation, of which $\mathsf{Z}\otimes\mathbb{1}$ and $\mathbb{1}\otimes\mathsf{Z}$ are particular instances. 

We should add that, through the technique of circuit randomization, coherent Markovian noise can be tailored into effective stochastic Pauli noise with the same error rate~\cite{Wallman2016noise,Ware2021experimental,Hashim2021randomized}. Coherent error rate, although not explicitly considered here, is well quantified by gate fidelity against stochastic Pauli errors. The technique was also experimentally observed to largely suppress signatures of non-Markovian errors~\cite{Ware2021experimental}.} 

\begin{figure}
  \centering
  	\includegraphics[width=8.5cm,height=10cm,keepaspectratio]{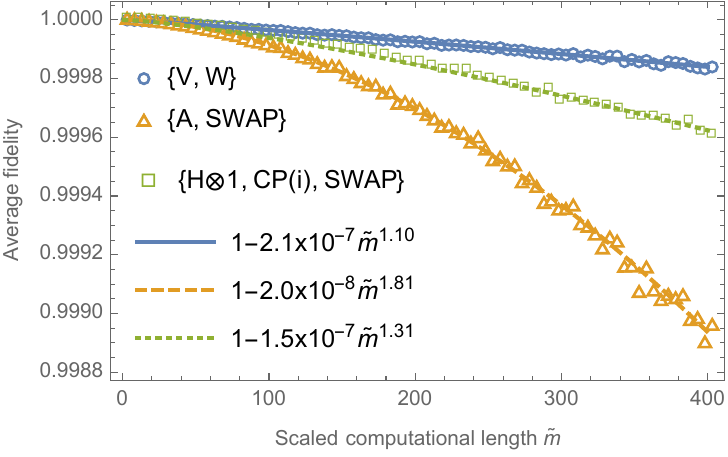}
  \caption{(Color on line.) Average fidelity $F_{\text{avg}}^{xyz}$, Eq.~\eqref{E:avg_fid_gaussian}, against the equiprobable Pauli error-set~\eqref{E:pauli_basis}, as a function of \emph{scaled} computational length $\tilde m$, for three strictly universal sets: $\{\mathsf{V} , \mathsf{W}\}$ (upper), $\{\mathsf{A},\mathsf{SWAP}\}$ (lower), and $\{\mathsf{H}\otimes \mathbb{1},\mathsf{CP}(i),\mathsf{SWAP}\}$ (middle). The couplings $\tau_a$ are normally distributed with standard deviation $10^{-4}$. The means are $\langle\tau_x\rangle = \langle\tau_z\rangle= 10^{-3}$, and $\langle\tau_y\rangle = 10^{-4}$. We find power-law best fits $1-F_{\text{avg}}^{xyz}=1-\alpha \tilde m^{\beta}$ with respective powers $\beta_1=1.10$,  $\beta_2=1.81$,  $\beta_3=1.31$. The scaled computational length takes account of scaling effects, as discussed in the text. The maximum value $\tilde m = 403$ corresponds to sequences of length $m=403$ for $\{\mathsf{V} , \mathsf{W}\}$ and $\{\mathsf{A},\mathsf{SWAP}\}$, and $m=253$ for Kitaev.}\label{F:benchmarking_gaussianXYZ}
\end{figure}

As a second figure of merit, let us consider fidelity against a fluctuating noise corresponding to the larger error set
\beq\label{E:pauli_basis}
\begin{aligned}
\mathcal{E}^{(1)}_{a}&=e^{-i\tau_a(\sigma_a\otimes\mathbb{1})}\\
\mathcal{E}^{(2)}_{b} &=e^{-i\tau_b(\mathbb{1}\otimes\sigma_b)}\\
\mathcal{E}^{\phantom{(1)}}_{ab} &=e^{-i\tau_a\tau_b(\sigma_a\otimes\sigma_b)},
\end{aligned}
\eeq
for $a,b\in\{x,y,z\}$, and normally distributed couplings $\tau_a , \tau_b$. The 15 corresponding generators constitute, along with the identity, a basis for all $4\times 4$ Hamiltonians. For now, these 15 errors are picked with equal probability $1/15$ to generate the noisy computations of Eq.~\eqref{E:faulty_sequence}, and the average fidelity 
\beq\label{E:avg_fid_gaussian}
F_{\text{avg}}^{xyz}(S,m)=\frac{1}{900}\sum_{r=1}^{900}\tilde F_{\text{avg}}^{xyz}(S,m,r)
\eeq
is evaluated as a function of computational length $m$. Here, $r$ enumerates 30 random generations of $\tau_x, \tau_y, \tau_z$ times 30 random noisy sequences for each generation. {In Fig.~\ref{F:benchmarking_gaussianXYZ} , we plot $F_{\text{avg}}^{xyz}$ as a function of the \emph{scaled} computational length $\tilde m$, defined below Eqn.~\eqref{E:avg_fid}.} If the interactions with the environment are such as to produce a stronger bias on $x$- and $z$-rotations, then the set $\{\mathsf{V} , \mathsf{W}\}$ is at an advantage. This is seen in Fig.~\ref{F:benchmarking_gaussianXYZ} where $\tau_x$ and $\tau_z$ have average $\langle\tau_x\rangle=\langle\tau_z\rangle=10^{-3}$, while $\tau_y$ has average one order smaller, $\langle\tau_y\rangle=10^{-4}$. All standard deviations are equal to $10^{-4}$. The power-law best fit $F_{\text{avg}}^{xyz}=1-\alpha \tilde m^{\beta}$ gives
\beq
F_{\text{avg}}^{xyz}\approx\begin{cases}
	1-2.1\times 10^{-7}  \tilde m^{1.10} & \text{ for }\{\mathsf{V} , \mathsf{W}\}\\
	1-2.0\times 10^{-8}  \tilde m^{1.81} & \text{ for }\{\mathsf{A},\mathsf{SWAP}\}\\
	1-1.5\times 10^{-7}  \tilde m^{1.31} & \text{ for }\{\mathsf{H}\otimes \mathbb{1},\mathsf{CP},\mathsf{SWAP}\}.
	\end{cases}
\eeq
\begin{figure}
  \centering
  	\includegraphics[width=8.5cm,height=10cm,keepaspectratio]{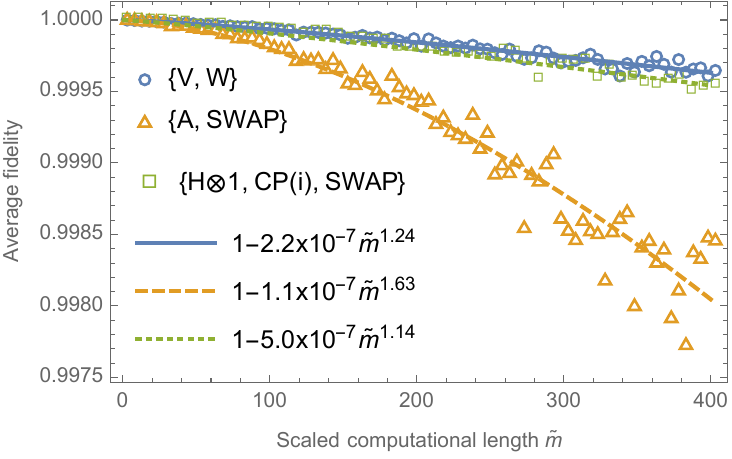}
  \caption{(Color on line.) Average fidelity $F_{\text{avg}}^{xyz}$, Eq.~\eqref{E:avg_fid_gaussian_2}, against the Pauli error-set~\eqref{E:pauli_basis}, as a function of \emph{scaled} computational length $\tilde m$, for three strictly universal sets: $\{\mathsf{V} , \mathsf{W}\}$ (upper), $\{\mathsf{A},\mathsf{SWAP}\}$ (lower), and $\{\mathsf{H}\otimes \mathbb{1},\mathsf{CP}(i),\mathsf{SWAP}\}$ (middle, just below upper). Errors without a $y$-rotation have probability $3/31$; other errors have probability $1/31$. The couplings $\tau_a$ are normally distributed without bias, and with standard deviation $10^{-3}$. The scaled computational length takes account of scaling effects, as discussed in the text. The maximum value $\tilde m = 403$ corresponds to sequences of length $m=403$ for $\{\mathsf{V} , \mathsf{W}\}$ and $\{\mathsf{A},\mathsf{SWAP}\}$, and $m=253$ for Kitaev.}\label{F:benchmarking_gaussianXYZ_2}
\end{figure}\\
For this degree of error anisotropy, the set $\{\mathsf{V} , \mathsf{W}\}$ presents an almost-linear decay of fidelity. 

As our third and last figure of merit, we once more consider fidelity against a fluctuating noise corresponding to the error set~\eqref{E:pauli_basis}, but we now assume that interactions with the environment are such as to make the system more prone to $x$- and $z$-rotations. For definiteness, the 8 errors $\mathcal{E}^{(1)}_{a}, \mathcal{E}^{(1)}_{a}, \mathcal{E}^{\phantom{(1)}}_{ab}$, for $a,b\in\{x,z\}$, are picked randomly with probability $3/31$, while the 7 remaining errors, each containing at least one $y$-rotation, are picked with probability $1/31$. The couplings $\tau_a$ are now identically distributed without bias, $\langle\tau_a\rangle=0$, and with standard deviation $10^{-3}$. The average fidelity is
\beq\label{E:avg_fid_gaussian_2}
F_{\text{avg}}^{xyz}(S,m)=\frac{1}{2500}\sum_{r=1}^{2500}\tilde F_{\text{avg}}^{xyz}(S,m,r),
\eeq
where $r$ enumerates 50 random generations of $\tau_x, \tau_y, \tau_z$ times 50 random noisy sequences for each generation. {In Fig.~\ref{F:benchmarking_gaussianXYZ_2}, we plot $F_{\text{avg}}^{xyz}$ as a function of the \emph{scaled} computational length $\tilde m$, defined below Eqn.~\eqref{E:avg_fid}.} The power-law best fit $F_{\text{avg}}^{xyz}=1-\alpha \tilde m^{\beta}$ gives
\beq
F_{\text{avg}}^{xyz}\approx\begin{cases}
	1-2.2\times 10^{-7} \tilde m^{1.24} & \text{ for }\{\mathsf{V} , \mathsf{W}\}\\
	1-1.1\times 10^{-7} \tilde m^{1.63} & \text{ for }\{\mathsf{A},\mathsf{SWAP}\}\\
	1-5.0\times 10^{-7} \tilde m^{1.14} & \text{ for }\{\mathsf{H}\otimes \mathbb{1},\mathsf{CP},\mathsf{SWAP}\}.
	\end{cases}
\eeq
In spite of the fact that the Kitaev set has a smaller $\beta$ exponent than $\{\mathsf{V} , \mathsf{W}\}$, {we find that $\{\mathsf{V} , \mathsf{W}\}$ is at an advantage, up to $\tilde m \sim 3500$, in the presence of a hard-$y$-axis anisotropy.} Evaluating the performance of $\{\mathsf{V} , \mathsf{W}\}$ on a larger error-set, generated by linear combinations of Pauli tensors, is the object of a future work. {The use of \emph{fractional} $\mathsf{V}$ and $\mathsf{W}$ operations is also likely to help reducing errors in variational algorithms, where small-angle rotations typically abound. Error-divisible gates implement these small-angle operations directly, without using long, noisy sequences of full rotations~\cite{Perez2021error-divisible}.}

Although we have been concerned with the short-$\tilde m$ stage of polynomial decay, it should be mentioned that for larger $\tilde m$, some of the curves plotted in Figs.~\ref{F:benchmarking_fixedXZ},~\ref{F:benchmarking_gaussianXYZ},~\ref{F:benchmarking_gaussianXYZ_2} present fidelity revivals (``echoes" in the Loschmidt echo~\cite{peres1984stability, Cucchietti2005decoherence, Wisniacki2012loschmidt}, not shown) before reaching the large-$\tilde m$ saturation stage.

\begin{figure}
  \centering
  	\includegraphics[width=8.5cm,height=10cm,keepaspectratio]{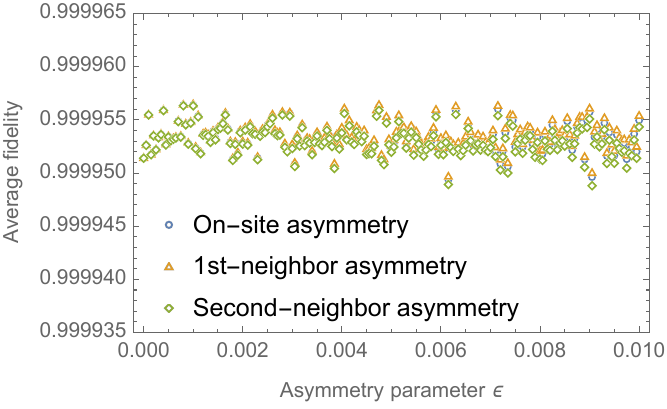}
  \caption{(Color on line.) Average fidelity $F_{\text{avg}}^{xz}$ against single-qubit $x$- and $z$-rotations as a function of asymmetry parameter $\epsilon$, for on-site energy asymmetry $A_0$, nearest-neighbor asymmetry $A_1$, and second-nearest-neighbor asymmetry $A_2$. ($A_0, A_1$, and $A_2$ are defined in Eqns.~\eqref{E:asym0}, \eqref{E:asym1_2}.) The computational length is $m=100$. The average is obtained over 200 iterations.}\label{F:asym_012}
\end{figure}

{For implementation purposes in realistic, non-ideal platforms, it is important to understand the effect of slightly breaking the $\omega$-invariance symmetry of the set $\{\mathsf{V} , \mathsf{W}\}$. For simplicity's sake, we consider once more four quantum dots arranged in a square, and perform fidelity simulations in the presence of a systematic asymmetry $A$ in the Hamiltonians. Specifically, $\omega$-invariant Hamiltonians $H$ are replaced with $H+A$ for on-site energy asymmetry, 
\beq\label{E:asym0}
A_0=\begin{pmatrix}\epsilon & 0 & 0 & 0\\
	0 & 0 & 0 & 0\\
	0 & 0 & 0 & 0\\
	0 & 0 & 0 & 0\end{pmatrix} \quad , \quad \epsilon\geq 0,
\eeq
and asymmetry in nearest-neighbor coupling strengths, and in second-nearest-neighbor coupling strengths, respectively
\beq\label{E:asym1_2}
A_1=\begin{pmatrix}0 & \epsilon & 0 & 0\\
	\epsilon & 0 & 0 & 0\\
	0 & 0 & 0 & 0\\
	0 & 0 & 0 & 0\end{pmatrix} \quad , \quad 
	A_2=\begin{pmatrix}0 & 0 & \epsilon & 0\\
	0 & 0 & 0 & 0\\
	\epsilon & 0 & 0 & 0\\
	0 & 0 & 0 & 0\end{pmatrix}.
\eeq 
For each asymmetry type, average fidelity $F_{\text{avg}}^{xz}$ against single-qubit $x$- and $z$-rotations is plotted as a function of $\epsilon$ in Fig.~\ref{F:asym_012}. We detect no singular effect of the asymmetries, with fluctuations well within $10^{-5}$, and a minor dependence on $\epsilon$.}

\subsection{Gauge potentials for $\omega$-classes}\label{S:gauge}
\begin{figure}
\includegraphics[width=8.7cm]{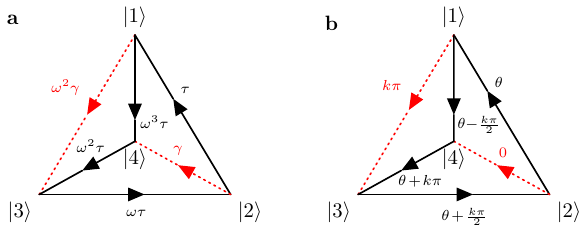}
\caption{(Color on line.) Core displayed as either planar or tetrahedral. The parameters are as in Fig.~\ref{F:model}, with $\tau = |\tau| e^{i\theta}$, $\gamma\in\mathbb{R}$, and $\omega = e^{ik\pi/2}$. \textbf{a} Transition amplitudes. The part of the Hamiltonian in $\text{span}(J_{\omega}, J_{\omega}^3)$ is proportional to $\tau$ (solid black). The part of the Hamiltonian in $\text{span}(J_{\omega}^2)$ is proportional to $\gamma$ (dotted red). \textbf{b} Gauge phases on links for $\gamma >0$.  If  $\gamma\in\mathbb{R}^{<0}$, there is an additional phase of $\pi$ on $\theta_{13}$ and $\theta_{24}$ (dotted red). If $\gamma=0$, then both $\theta_{13}$ and $\theta_{24}$ vanish.}\label{F:model_tetrahedral}
\end{figure}

\begin{figure}
\includegraphics[width=8.7cm]{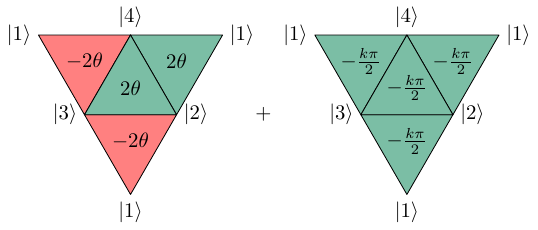}
\caption{(Color on line.) Flux structure modulo $2\pi$ for $H\in\text{span}(J_{\omega}, J_{\omega}^3)$. The parameters are $\tau = |\tau| e^{i\theta}$, and $\omega = e^{ik\pi/2}$. The first structure depends on the coupling $\tau$. The second structure is topological, and depends only on the class $\omega$.}\label{F:flux_structure}
\end{figure}

\begin{figure}
\includegraphics[]{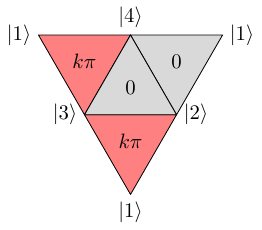}
\caption{(Color on line.) Flux structure modulo $2\pi$ for $H\in\text{span}(J_{\omega}^2)$. The parameters are $\gamma\in\mathbb{R}^{> 0}$, and $\omega = e^{ik\pi/2}$. If $\gamma <0$, there is an additional flux $\pi$ in each plaquette. If $\gamma=0$, all fluxes vanish.}\label{F:flux_structure2}
\end{figure}

We now provide a ``lattice gauge field" description of $\omega$-rotation invariance. In Fig.~\ref{F:model_tetrahedral} the core is displayed so that it can be visualized as either planar (as shown) or tetrahedral (by raising the central point $|4\rangle$). The complex hopping parameters linking the sites (the \textit{link variables}) have the form $h_{jk}=|h_{jk}|e^{i\theta_{jk}}$. In the lattice picture, vertices $|j\rangle$ stand for the matter field, and the phases on the links $|j\rangle \overset{\theta_{jk}} {\longrightarrow} |k\rangle$ correspond to a $\mathbf{U}(1)$ gauge potential $\theta_{jk}=\int_j^k \mathbf{A}\cdot d\mathbf{r}$. Paths around elementary triangular plaquettes yield gauge-invariant plaquette fluxes,
\beq
\Phi_j =\oint_{\bigtriangleup_j} \mathbf{A}\cdot d\mathbf{r} \noquad ,
\eeq
which may be written $\oint_{\bigtriangleup_j} \mathbf{A}\cdot d\mathbf{r}=\iint_{\triangle_j}(\nabla\times\mathbf{A})\cdot d\mathbf{s}$. A Hermitian Hamiltonian has $\theta_{jk}=-\theta_{kj}$, and it is straightforward to check directly that the field $\mathbf{B}=\nabla\times\mathbf{A}$ is divergence-free in the tetrahedron,
\beq
\sum_j\Phi_j = 0.
\eeq
As a consequence, only three plaquette fluxes are linearly independent, and the planar and tetrahedral models are completely equivalent. (The 4-level quansistor is essentially 2-dimensional. This is in contrast to higher-level quansistors, which are intrinsically higher-dimensional, as explained in the Outlook.) Of the six gauge phases $\theta_{jk}$, $j<k$, three are independent and generate a manifold of Hamiltonians for each given flux structure. It is convenient to distinguish $\omega$-circulant Hamiltonians by their flux structure or ``magnetic" field, whether fundamental or synthetic. Hamiltonians with different flux structures belong to gauge-inequivalent classes, and are measurably different. Fig.~\ref{F:model_tetrahedral} displays (a)~transition amplitudes, and (b)~gauge phases. As before, the parameters are $\tau = |\tau| e^{i\theta}$, $\gamma\in\mathbb{R}$, and $\omega = e^{ik\pi/2}$.
The corresponding flux structures (modulo $2\pi$) are represented in Figs.~\ref{F:flux_structure} and~\ref{F:flux_structure2}. For any Hamiltonian of class $X$, the flux structure is as in the left diagram of Fig.~\ref{F:flux_structure}. Therefore, the topological flux structure $\Phi_j\equiv -\frac{\pi}{2}$ observed in the right diagram of Fig.~\ref{F:flux_structure}, characteristic of Hamiltonians $H\in\text{span}(Y , Y^3)$, cannot be realized by any Hamiltonian of class $X$. Symbolically, $\text{span}(X , X^2, X^3)\ncong \text{span}(Y , Y^3)$. Similarly, the topological flux structure observed in Fig.~\ref{F:flux_structure2} cannot be realized by Hamiltonians of class $X$, hence $\text{span}(X , X^2, X^3)\ncong \text{span}(Y^2)$. On the other hand, by combining the diagrams of Figs.~\ref{F:flux_structure} and~\ref{F:flux_structure2} we see that $\text{span}(X , X^2, X^3)\cong \text{span}(Y , Y^2, Y^3)$. Indeed, a matrix of class $X$ with $X^1$ coefficient $\tau=|\tau|e^{i\theta}$ is gauge-equivalent to a matrix of class $Y$ with $Y^1$ coefficient $\tau' = |\tau'| e^{i\theta'}$ if and only if $\theta\equiv \theta' -\pi/4$ mod $2\pi$. In particular, the Hamiltonians $H_1$ and $\tilde H$ from the universality proof, Eqs.~\ref{E:H_1_universality} and~\ref{E:Q_universality}, belong to inequivalent flux structures (although it can be shown that this inequivalence is not generic).

\subsection{Physical implementation}\label{S:implementation}
In the previous section, we have argued that Hamiltonians from different $\omega$-classes may have different flux structures, with three linearly independent plaquette fluxes. They could therefore be realized by applying magnetic fields onto 2-dimensional or 3-dimensional charged systems with initial Hamiltonians in the form of the off-mode Hamiltonian, Eq.~\eqref{E:H_off_mode_intro}. The topological (rightmost) flux structure from Fig.~\ref{F:flux_structure}, for instance, could be produced from a very long and thin solenoid penetrating a tetrahedron through one face, and isotropically releasing a flux of $2\pi$ at the center of the tetrahedron. This flux structure properly belongs to class $Y$, and cannot be realized in class $X$.

In this section we sketch how the classes $X$ and $Y$ could be implemented in a wide range of physical systems, comprised of either charged or neutral levels, using the techniques of synthetic gauge fields. The appearance of gauge structures in systems with parameter-dependent Hamiltonians~\cite{berry1984quantal,simon1983holonomy} or time-periodic Hamiltonians~\cite{sorensen2005fractional} is well known. In the former case, and when the adiabatic approximation holds, the dynamics of an adiabatically evolving particle can be projected onto the subspace spanned by the $m$th eigenstate $\psi_m (\mathbf{R}(t))$. The resulting effective Schr{\"o}dinger equation for $\psi_m (\mathbf{R}(t))$ involves a \textit{Berry connection} $\mathbf{A(\mathbf{R})}$ playing the role of a gauge potential, through the substitution $\mathbf{p}\to\mathbf{p}-\mathbf{A}$ in the effective Hamiltonian, or equivalently, as a geometric phase $\exp i\int d\mathbf{R}\cdot \mathbf{A}$ acquired by $\psi_m (\mathbf{R}(t))$ over the displacement. This has been shown to occur in mechanical systems~\cite{wilczek1984appearance}, molecular systems~\cite{mead1992geometric}, and condensed matter systems~\cite{xiao2010berry}. Similarly, for systems driven by fast time-periodic modulations (Floquet engineering), one may consider the evolution at stroboscopic times $t_N = NT$, where $T$ is the driving period~\cite{grifoni1998driven,aidelsburger2018artificial}. Here again, the resulting effective dynamics has been shown to yield non-trivial gauge structures in different platforms such as condensed matter systems~\cite{dunlap1986dynamic,castro-neto2007electronic}, photonics~\cite{hafezi2013synthetic,ozawa2019topological}, ultracold atoms in optical lattices~\cite{anderson2011synthetic,struck2012tunable,hauke2012non-abelian,aidelsburger2013realization,galitski2013spin-orbit,galitski2019artificial}, and ions in micro-fabricated traps~\cite{bermudez2011synthetic,bermudez2012photon-assisted}.  In a lattice with coordination number $d$, nearest-neighbor hopping terms
$K_{\mathbf{m},\mathbf{m}+\mathbf{u}}|\mathbf{m}\rangle\langle \mathbf{m}+\mathbf{u}|$ act on wavefunctions as
\beq
\psi(\mathbf{m}) \to K_{\mathbf{m},\mathbf{m}+\mathbf{u}}\psi(\mathbf{m}+\mathbf{u})=K_{\mathbf{m},\mathbf{m}+\mathbf{u}}e^{-i\mathbf{u}\cdot\mathbf{p}}\psi(\mathbf{m}),
\eeq
where naturally $\mathbf{p}$ is the momentum operator and $\mathbf{u}$ is a vector of unit norm in $\mathbb{Z}^d$. In the presence of an effective gauge potential $\mathbf{A}(\mathbf{m})$, the \textit{Peierls substitution} $\mathbf{p}\to\mathbf{p}-\mathbf{A}(\mathbf{m})$ amounts to the complexification of real hopping parameters
\beq
K_{\mathbf{m},\mathbf{m}+\mathbf{u}} \to K_{\mathbf{m},\mathbf{m}+\mathbf{u}}e^{i\mathbf{u}\cdot\mathbf{A}(\mathbf{m})} =
K_{\mathbf{m},\mathbf{m}+\mathbf{u}}e^{i\theta_{\mathbf{m},\mathbf{m}+\mathbf{u}}}.
\eeq
The Peierls phases $\theta_{\mathbf{m},\mathbf{m}+\mathbf{u}}$ may also depend on internal degrees of freedom (pseudospin) and can then be thought of as resulting from an artificial or synthetic non-abelian gauge field~~\cite{aidelsburger2018artificial}.
For the implementation of the classes $X$ and $Y$, we need to realize the gauge-invariant flux structures described in Section~\ref{S:gauge}, whether fundamental or artificial. One possibility is to Floquet engineer Peierls phases as in the Hamiltonians~\eqref{E:H_position} and~\eqref{E:H_position_i}. In the former, we have Peierls phases $\theta_{j,j+1}\equiv \theta$, and all others zero. In the latter, we have instead $\theta_{j,j+1}=\theta + (\pi/2)^{j-1}$ and $\theta_{13}=\pi$, and all others zero.

In~\cite{hauke2012non-abelian}, for instance, lattice shaking is used to prompt a fast periodic modulation of the on-site energies of a tight-binding Hamiltonian analogous to our off-mode Hamiltonian, Eq.~\eqref{E:H_off_mode_intro}: 
\beq
H(t)=-\sum_{\langle ij \rangle} K_{ij}a_i^{\dagger}a_j + \sum_i (\epsilon_i + v_i (t))a_i^{\dagger}a_i
\eeq
where $K_{ij}>0$, $v_i (t) = v_i (t+T)$, and $\langle v_i\rangle_T=\frac{1}{T}\int_0^T dt\; v_i(t) = 0$. Using Floquet analysis, the resulting effective time-independent Hamiltonian proves to be of the form
\beq
H_{\text{eff}}=-\sum_{\langle ij \rangle} |K_{ ij }^{\text{eff}}|e^{i\theta_{ij}}a_i^{\dagger}a_j + \epsilon^{\text{eff}}\sum_i  a_i^{\dagger}a_i \noquad ,
\eeq
with complex tunneling amplitudes
\beq
|K_{ ij }^{\text{eff}}|e^{i\theta_{ij}}=K_{ij}\langle e^{i(w_j-w_i)/\hbar}\rangle_T \noquad ,
\eeq
where $w_i(t)=-\int_{t_0}^t dt' v_i(t') + \langle \int_{t_0}^t dt' v_i(t') \rangle_T$. As long as the driving functions break certain symmetries, the Peierls phases can be varied smoothly to any value between 0 and $2\pi$. Producing non-trivial Peierls phases that cannot be gauged away may require additional static structure, like large energy offsets $|\epsilon_j - \epsilon_i|  \gg K_{ij}$~\cite{hauke2012non-abelian}. In our setup, these large energy offsets are already present in the off-mode Hamiltonian to effectively suppress spontaneous transitions between logical states (position eigenstates).

\section{Coupling to leads}\label{S:envcoupling}
We now consider the effect of the semi-infinite leads on the core system. As indicated in the Hamiltonian~\eqref{E:H_manyparticles} and in Fig.~\ref{F:model}, each site (for example, a quantum dot) is tunnel-coupled to its own lead (which could be, for example, a semi-infinite spin chain) but the parameters and coupling constants of the four leads are chosen to be identical. The transition amplitudes in the leads are set to unity, and the lead-to-site coupling $t_c$ can be chosen real and positive with no loss of generality. Coupling the core to the leads may serve to model the core's immersion in its immediate environment, and that is the point of view adopted in Section~\ref{S:protected_gates}. Alternatively, the leads may represent designed transmission wires between the core and distant devices. This perspective is explored in Section~\ref{S:leads_transmission}.

 It is shown in Appendix~\ref{S:effective_H_appendix} that the effect of the leads on the core Hamiltonian~\eqref{E:isolated_hamiltonian} can be summarized in an effective, energy-dependent diagonal offset:
\beq\label{E:H_pos_E_omega}
H_{\infty}(E)= H^{\text{pos}}(\mathbf{g},\omega) +t_c^2\Sigma (E) \mathbb{1},
\eeq
where $\Sigma (E)$ is the surface Green's function of a semi-infinite lead:
\beq
\Sigma (E) =\frac{E}{2}-\frac{\sqrt{(E+i0^+)^2 - 4}}{2}\; .
\eeq
It follows that $\omega$-circulation is preserved. For instance, when $\omega=1$ (class $X$) we have the circulant effective Hamiltonian
\beq
H_{\infty}(E)=\begin{bmatrix}
\epsilon_{\infty}(E) & \tau & \gamma & \phantom{*}\tau^{\dagger}\\
 \phantom{*}\tau^{\dagger} & \epsilon_{\infty}(E) & \tau & \gamma\\
\gamma &  \phantom{*}\tau^{\dagger} & \epsilon_{\infty}(E) & \tau\\
\tau & \gamma &  \phantom{*}\tau^{\dagger} & \epsilon_{\infty}(E)\end{bmatrix}
\eeq
with effective self-energies
\beq\label{E:epsilon_complexB}\begin{aligned}
\epsilon_{\infty} (E)&=\epsilon + t_c^2\Sigma (E)\\
	&= \epsilon + t_c^2\left(\frac{E}{2}-\frac{\sqrt{(E+i0^+)^2 - 4}}{2}\right).
\end{aligned}\eeq
Because $\omega$-rotation invariance is preserved, the eigenstates are \textit{energy-independent}, and still given by~\eqref{E:phi_k}. The corresponding effective eigenvalues are obtained from the isolated levels $\lambda_k$,  \eqref{E:bare_lambda_k}, by the replacement $\epsilon\to\epsilon_{\infty}(E)$:
\beq\label{E:lambda_E_phi}
\begin{aligned}
\lambda_{k,\infty}(E) &= \epsilon_{\infty}(E) + 2|\tau| \cos (\theta+\tfrac{k\pi}{2}) +(-1)^k \gamma\\
				 &=t_c^2\Sigma (E) +\lambda_k
\end{aligned}
\eeq
for $k=1,\dots , 4$. But these are not effective eigenenergies as can be seen from the Green's function:
\beq\label{E:Green_E_phi}
G_{\text{core}}(E)=(E-H_{\infty}(E))^{-1}=\sum_k \frac{|\phi_k\rangle\langle\phi_k |}{E-\lambda_{k,\infty}(E)}\noquad ,
\eeq
the effective energy levels of the core-with-leads are fixed points  $E_k^{\star} = \lambda_{k,\infty}(E_k^{\star})$. (From now on the symbol ``$\star$'' will always indicate an effective energy due to the presence of the leads.) From \eqref{E:epsilon_complexB}, and the convention used for the definition of the complex square root,
\beq\label{E:fx_pts_E_1}
\begin{aligned}
E_k^{\star}&=\frac{1}{1-t_c^2}\left[\left(1-\tfrac{t_c^2}{2}\right)\lambda_k- \tfrac{t_c^2}{2}\sqrt{\lambda_k^2-4(1-t_c^2)}\right]\\
	&=\frac{1}{1-t_c^2}\left[\left(1-\tfrac{t_c^2}{2}\right)\lambda_k- i\tfrac{t_c^2}{2}\sqrt{4(1-t_c^2)-\lambda_k^2}\right] .
\end{aligned}
\eeq
Since each $k$-mode is decoupled from the others, we have chosen $\lambda_k\geq 0$ with no loss of generality. Eq.~\eqref{E:fx_pts_E_1} and the corresponding expression for negative $\lambda_k$ is obtained in Appendix~\ref{S:solution_dot_lead_dot_appendix} by analytically solving the core-with-leads Schr{\"o}dinger equation.  Because the effective eigenstates do not depend on the scattering energy, it is easy to define a first-order effective core Hamiltonian which is energy-independent
\beq\label{E:Green2_E_phi}
G_{\text{core}}(E)\approx (E-H_{\text{eff}})^{-1}=\sum_k \frac{|\phi_k\rangle\langle\phi_k |}{E-E_k^{\star}}\noquad .
\eeq
In the ordered eigenbasis $\{|\phi_1\rangle,|\phi_2\rangle,|\phi_3\rangle,|\phi_4\rangle\}$ we have $\langle\phi_j | H_{\text{eff}} |\phi_k\rangle = E_j^{\star}\delta_{jk}$. All the results from Section \ref{S:isolated} can now be modified by the replacement $\lambda_k \to E_k^{\star}$.
Note that $E_k^{\star}$ is real if and only if $|\lambda_k|\geq 2\sqrt{1-t_c^2}+O(t_c^4)$. (More precisely, $|\lambda_k|\geq 2-t_c^2$, as shown in the Appendix. See~\eqref{E:fixed_point_solution_appendix},\eqref{E:fixed_point_solution_appendix_negE}.) Since any path in $(\lambda_1, \lambda_2 , \lambda_3, \lambda_4)$-space corresponds to a unique path in parameter space $(\epsilon , \alpha , \beta , \gamma )$, each $E_k^{\star}$ can be made real or complex independently of the other three. Thus, each eigenstate can be made to evolve unitarily or not by adjusting the internal parameters of the core, permitting exquisite control over (partial) decoherence~\cite{aharony2012partial}. 

For class $Y$ (i.e. $\omega=e^{i\pi/2}$), expressions identical to~\eqref{E:fx_pts_E_1},\eqref{E:Green2_E_phi} hold with the replacements $\phi_k\to\chi_k$~\eqref{E:chi_k} and $\lambda_k\to\lambda'_k$~\eqref{E:lambda_para_i}. Again, full control over the core energies allows to make each $E_k^{\star}$ real or complex independently of the other three.

The same is true, with a caveat, when the core is in the nonsymmetric off mode,
\beq
H_{\text{off}}=-\sum_{\langle ij \rangle} K_{ij}a_i^{\dagger}a_j + \sum_i \epsilon_i a_i^{\dagger}a_i \noquad ,
\eeq
where large energy offsets $|\epsilon_i - \epsilon_j|\gg K_{ij}>0$ effectively suppress spontaneous transitions, so that position $i$ is almost a good quantum number. Then again $G_{\text{core}}(E)\approx (E-H_{\text{eff}})^{-1}=\sum_m \frac{| m \rangle\langle m |}{E-\epsilon_m^{\star}}$ with $\epsilon_m^{\star}=\frac{1}{1-t_c^2}\left[\left(1-\tfrac{t_c^2}{2}\right)\epsilon_m- \tfrac{t_c^2}{2}\sqrt{\epsilon_m^2-4(1-t_c^2)}\right]$. This time care must be taken to maintain the large-offset condition when lowering the $\epsilon_m$'s below the escape threshold, in order to prevent spurious logical transitions.

\subsection{Leakage-free logical operations}\label{S:protected_gates}
In this section we show that, even in the presence of leads, all logical gates can be realized without leakage, and are still symmetry-protected from unbiased parameter noise. It is sufficient to consider single-pulse, $\omega$-circulant gates, since they are universal for quantum computation. We could use only the gates $\{\mathsf{V},\mathsf{W}\}$ from Section~\ref{S:universality}, for instance, which are especially resilient against $x$- and $z$-rotation errors. Let a single pulse producing the gate $U$ be given in the time interval $t\in [0,T]$ by the path $(\lambda_1 (t),\dots , \lambda_4 (t))$ in the $\mathbb{R}_4$ manifold of eigenenergies of the bare core. With a lead coupled to each site, eigenenergies $\lambda_k$ are modified to possibly complex effective eigenenergies $E_k^{\star}$. We must make sure that $|\lambda_k|\geq 2-t_c^2$ at all times to keep $E_k^{\star}$ real and prevent escape through the leads. If not, rescaling the energies by $n$ and the time by $1/n$ leaves unchanged the unitary $U=\exp \left(-i\int_0^T dt\, H_{\text{core}}(t)\right)$. Avoiding the energy band of the leads is therefore not an issue. Moreover, for these values of $\lambda_k$, the function $E_k^{\star} (\lambda_k)$ increases monotonically, and hence is one-to-one. As an immediate consequence, there is a (unique) path $(p_1 (t),\dots , p_4 (t))$ lying entirely outside the band of the leads such that
\beq
E_k^{\star} (p_k (t)) = \lambda_k (t) \hspace{1cm} k\in\{1,\dots , 4\}, \; t\in [0,T].
\eeq
We thus obtain
\beq
U_{\text{eff}}=\exp \left(-i\int_0^T dt\, H_{\text{eff}}(t)\right)=U.
\eeq
Having reproduced the ideal gate $U$ in the presence of the leads, and recalling that identical leads preserve $\omega$-rotation invariance, we conclude that the effective gate $U_{\text{eff}}$ is symmetry-protected from unbiased noise in the \textit{effective} parameters $E_k^{\star}$. And since the functions $E_k^{\star} (\lambda_k)$ are very nearly linear outside the band, unbiased noise in $E_k^{\star}$ is equivalent to unbiased noise in the bare parameters. This completes our claim that, even in the presence of leads, logical gates can be realized without leakage and are still protected against unbiased noise in the parameters (whether bare or effective).

\subsection{Core as quantum memory}\label{S:leads_transmission}
Within each $\omega$-class the effective eigenenergies can be chosen real or complex independently of one another, and as a consequence each energy eigenstate can independently be made to dissipate in the leads or remain stationary. (The off mode offers comparatively less flexibility because of the condition $|\epsilon_i - \epsilon_j|\gg K_{ij}>0$, although crossing levels is ill-advised in any mode.) The dissipation of the $k$-th eigenstate is characterized by the \textit{tunable} dynamical rate
\beq
\tau^{-1}(\lambda_k)=|\text{Im}(E_k^{\star})|,
\eeq
which is found from~\eqref{E:fx_pts_E_1} to be a continuous function of the (fully controllable) energy $\lambda_k$, with values in the interval $[0,2/3\sqrt{3}]\approx [0,0.385]$. The ability to prevent the eigenstates from escaping to the leads allows us to consider the core as a versatile quantum memory unit that can protect a state $\sum_k a_k |\phi_k\rangle$ for a long time, and then release it entirely or partially at a later time. (In what follows $|\phi_k\rangle$ will stand for an eigenstate of either class $X$ or $Y$, unless specified otherwise.)  This is the perspective that we adopt in this section, and to do so it is convenient to go beyond the first-order Green's function analysis that we have employed so far, which loses track of effective core eigenstates as they escape, with no possibility of ever coming back. We emphasize that our proposal assumes nothing other than the existence of a flat symmetry class in the physical support of information, an aspect which to the best of our knowledge has not been exploited in quantum memory technologies~\cite{nilsson2006nanowire,aharon2016multi-bit,heshami2015quantum,saglamyurek2018coherent,koong2020multiplexed}. 

We consider the following \textit{finite} system: it consists of two $\omega$-circulant 4-level cores standing face to face, and connected by four identical leads, each comprised of $L$ sites. One may think of it as a square prism of height $L$, with the cores as top and bottom faces. The cores act as memory storage units. We will identify when and why a state localized on one core will scatter within characteristic time $\tau_s$, eventually reaching the other core. Alternatively, we discuss how the localized state can be protected from scattering over a timescale $\tau_b \gg \tau_s$. The index $b$ stands for \textit{bound states}, whose presence or absence determines the dissipation regime.

\begin{figure*}
  \centering
  	\includegraphics[]{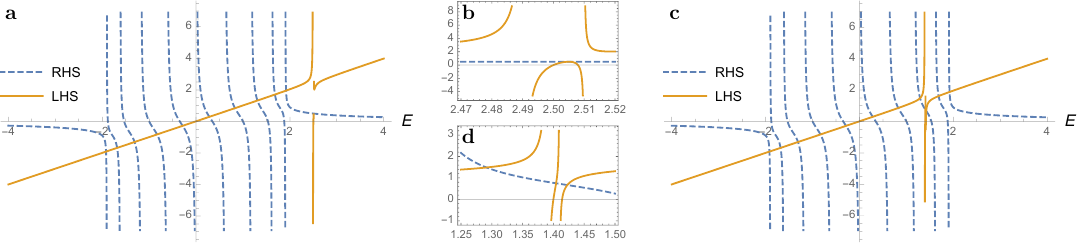}
  \caption{(Color on line.) Left-hand (LHS) and right-hand (RHS) sides of the energy constraint equation \eqref{E:E_constraint} as a function of the dimensionless energy $E$ for $(L,t_{c,1},t_{c,2})=(10,0.1,0.1)$. The LHS curve crosses the RHS curve $L+2$ times. \textbf{a} For $\lambda=\mu=2.5$, the crossings correspond to $L$ continuum states in the band, plus two bound states near $\lambda$. \textbf{b} Zoom-in of the neighborhood of $E = \lambda$ for $\lambda=\mu=2.5$. The nearly degenerate bound state energies $E=2.5049765$ and $E=2.5049904$ are not yet resolved, but we plot these states in Fig.~\ref{F:evenNgraph_boundstates}.  \textbf{c} For $\lambda=\mu=1.4$ the LHS curve crosses the RHS curve $L+2$ times within the band. \textbf{d} Zoom-in of the neighborhood of $E = \lambda$ for $\lambda=\mu=1.4$, showing one continuous scattering mode (leftmost) and two hybridized scattering modes with energy separation $\sim 10^{-2}$.}\label{F:evenN_ab_maintext}
\end{figure*}

In Appendix~\ref{S:solution_dot_lead_dot_appendix} we show that the $\omega$-circulant system decouples into four identical modes, each in the form of two sites of self-energies $\lambda , \mu$ connected by a finite lead of $L$ sites. The single-particle Hamiltonian of one mode is
\beq
H_{1P}=\left[\begin{array}{c|ccc|c}
\lambda & t_{c,1} &  & & \\
\hline
t_{c,1}^* & 0 & 1 &  & \\
 & 1 & 0 & \ddots & \\
 &   & \ddots & \ddots & t_{c,2}\\
\hline
 &  &  & t_{c,2}^* & \mu
\end{array}\right].
\eeq
(The argument generalizes in a straightforward manner if the 4-level cores are replaced by $N$-level cores.)  The corresponding Schr{\"o}dinger equation is easily solved, yielding eigenenergies $E$ in implicit form \beq\label{E:E_constraint}
\frac{\Delta\tilde\Delta E - \Delta - \tilde\Delta}{\Delta\tilde\Delta - 1}=\frac{U_{L-2}(E/2)}{U_{L-1}(E/2)},
\eeq
where 
\beq\label{E:Delta_Delta_tilde}
\Delta=\frac{E -\lambda}{|t_{c,1}|^2},\qquad
\tilde\Delta=\frac{E -\mu}{|t_{c,2}|^2},
\eeq
and $U_n (x)$ is a Chebyshev polynomial of the second kind.
The $L+2$ solutions of~\eqref{E:E_constraint} are the system's eigenenergies. Unsurprisingly, this equation cannot be solved analytically; a graphical solution is displayed in Fig.~\ref{F:evenN_ab_maintext}. Note that the RHS (blue curves) is independent of the couplings to the cores and pertains to the spectrum of the leads whereas the LHS (yellow curves) is independent of the lead parameters and pertains to the coupling. $L$ solutions always belong to the energy band $[-2,2]$, and form the (perturbed) continuous spectrum of the leads. The two remaining solutions may lie outside the band (bound states) or within the band (hybridized scattering states). We now discuss these cases in turn.

Let us write single-particle states as
\beq\label{E:single-particle_eigenstate}
|E\rangle = \frac{\beta_0}{t_{c,1}^{*}} |0\rangle + \sum_{j=1}^L\beta_{j}|j\rangle + \frac{\beta_{L+1}}{t_{c,2}}|L+1\rangle,
\eeq
where $|j\rangle$ is the state with one particle on site $j$. The coefficients of bound states, with energies $|E|>2$, are given by 
\beq
\beta_{j}=(\pm 1)^j \mathcal{N}\left[\tfrac{|E|\pm\lambda}{|t_{c,1}|^2}\sinh j\xi - \sinh (j-1)\xi\right],
\eeq
for $0\leq j\leq L+1$.
Here $\pm = \text{sgn}(E)$, $\xi=\cosh^{-1} (|E|/2)$ and $\mathcal{N}$ is a normalization factor. These states are localized around both endpoints, decaying exponentially over the characteristic length scale $\xi^{-1}$ from the endpoints.

\begin{figure}
  \centering
	\includegraphics[]{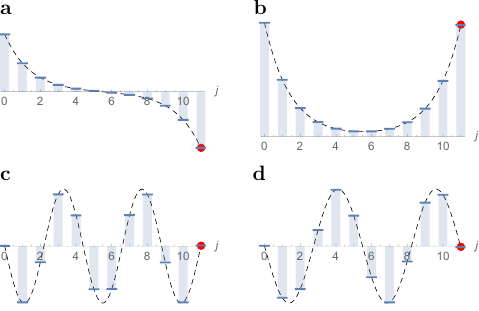}  		
  \caption{(Color on line.) The two bound states (\textbf{a,b}) and two states from the continuum (\textbf{c,d}) for $(L,t_{c,1},t_{c,2})=(10,0.1,0.1)$ and $\lambda=\mu=2.5$. \textbf{a}~Antisymmetric bound state, $E=2.5049765$. \textbf{b}~Symmetric bound state, $E=2.5049904$. The energy separation is $\sim 10^{-5}$. A generic feature of bound states is their large amplitude at the endpoints (sites 0 and $L+1$). \textbf{c}~A symmetric continuum state, $E=0.282$. \textbf{d}~An antisymmetric continuum state, $E=0.827$. A generic feature of continuum states is their small amplitude at the endpoints. In each graph, the red dot represents a consistency condition on $\beta_{L+1}$. See Appendix~\ref{S:solution_dot_lead_dot_appendix}, Eq.~\eqref{E:bdr2} for details.}\label{F:evenNgraph_boundstates}
\end{figure}

Throughout this section, all calculations will be done using $L=10$ and $t_{c,1}=t_{c,2}\equiv t_c =0.1$. The bound states for $\lambda=\mu=2.5$ are displayed in Fig.~\ref{F:evenNgraph_boundstates} \textbf{a,b}. 
We see that there is a symmetric state $|b_S\rangle$ and an antisymmetric state $|b_A\rangle$, a consequence of the Schr{\"o}dinger equation symmetry $j\leftrightarrow L+1-j$ resulting from $\Delta = \tilde\Delta$ (see Appendix~\ref{S:solution_dot_lead_dot_appendix}). When $L\to\infty$, the limiting expression for $\beta_{j}$  describes a bound state localized at the left endpoint and decaying exponentially with distance. Similarly, the limiting expression for $\beta_{L+1-j}$ describes a state localized at the right endpoint. In that limit, the eigenvalue equation \eqref{E:E_constraint} is equivalent to the fixed-point relations $E=\lambda \pm |t_{c,1}|^2 \Sigma(E)$  and $E=\mu \pm |t_{c,2}|^2 \Sigma(E)$ for the states localized on the left and right, respectively. In the Green's function treatment of the core with semi-infinite leads (see Appendix~\ref{S:effective_H_appendix}), the same fixed-point relation appears as the effective self-energy of the core once the leads are traced out. In the finite-$L$ case, states localized around a single end of the lead will only be approximately stationary. If the system evolves for a long time, the state localized on one end will eventually tunnel through the lead.

All other single-particle solutions, Eq.~\eqref{E:single-particle_eigenstate}, fall within the band of the leads, $E \in [-2,2]$, with coefficients given by
\beq
\beta_{j}=\mathcal{N}\Big[\tfrac{E-\lambda}{|t_{c,1}|^2}\sin j\theta-\sin (j-1)\theta\Big],
\eeq
where $\theta=\cos^{-1}(E/2)$. For any finite $L$, there are $L$ scattering states from the (perturbed) continuous spectrum of the leads. A generic feature of these $L$ states is their small amplitude at the endpoints. Two continuum states for $\lambda=\mu=2.5$ are displayed in Fig.~\ref{F:evenNgraph_boundstates} \textbf{c,d}.

\begin{figure}
  \centering
  	\includegraphics[]{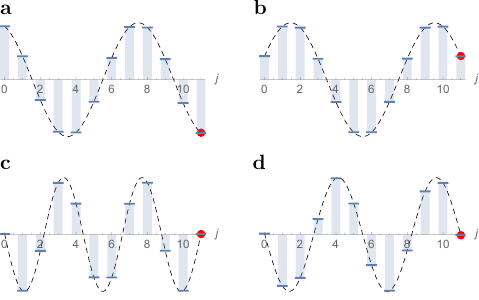} 
  \caption{(Color on line.) The two hybridized scattering states (\textbf{a,b}) and two states from the continuum (\textbf{c,d}) for $(L,t_{c,1},t_{c,2})=(10,0.1,0.1)$ and $\lambda=\mu=1.4$. \textbf{a}~Antisymmetric hybridized state, $E=1.4044$. \textbf{b}~Symmetric hybridized state, $E=1.4226$. The energy separation is $\sim 10^{-2}$. A generic feature of hybridized states is their relatively large amplitude at the endpoints (sites 0 and $L+1$). \textbf{c}~A symmetric continuum state, $E=0.282$. \textbf{d}~An antisymmetric continuum state, $E=0.827$. Continuum states are visually indistinguishable from those of Fig.~\ref{F:evenNgraph_boundstates}.}\label{F:evenNgraph_hybrid}
\end{figure}

When bound states are not present, in addition to the continuum states there will be two hybridized scattering states: one symmetric $|h_S\rangle$ and one antisymmetric $|h_A\rangle$. A generic feature of these states is their relatively large amplitude at the endpoints. Such states are displayed in Fig.~\ref{F:evenNgraph_hybrid} \textbf{a,b} for $\lambda=\mu=1.4$; for this same case example scattering states with $E \in [-2,2]$ are displayed in Fig.~\ref{F:evenNgraph_hybrid} \textbf{c,d}.

Decay rates in the presence of bound states $|b_S\rangle$ and $|b_A\rangle$ are compared with decay rates in the presence of hybridized states $|h_S\rangle$ and $|h_A\rangle$. Let the state be
\beq
|\psi\rangle = \sum_1^L a_n |n\rangle + c_S|\varphi_S\rangle +c_A|\varphi_A\rangle,
\eeq
where $|n\rangle$ is a state from the continuous spectrum, and $|\varphi_{S,A}\rangle$ is bound or hybridized. Then
\beq
\langle\psi | e^{-itH}|\psi\rangle = \sum_{n=1}^L |a_n|^2 e^{-itE_n} + |c_S|^2e^{-itE_S}+|c_A|^2e^{-itE_A}.
\eeq
Let $f(E_m)=|a_m|^2$ and assume $E_m-E_n \approx (m-n)\epsilon$ where $\epsilon = 4/(L-1)$. In the continuous limit,
\beq
\langle\psi | e^{-itH}|\psi\rangle \sim \tilde f(t) + |c_S|^2e^{-itE_S}+|c_A|^2e^{-itE_A},
\eeq
where $\tilde f$ is the inverse Fourier transform of $f$, and
\beq
\begin{aligned}
|\langle\psi | &e^{-itH}|\psi\rangle|^2 \sim  \\
&|\tilde f( t) |^2 + 2 \tilde f( t) \big(|c_S|^2\sin E_S t + |c_A|^2\sin E_A t\big)\\
 							&+|c_S|^4 +|c_A|^4 + 2|c_S|^2|c_A|^2 \cos (E_S - E_S)t.
\end{aligned}
\eeq
If $f(E)$ has support of width $\Delta E$ (in $[-2,2]$), its inverse Fourier transform $\tilde f (t)$ will decay within time $\Delta t \sim O(1/\Delta E)$. A small decay rate implies that $|\psi\rangle$ has overlap almost zero with most states of the continuous spectrum. This is possible for a \textit{localized} $|\psi\rangle$ only if bound states $|b_S\rangle$ and $|b_A\rangle$ are available, e.g., if $\lambda = \mu \notin [-2,2]$. An example is $|\psi\rangle = (|b_S\rangle + |b_A\rangle)/\sqrt{2}$, which is well localized around the left-hand dot. Then
\beq
|\langle\psi | e^{-itH}|\psi\rangle|^2 =\frac{1}{2}\big(1+\cos (E_S^b - E_A^b)t\big).
\eeq
With the canonical parameter values $(L,t_c)=(10,0.1)$, this goes slowly to zero with rate $\tau_b^{-1}=E_S^b-E_A^b \sim 10^{-5}$ (and oscillates back and forth unless $L$ is infinite).

If bound states are not available (e.g., if $\lambda = \mu \in [-2,2]$), then a localized state necessarily has an $f(E)$ with large support $\Delta E$, and $|\langle\psi | e^{-itH}|\psi\rangle|^2$ will decay within time $\Delta t \sim O(1/\Delta E)$ to the oscillating steady-state
\beq
|\langle\psi | e^{-itH}|\psi\rangle|^2 \sim |c_S|^4 +|c_A|^4 + 2|c_S|^2|c_A|^2 \cos (E_S^h - E_A^h)t.
\eeq
With the same canonical parameter values, this oscillates with rate $\tau_s^{-1}=E_S^h-E_A^h \sim 10^{-2}$. With these values, the characteristic escape time of localized states is reduced by a factor of $10^3$ when effective eigenenergies become complex and bound states are no longer available. The overall rate of decay, $\text{max}(\Delta E , E_S^h-E_A^h )\leq 4$, can be as large as $O(1)$. The above analysis shows that the core in either $\omega$-class has the ability to receive and release states over the timescale $\tau_s$ or shorter, and to store a state over the much larger timescale $\tau_b$. Switching between these two coupling regimes is performed by tuning the internal parameters of the core. The same procedure is also possible in the nonsymmetric off mode, with somewhat less flexibility due to the large-offset condition.

\subsection{Qubit initialization and readout}\label{S:qubit_measurement}
Initializations and measurements are naturally performed through position measurements or energy measurements of either $\omega$-class. Position eigenstates correspond to binary-code logical states, as given in Table~\ref{T:logical_bases},
whereas energy eigenstates of classes $X$ and $Y$ correspond to columns of $\mathcal{F}^{\dagger}$ and $(\mathcal{FD})^{\dagger}$, respectively:
\beq
|\phi_k\rangle = \sum_{\ell}\mathcal{F}^{\dagger}_{\ell , k} |\ell\rangle \noquad , \quad |\chi_k\rangle = \sum_{\ell}(\mathcal{FD})^{\dagger}_{\ell , k} |\ell\rangle .
\eeq
Define, for instance, the POVM $\{E_0 , E_1\}$ with elements the joint-position projectors
\beq
\begin{aligned}
E_0 &= |1\rangle\langle 1 | + |2\rangle\langle 2|\\
E_1 &= |3\rangle\langle 3 | + |4\rangle\langle 4 |.
\end{aligned}
\eeq
These operators correspond to the measurement/initialization of the \textit{first} qubit only:
\beq
\begin{aligned}
E_0 &= |00\rangle\langle 00 | + | 01\rangle\langle 01 |=|0\rangle\langle 0 | \otimes \mathbb{1}\\
E_1 &= | 10\rangle\langle 10 | + |11\rangle\langle 11 |= |1\rangle\langle 1 | \otimes \mathbb{1}.
\end{aligned}
\eeq
Similarly, the POVM $\{E^0 , E^1 \}$ with elements
\beq
\begin{aligned}
E^0 &= |1\rangle\langle 1 | + |3\rangle\langle 3|\\
E^1 &= |2\rangle\langle 2 | + |4\rangle\langle 4 |
\end{aligned}
\eeq
corresponds to the measurement/initialization of the \textit{second} qubit:
\beq
\begin{aligned}
E^0 &= |00\rangle\langle 00 | + | 10\rangle\langle 10 |= \mathbb{1} \otimes |0\rangle\langle 0 |\\
E^1 &= | 01\rangle\langle 01 | + |11\rangle\langle 11 |= \mathbb{1} \otimes |1\rangle\langle 1 | .
\end{aligned}
\eeq
Note that Gray code produces the same output.

\section{Scalability}\label{S:scalability}
We now propose a scalable technology for universal quantum computation. The idea is strongly reminiscent of classical computer architecture, in which operations are decomposed into elementary two-bit steps to be performed on large arrays of transistors. Here we use the fact that any unitary $U$ on $d$ qubits can be approximated to arbitrary accuracy by finite products of elementary two-qubit unitaries, i.e., operations of the form $S=\mathbb{1}^{\otimes m}\otimes K \otimes \mathbb{1}^{\otimes d-m-2}$, where $K$ is a $4\times 4$ unitary~\cite{divincenzo1995twobit}.  For any $\epsilon>0$ there exists a finite sequence of such two-qubit unitaries $S_1 , \dots , S_k$ achieving
\beq
\underset{|\psi\rangle}{\text{max}}\;  \lVert (U  -  S_1 S_2 \cdots S_k )|\psi\rangle \rVert < \epsilon,
\eeq
where $|\psi\rangle$ is any normalized $d$-qubit state. We thus consider the possibility of realizing quantum computation on scalable grids of 4-level cores. By analogy with the role played by transistors in classical computation, we may consider the cores to be quantum-computational transistors, or more succinctly, \textit{quansistors}.

\subsection{Quansistors}\label{S:quansistors}
The core, or quansistor, is a four-level tight-binding system with the ability to become $\omega$-rotation-invariant for $\omega=1,e^{i\pi/2}$ (classes $X,Y$). Until now we have mostly considered the quansistor in its symmetric form, performing computation on its double qubit. From the universality proof of Section~\ref{S:universality}, we know that the set of gates $\{\mathsf{V} , \mathsf{W}\}$, constructed from the Hamiltonians~\eqref{E:H_1_universality},\eqref{E:Q_universality}, is universal on two qubits. This set contains one representative from each class, $X$ and $Y$, and these representatives realize inequivalent flux structures. The next step is to allow interactions between quansistors. We choose the most basic two-quansistor interaction, involving one qubit from each quansistor. To that end, we need to materialize the tensor product structure inherent to the quansistor logical basis $\{|00\rangle, |01\rangle, |10\rangle, |11\rangle\}$. Thus far, these qubit states merely label the states of the quansistor, and need to be factored into pairs of \textit{spatially separable} qubits before they can be shared with distinct target quansistors. Let us devote some attention to the nonsymmetric form of the quansistor, which is also the off mode for computation. Note that the off-mode Hamiltonian cannot be the $\omega$-circulant matrix $\epsilon \mathbb{1}$, because the degenerate eigenstates of the latter are unstable to perturbations.

\begin{figure*}
\includegraphics[]{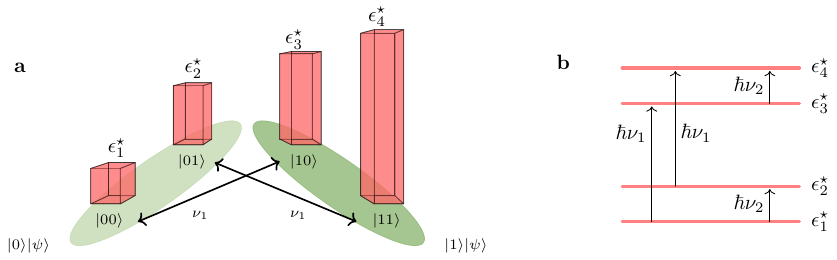}
\caption{(Color on line.) Quansistor's off mode. \textbf{a} Real effective on-site energies $\epsilon_i^{\star}$ and negligible transition amplitudes $0<K_{ij}\ll |\epsilon_i^{\star} - \epsilon_j^{\star} |$ (not shown). Position is almost a good quantum number, and logical states (position eigenstates) are almost stationary. Resonance at the qubit-splitting frequency $\nu_1$ corresponds to 1st qubit oscillation. The quansistor is then effectively a single qubit with basis states $|0\rangle |\psi\rangle$ and $|1\rangle |\psi\rangle$  (light green and dark green, respectively). \textbf{b} Energy diagram satisfying the qubit-splitting conditions~\eqref{E:e_difference_1},~\eqref{E:e_difference_2}. }\label{F:offmode}
\end{figure*}

\begin{figure}
\includegraphics[]{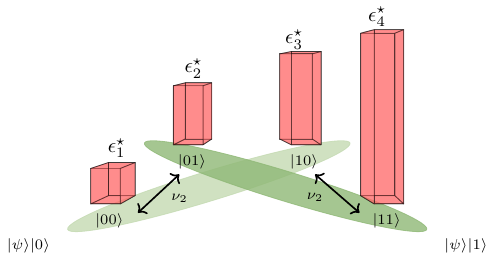}
\caption{(Color on line.) Quansistor's off mode. Resonance at the qubit-splitting frequency $\nu_2$ corresponds to 2nd qubit oscillation. The quansistor is then effectively a single qubit with basis states $|\psi\rangle |0\rangle$ and $|\psi\rangle |1\rangle$  (light green and dark green, respectively).}\label{F:offmode2}
\end{figure}

In the off mode, the Hamiltonian is simply
\beq\label{E:H_off_mode}
H_{\text{off}}=-\sum_{\langle ij \rangle} K_{ij}a_i^{\dagger}a_j + \sum_i \epsilon_i^{\star} a_i^{\dagger}a_i \noquad ,
\eeq
where $K_{ij}>0$, and large energy offsets $|\epsilon_i^{\star} - \epsilon_j^{\star}|\gg K_{ij}$ effectively suppress spontaneous transitions, so that position $i$ is almost a good quantum number in the off mode. (As before, the symbol ``$\star$'' indicates an effective energy due to the presence of the leads.) Logical states $\{|00\rangle , |01\rangle , |10\rangle , |11\rangle\}$ coincide with position eigenstates $\{|1\rangle , |2\rangle , |3\rangle , |4\rangle\}$, respectively. Section~\ref{S:implementation} illustrates how the system can be switched from~\eqref{E:H_off_mode} to $\omega$-circulant Hamiltonians of class $X$ or $Y$ and back using electrically charged levels and magnetic fields on the one hand, and neutral levels and synthetic gauge fields on the other. Notice that
\beq\label{E:span_12_34}
\begin{aligned}
\text{span}\{|1\rangle , |2\rangle\}&=\{|0\rangle |\psi\rangle \, |\, \psi\text{ any 2nd qubit state }\}\\
\text{span}\{|3\rangle , |4\rangle\}&=\{|1\rangle |\psi\rangle \,|\, \psi\text{ any 2nd qubit state }\}
\end{aligned}
\eeq
(see Fig.~\ref{F:offmode}). Similarly,
\beq\label{E:span_13_24}
\begin{aligned}
\text{span}\{|1\rangle , |3\rangle\}&=\{|\psi\rangle  |0\rangle \,|\, \psi\text{ any 1st qubit state }\}\\
\text{span}\{|2\rangle , |4\rangle\}&=\{|\psi\rangle |1\rangle \, |\, \psi\text{ any 1st qubit state }\}
\end{aligned}
\eeq
(see Fig.~\ref{F:offmode2}). We now demand that the off-mode on-site energies satisfy
\ba
\epsilon_3^{\star} - \epsilon_1^{\star} &= \hbar\nu_1 = \epsilon_4^{\star} - \epsilon_2^{\star},\label{E:e_difference_1}\\
\epsilon_2^{\star} - \epsilon_1^{\star} &= \hbar\nu_2 = \epsilon_4^{\star} - \epsilon_3^{\star}\label{E:e_difference_2}
\ea
Eqs.~\eqref{E:span_12_34} and~\eqref{E:e_difference_1} imply that setting the quansistor into resonance at frequency $\nu_1$ prompts the onset of 1st-qubit oscillations $|0\rangle |\psi\rangle \leftrightarrow |1\rangle |\psi\rangle$ with frequency $\nu_1$. By the same token, Eqs.~\eqref{E:span_13_24} and~\eqref{E:e_difference_2} imply that quansistor resonance at frequency $\nu_2$ corresponds to 2nd-qubit oscillations $|\psi\rangle |0\rangle \leftrightarrow  |\psi\rangle|1\rangle$ at frequency $\nu_2$. For this reason, the frequencies $\nu_1$ and $\nu_2$ may be called \textit{qubit splitting}, and will be used to exchange single qubits between distant quansistors. When coupled to a single-mode resonator, the quansistor resonating at one of these frequencies $\nu_q$ will effectively look like a single qubit coupled to the oscillator as described by the Jaynes-Cummings Hamiltonian
\beq
H_{\text{JC}} =  \hbar \nu_r a^{\dagger}a + \frac{\hbar \nu_q}{2}\sigma_z + \hbar g (a^{\dagger}\sigma^{+}+a \sigma^{-}),
\eeq
where $\sigma_z=|1\rangle\langle 1| -|0\rangle\langle 0|$. The frequencies $\nu_1 \pm \nu_2$, on the other hand, are not qubit splitting, and correspond to the oscillations $|00\rangle \leftrightarrow |11\rangle$ and $|01\rangle \leftrightarrow |10\rangle$, respectively. The same results can be obtained in Gray code with minor modifications.

\subsection{Scalable architecture}\label{S:architecure}
Interactions between quansistors are to be performed by bringning their qubit-splitting frequencies into resonance. It seems desirable to mediate the coupling with single-mode resonators, allowing distributed circuit elements, and to work in the dispersive regime where two quansistors $A,B$ are mutually resonant, but far-detuned from the resonator:
\beq
\nu_A = \nu_B\neq \nu_r .
\eeq
The interaction then proceeds through virtual photon exchange, as opposed to real photons in the resonant regime, alleviating the major drawback of the latter, namely the resonator-induced decay due to energy exchange with the resonator (Purcell effect)~\cite{blais2020circuit}. In the rotating wave approximation, the Hamiltonian in the absence of direct coupling between the quansistors is
\beq
H=\hbar \nu_r a^{\dagger}a +\sum_{j=A,B} \frac{\hbar \nu_{q}}{2}\sigma_{z,j} + \sum_{j=A,B} \hbar g_j (a^{\dagger}\sigma_j^{-}+a \sigma_j^{+}),
\eeq
where $\nu_q$ is the common frequency of $A$ and $B$. We use the Baker-Campbell-Hausdorff formula $e^S H e^{-S}=H + [S,H]+\frac{1}{2!}[S,[S,H]]+\dots$ to perform the unitary transformation $U=\exp \sum_j \frac{g_j}{\Delta}(a^{\dagger}\sigma_j^{-}-a\sigma_j^{+})$, with detuning $\Delta=|\nu_r - \nu_q|$. To first order in $g_j / \Delta$ we get
\beq
H= \hbar \nu_r a^{\dagger}a + \sum_{j=A,B} \frac{\hbar \nu_{q}}{2}\sigma_{z,j} + \hbar K (\sigma_A^{+}\sigma_{B}^{-}+\sigma_A^{-}\sigma_B^{+}),
\eeq
where $K=2\hbar g_A g_B / \Delta$. The qubit-cavity interaction terms cancel out exactly, leaving only an effective qubit-qubit interaction with coupling $K$. This term, when evolved for a time $\pi/4K$, generates the $\sqrt{i\mathsf{SWAP}}$ gate, which is entangling and equivalent to the $\mathsf{CNOT}$ gate~\cite{loss1998quantum,blais2020circuit}, up to single-qubit operations already available within the quansistors. The dispersive regime thus allows for the possibility of long-distance entangling interactions between quansistors.

\begin{figure}
\centering
\includegraphics[width=8cm]{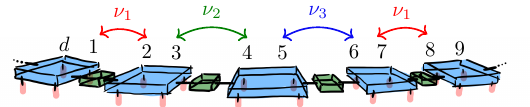}
\caption{(Color on line.) A $d$-quansistor linear array for universal computation on $2d$ qubits. The numbers schematically represent the qubits. Quansistors (larger, blue) perform two-qubit universal operations on qubit pairs $(2j, 2j +1)$, $j=0,\dots , d-1 \, (\text{mod } d)$. Couplers (smaller, green) perform entangling $\sqrt{i\mathsf{SWAP}}$ gates on qubit pairs $(2j-1, 2j)$, $j=0,\dots , d-1 \, (\text{mod } d)$. Leads are attached to the lower faces of the quansistors. With only three qubit-splitting frequencies in the system, $\nu_1 , \nu_2 , \nu_3$, untimely second-nearest-quansistor interactions occur at higher orders. (For instance, pairs (2,3) and (6,7) both respond to $\nu_1$.) More frequencies may be added to suppress these unwanted exchanges. The required physical resources scale linearly in the number of qubits.}\label{F:linear_array}
\end{figure}

\begin{figure}
\begin{tikzpicture}
\node[inner sep=0pt] (cluster) at (0,0)
    {\includegraphics[trim={0 5.5cm 0 5cm},clip, width=.4\textwidth]{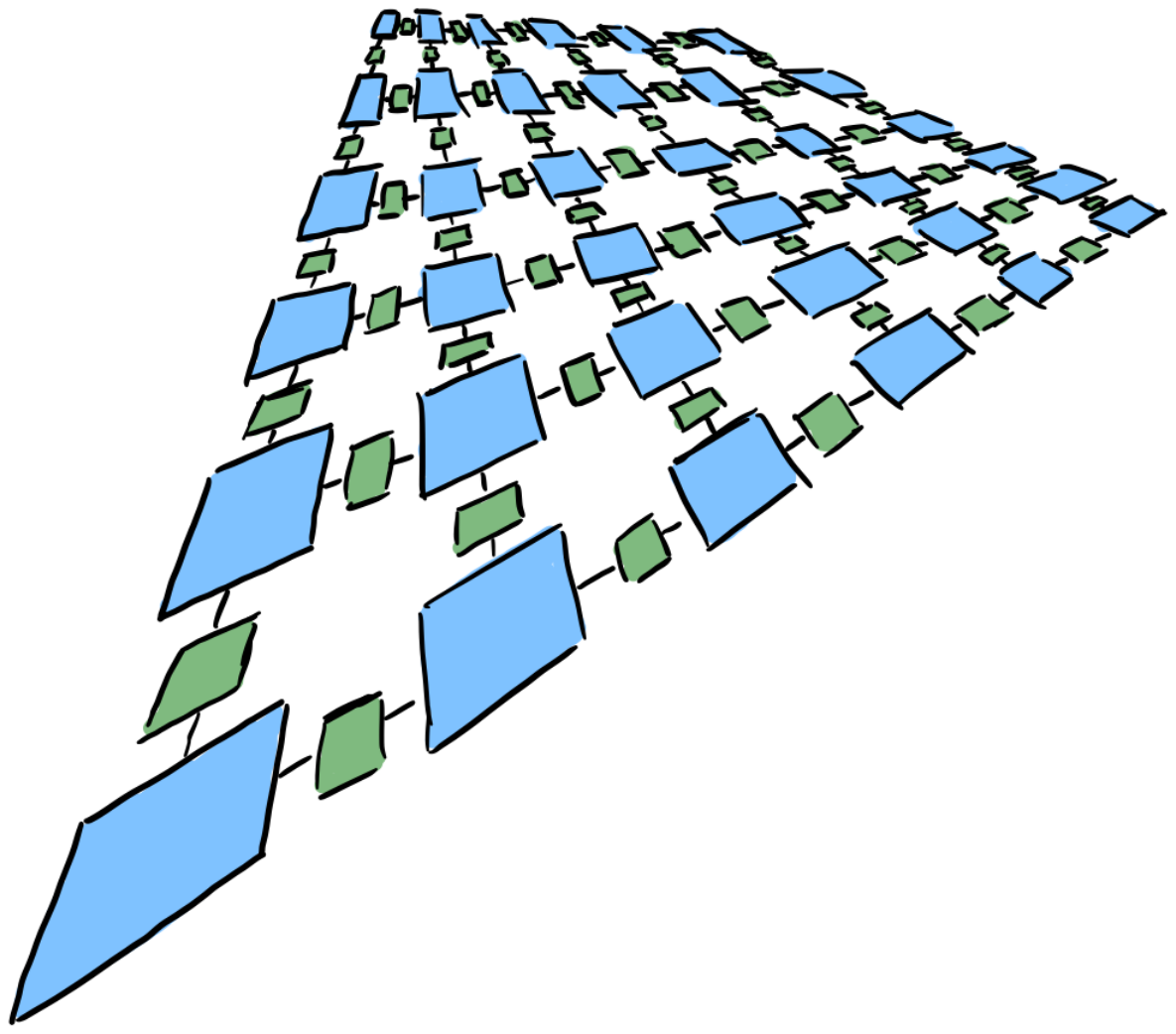}};
\end{tikzpicture}
\caption{(Color on line.) A $36$-quansistor grid for universal computation on $72$ qubits. Quansistors (large blue squares) perform universal two-qubit operations. Couplers (smaller green squares) perform entangling $\sqrt{i\mathsf{SWAP}}$ between qubits of adjacent quansistors. Each quansistor is coupled to four identical leads (not shown) for initialization and readout. The required physical resources scale linearly in the number of qubits.}\label{F:grid}
\end{figure}

One possible architecture for universal computation on $2d$ qubits consists in a closed linear array of $d$ quansistors coupled through resonators. Each quansistor represents a pair of qubits, and every qubit is represented exactly once (see Fig.~\ref{F:linear_array}). Additional qubit-splitting frequencies may be added to prevent untimely higher-order, $m$th-nearest-neighbor couplings. To cut down space and running time costs, a physical implementation of the array would likely be folded, and a resonator would couple every adjacent pair of quansistors, thus reducing considerably the need for qubit-shuffling operations. Fig.~\ref{F:grid} schematically depicts a planar-grid computer. Each quansistor would be coupled to four identical leads (not shown in the figure) for initialization and readout.

\section{Outlook}
Error-correcting codes are a promising avenue to make quantum computation fault-tolerant. In our technology, quansistors are the support of the \textit{physical} qubits, whose states and interactions are symmetry-protected, to some extent, against external influences. Once we identify the dominant errors affecting them, we can find $k$-qubit states (typically $k=5,7,9$) that are invariant under those errors. These would be encoded \textit{logical} qubits. (In the standard Hamming notation, we get a code of type $[kn,n]$ using $kn$ physical qubits, and having $2^n$ logical codewords~\cite{steane1996multiple}.) Having \textit{physical} universality at hand gives the freedom to encode qubits and operations yielding \textit{encoded} universality. Because our technology is scalable ($Q$ quansistors making $2Q$ qubits), the encoding uses $k$ identical `processors' instead of one.

Crucially, we must determine whether the dominant errors are constrained by the symmetry of the quansistors. We have already observed a significant robustness of the universal set $\{\mathsf{V}, \mathsf{W}\}$ against single-qubit and double-qubit $x$- and $z$-rotation errors. As a second source of decoherence, we considered the coupling to semi-infinite leads, and have omitted other factors such as the effect of a heat bath on the system, the types of errors that it would produce, and the extent to which it would destroy symmetry.

Throughout, we have used $\omega$-rotation invariance as the prototype of a symmetry of flat classes, universal for quantum computation, and realistically implementable physically. Other flat classes would perform equally well at protecting information and operations, as long as leads (or any other immediate environment of the clusters) do not break the corresponding symmetry. Universal sets of gates originating from nondegenerate Hamiltonians are symmetry-specific, but should not be too difficult to find given the relative scarcity of non-universal sets and the completeness of flat classes. The possibilities of physical implementation, on the other hand, will strongly depend on the chosen symmetry and would have to be found on a case-by-case basis. 

There might be additional value to using larger qubit clusters, i.e. $k$-qubit quansistors realized as $2^k$ sites with symmetries, for $k=3,4,5$. All observations from the previous point regarding protection, universality, and implementation, apply here as well. Larger clusters and symmetry classes would offer protection to $k$-qubit operations. It could also allow the encoding of a logical qubit within a \textit{single} $k$-qubit quansistor. Encodings with $k=3$ (8 sites) can already correct some single-qubit errors, while some encodings with $k=5$ (32 sites) can correct \textit{any} single-qubit error~\cite{nielsen2010quantum}. 

There is provable surplus value to using higher alphabets (ququarts, specifically), instead of qubit pairs, in encrypted communications~\cite{bechmann-pasquinucci1999quantum}. Since $\omega$-circulant 4-level quansistors are universal in $\mathbf{U}(4)$, i.e.~universal for single-\textit{ququart} operations, and have the ability to dynamically decouple from the leads, a quansistor-with-leads could be a versatile memory unit for ququarts, and become an essential component of quantum-secure communications. Generalizations to $d$-level quansistors (qudits), with $d>4$, is also conceivable, as argued in the previous point. Error-correcting codes, notably, have been developed for qudits, that use $d\times d$ Weyl matrices (or equivalently the $d\times d$ version of our matrices $X$ and $Y$)~\cite{gottesman1999fault, gottesman2001encoding}. Note, however, that the diagram of transition amplitudes of the general qudit ($K_d$ in graph theory terminology) is non-planar for $d\geq 5$~\cite{rosen2002discrete}, in contrast to the essentially planar diagram $K_4$ of the 4-level system, Fig.~\ref{F:model_tetrahedral}. A gauge potential implementation of $\omega$-rotation invariant qudits, in the spirit of Sections~\ref{S:gauge} and~\ref{S:implementation}, might then be essentially three-dimensional. 

For the purpose of reconstructing the final state of an $N$-qubit system, a symmetry based on Pauli tensors of dimension $d=2^N$ is better suited than $\omega$-rotation invariance (based on  Weyl matrices) as it allows for the maximal number ($2^N +1$) of mutually unbiased bases, and complete state characterization via state tomography~\cite{bandyopadhyay2001new, lawrence2001mutually}. By contrast, the 4-level quansistor ($N=2$) based on $\omega$-rotation invariance has only \textit{three} unbiased bases: $\{|m\rangle\}$, $\{|\phi_k\rangle\}$, $\{|\chi_k\rangle\}$, the respective eigenbases of $Z, X, Y$. As for reconstructing the final state of an $N$-qu\textit{p}it system (where $p$ is an odd prime), the Weyl-based scheme of dimension $d=p^N$ allows for the maximal number ($p^N +1$) of mutually unbiased bases, and complete state characterization via quantum tomography~\cite{bandyopadhyay2001new, durt2010mutually}. This could be of value for encrypted communication using a higher (prime) alphabet (see previous point), and could be based on $p$-site quansistors with $\omega$-rotation invariance. 

To perform inter-quansistor interactions, we have used the simplest possible scenario involving a single qubit from each quansistor. It would be worth investigating whether a combined use of resonators and symmetry could make possible the implementation of robust 3- and 4-qubit gates, or even interactions soliciting three quansistors or more. However, this is beyond the scope of this work.

\section{Conclusion}
In this work, we have put forward a blueprint for scalable universal quantum computation based on 2-qubit clusters (quansistors) protected by symmetry ($\omega$-rotation invariance). We find a significant robustness of the proposed universal set against single-qubit and double-qubit $x$- and $z$-rotation errors. Embedding in the environment, initialization and readout are achieved by tunnel-coupling each quansistor to four identical semi-infinite leads. We show that quansistors can be dynamically decoupled from the leads by tuning their internal parameters, giving them the versatility required to act as controllable quantum memory units. With this dynamical decoupling, universal 2-qubit logical operations within quansistors are also symmetry-protected against unbiased noise in their parameters. Two-quansistor entangling operations are achieved by resonator-coupling their qubit-splitting frequencies to effectively carry out the $\sqrt{i\mathsf{SWAP}}$ gate, with one qubit coming from each quansistor. We have also identified a variety of platforms that could implement $\omega$-rotation invariance. 

The complete tunability of $\omega$-circulant quansistors can be exploited to build highly expressive and trainable parameterized quantum circuits, to be used as the noisy intermediate-scale quantum (NISQ) component of a quantum-classical hybrid machine learning model~\cite{Benedetti2019parameterized,sim2019expressibility,holmes2022connecting}. These ideas will be explored in detail in a future publication.

\section{Acknowledgements}
This work was supported in part by the Natural Science and Engineering Research Council of Canada and by the Fonds de Recherche Nature et Technologies du Qu{\'e}bec via the INTRIQ strategic cluster grant. C.B. thanks the Department of National Defence of Canada for financial support to facilitate the completion of his PhD.

\section{Appendices}
\appendix
\section{Mathematical framework}\label{S:math}
This section describes some mathematical aspects of $\omega$-rotation invariance. Although this paper has focused on a four-level system, many of the results are easily generalized. Here we consider the more general case of an $N$-level system. For multiple reasons it may be desirable to have full control over the $N$ eigenenergies of the system. In what follows we consider what we propose to call \textit{flat classes} of Hamiltonians: classes of Hermitian matrices $\{H(\mathbf{g})\mid \mathbf{g}\in\mathbb{R}^N\}$ with a common eigenbasis, and real eigenenergies $\lambda_1 (\mathbf{g}),	\dots , \lambda_N (\mathbf{g})$ in one-to-one linear correspondence with the values of the parameters, that is $\lambda_r (\mathbf{g})=\sum_s g_s\lambda_{sr}$ with $\det [\lambda_{sr}]\neq 0$. (The term \textit{flat} refers to vanishing Berry curvature in $\mathbf{g}$-space.) A class of Hamiltonians is flat in this sense if and only if it can be represented as a sum of $N\times N$ Hermitian matrices,
\beq\label{E:flatclass}
H(g_1,\dots ,g_N) = \sum_{s=1}^N g_{s} H_{s} \hspace{0.5cm},\hspace{0.5cm} [H_{s} , H_{r}]=0 \quad(\forall s , r)
\eeq
and the $N\times N$ matrix $[\lambda_{sr}]$ of all eigenvalues of the $H_{s}$'s is nonsingular, $\det [\lambda_{sr}]\neq 0$. (The key observation is the action of the diagonalizing unitary: $\mathcal{U}^{-1}(\sum_s g_s H_s)\mathcal{U}=\sum_s g_s \text{diag}(\lambda_{s1},\dots , \lambda_{sN})=\text{diag}(\lambda_1 (\mathbf{g}),\dots , \lambda_N (\mathbf{g}))$.) Flat classes therefore coincide with $N$-dimensional commutative algebras of $N\times N$ Hermitian matrices, the nonsingularity of $[\lambda_{sr}]$ being equivalent to the linear independence of the $H_{s}$'s. Because each flat class is diagonalized by a common unitary $\mathcal{U}$ (unique up to permutations), the set of all flat classes that correspond to the same $[\lambda_{sr}]$ is in one-to-one correspondence with unitaries modulo permutations. (In this paper, we do not consider non-unitarily diagonalizable matrices, like non-Hermitian $\mathcal{PT}$-symmetric Hamiltonians, for instance.) The exponentials of a flat class also share the common eigenbasis of the class, and their eigenenergies are in (nonlinear) one-to-one correspondance with the values of the parameters $\mathbf{g}$. There seems to be an interesting connection between flat classes, on the one hand, and commuting bases of unitary matrices~\cite{bandyopadhyay2001new} and stabilizers of quantum error-correcting codes~\cite{gottesman1999fault}, on the other. 

For the purpose of quantum computation a single flat class is clearly not enough because it is commutative. The $\omega$-circulant matrices defined in \eqref{E:omega_circ} are of the form \eqref{E:flatclass}, and constitute a flat class for each $\omega$. Independent control over the energy levels is a prime motivation for using $\omega$-rotation invariance, but we stress again that this choice of symmetry is far from unique. Starting from any nonsingular matrix $[\lambda_{sr}]$, defining the functions $\lambda_r (\mathbf{g})=\sum_s g_s\lambda_{sr}$, and applying any unitary $\mathcal{U}$ to the matrix class $\{ \text{diag}(\lambda_1(\mathbf{g}),\dots, \lambda_N(\mathbf{g}))\mid \mathbf{g}\in\mathbb{R}^N\}$  will produce a commutative family of (Hermitian) Hamiltonian matrices $\mathcal{H}(\mathbf{g})$ with eigenstates independent of $\mathbf{g}$, and linearly controlled eigenenergies $\lambda_r (\mathbf{g})$, i.e.~a flat class.  If symmetries other than $\omega$-rotation invariance were preferred for practical reasons, one would replace the diagonalizing unitaries  $\mathcal{F}$ and $\mathcal{FD}$, defined in Eqs.~\eqref{E:regularFourier},\eqref{E:Fourier_D}, with the appropriate operations.

It is instructive to recast $\omega$-rotations in terms of Sylvester's clock-and-shift matrices (also called Weyl's matrices or generalized Pauli matrices)
\beq
Z_4=\begin{pmatrix}
1 	& 		& 			& 	 	\\
  	& \omega	&			&		\\
	&		& \omega^2 	&		\\
	&		&			& \omega^3 	\\
	\end{pmatrix} \noquad , \quad
X_4=\begin{pmatrix}
0 	& 1		& 	0		& 	0 	\\
0  	& 0	&	1		&	0	\\
0	&0		&0 	&1		\\
1	&0		&0			& 0 	\\
	\end{pmatrix},
\eeq
with $\omega=e^{i\pi/2}$. For $k=0,\dots ,3$ we have the identity $Z_4^k X_4  = J_{e^{ik\pi/2}}$. The matrices $\{Z_4^k X_4^{j}\}_{k,j = 0,\dots ,3}$ constitute a non-Hermitian trace-orthogonal basis for $\mathfrak{gl}(4,\mathbb{C})$, and in fact the same is true of their obvious $N$-dimensional generalization, with $\omega=e^{i2\pi/N}$, which span $\mathfrak{gl}(N,\mathbb{C})$ and are orthogonal under the Hilbert-Schmidt inner product. If $N=2$, they reduce (up to a factor) to the Pauli matrices. The matrices $X_N$ and $Z_N$ are central to Weyl's formulation of periodic finite-dimensional quantum mechanics where they respectively correspond to finite position and momentum shifts:
\beq
\begin{aligned}
X_N&=e^{i(2\pi/N)\hat{p}}\noquad , \quad X_N | x\rangle = | x-1 \text{ mod } N\rangle\\
Z_N&=e^{i(2\pi/N)\hat{x}}\noquad , \quad Z_N | \hspace{1pt}p\rangle = | p+1 \text{ mod } N\rangle
\end{aligned}
\eeq
where of course $| x\rangle$ and $| \hspace{1pt}p\rangle$ are position and momentum eigenstates, respectively. Thus $\omega$-rotation invariance is a symmetry in quantum (or optical) phase space, and is not found in other, internal-space generalizations of Pauli matrices like Pauli tensor products or Gell-Mann matrices. 

In any dimension $\geq 2$, the eigenbases of $Z_N, X_N$, and $Z_N X_N$ are \textit{mutually unbiased}: a measurement in one (orthonormal) basis $\{|\psi_r\rangle\}$ provides no information about measurements in another (orthonormal) basis $\{|\eta_s\rangle\}$ because $|\langle \psi_r|\eta_s\rangle|=\frac{1}{\sqrt{N}}$ for any $r,s$. In particular when $N=4$, the flat classes of $Z_4$, $X_4$ (the matrix $X$ in the main text), and $Y_4$ (the matrix $Y$ in the main text) have mutually unbiased eigenbases $\{|m\rangle\} , \{|\phi_k\rangle\}$, and $\{|\chi_k\rangle\}$ respectively. (See \eqref{E:phi_k} and \eqref{E:chi_k}.)

\section{Effective core Hamiltonian}\label{S:effective_H_appendix} 
\noindent Consider an $N$-level core tunnel-coupled to $N$ identical semi-infinite leads:
\beq\label{E:H_N_leads_appendix}
\begin{aligned}
&\mathcal{H}=\mathcal{H}_{\text{core}} + \mathcal{H}_{\text{int}} + \mathcal{H}_{\text{lead}}\\
 &=\frac{1}{2}\sum_{i,j = 1}^4 h_{ij}a_i^{\dagger}a_j  + \sum_{i=1}^4 t_{c,i} a_i^{\dagger}b_{i,1}^{\phantom{\dagger}}  + \sum_{i=1}^4 \sum_{j=1}^{\infty} b_{i,j}^{\dagger}b_{i,j+1}^{\phantom{\dagger}} +\text{ h.c.}
\end{aligned}
\eeq
For the time being, the matrix $h$ is not required to be Hermitian, and the couplings $t_{c,i}$ need not be equal, but could be chosen real positive with no loss of generality since the Hamiltonian is invariant under $t_{c,i}\to t_{c,i}e^{i\theta_i}$, $b_{i,j}\to b_{i,j}e^{-i\theta_i}$. We let them be complex anyways. Let us restrict the system's dynamics to the single-particle sector of Hilbert space. For illustration purposes, our examples below will use a core with $N=3$ levels, but all the results go over to general $N$. The one-particle Hamiltonian matrix $H$ with $N=3$ is
\beq
\left[\begin{array}{ccc|ccc|ccc|ccc}
h_{11} & h_{12} & h_{13} & t_{c,1} &  &  &  &  &  &  &  & \\
h_{21} & h_{22} & h_{23} &  &  &  & t_{c,2} &  &  &  &  &\\
h_{31} & h_{32} & h_{33} &  &  &  &  &  &  & t_{c,3} &  &\\
\hline
t_{c,1}^* &  &  & 0 & 1 &  &  &  &  &  &  & \\
 &  &  & 1 & 0 & \ddots &  & & &  &  & \\
 &  &  &  & \ddots & \ddots &  &  & &  &  &\\
 \hline
  & t_{c,2}^* &  &  &  &  & 0 & 1 &  &  &  & \\
  &  &  &  &  &  & 1 & 0 & \ddots &  &  & \\
 &  &  &  &  &  &  & \ddots & \ddots &  &  & \\
 \hline
  &  & t_{c,3}^* &  &  &  &  &  &  & 0 & 1 & \\
  &  &  &  &  &  &  &  &  & 1 & 0 & \ddots\\
  &  &  &  &  &  &  &  &  &  & \ddots & \ddots
\end{array}\right]
\eeq
which we write as
\beq
H=\left[\begin{array}{c|c}
h & V\\
\hline
V^{\dagger} & \mathbb{1}_3 \otimes h_{\text{lead}}
\end{array}\right]
\eeq
in obvious notation. The corresponding Green's function is
\beq
\begin{aligned}
G(E)&= (E\mathbb{1} - H)^{-1}\\
&=\left[\begin{array}{c|c}
E\mathbb{1}_3-h & -V\\
\hline
-V^{\dagger} & \mathbb{1}_3 \otimes (E\mathbb{1}_{\text{lead}}-h_{\text{lead}})
\end{array}\right]^{-1},
\end{aligned}
\eeq
where $\mathbb{1}_{\text{lead}}$ is the identity on the single-lead space. The top-left $3\times 3$ block of the Green's function, $G_{\text{core}}(E)$, can be obtained using the blockwise inversion formula
\beq
\left[\begin{array}{c|c}
A & B\\
\hline
C & D
\end{array}\right]^{-1}=\left[\begin{array}{c|c}
\left(A-BD^{-1}C\right)^{-1} & \cdots\\
\hline
\cdots & \cdots
\end{array}\right].
\eeq
We obtain
\beq\label{E:G_core_appendix}
\begin{aligned}
G_{\text{core}}(E) &=\left(E\mathbb{1}_3 - h - V\left[\mathbb{1}_3 \otimes (E\mathbb{1}_{\text{lead}}-h_{\text{lead}})\right]^{-1}V^{\dagger}\right)^{-1}\\
				&=\left(E\mathbb{1}_3 - h_{\infty}(E)\right)^{-1}.
\end{aligned}
\eeq
Noticing that
\beq
(\mathbb{1}_3 \otimes (E\mathbb{1}_{\text{lead}}-h_{\text{lead}}))^{-1}=\mathbb{1}_3\otimes G_{\text{lead}}(E),
\eeq
$G_{\text{lead}}(E)$ being the Green function of a single lead, we obtain from~\eqref{E:G_core_appendix}
\beq
h_{\infty}(E)=h+V\left(\mathbb{1}_3 \otimes G_{\text{lead}}(E) \right)V^{\dagger}.
\eeq
A straightforward calculation yields
\beq\label{E:h(E)_appendix}
h_{\infty}(E)=h+\begin{bmatrix}
|t_{c,1}|^2 		& 			& 					\\
			& |t_{c,2}|^2 	&					\\
			&			& |t_{c,3}|^2 \end{bmatrix} \Sigma (E),
\eeq
where $\Sigma (E)$ is the surface Green's function of a single lead :
\beq
\Sigma (E) = (G_{\text{lead}})_{00}(E)=\frac{E}{2}-\frac{\sqrt{(E+i0^+)^2 - 4}}{2}.
\eeq
The analogous version of~\eqref{E:h(E)_appendix} for general $N$ is now obvious. Remarkably, if $h$ is $\omega$-circulant and $|t_{c,i}|=t_c$ for all $i$, the effective Hamiltonian $h_{\infty}(E)$ is also $\omega$-circulant:
\beq
h=\sum_{s=0}^3 z_s J_{\omega}^s \hspace{0.1cm}\longrightarrow \hspace{0.1cm} h_{\infty}(E)= \left(z_0+t_c^2 \Sigma (E)\right)\mathbb{1} + \sum_{s=1}^3 z_s J_{\omega}^s \noquad .
\eeq
In particular, when $h$ is $4\times 4$ Hermitian we obtain expression~\eqref{E:H_pos_E_omega} in the main text. Again, the fact that $h_{\infty}(E)$ is unaffected by the presence of phase factors, dynamical or stochastic, in the core-to-lead couplings $t_{c,i}$ is a consequence of~\eqref{E:H_N_leads_appendix} being invariant under $t_{c,i}\to t_{c,i}e^{i\theta_i}$, $b_{i,j}\to b_{i,j}e^{-i\theta_i}$.

\section{Analytical solution: Two cores connected by finite leads}\label{S:solution_dot_lead_dot_appendix} 
\begin{figure*}
  \centering
  	\includegraphics[]{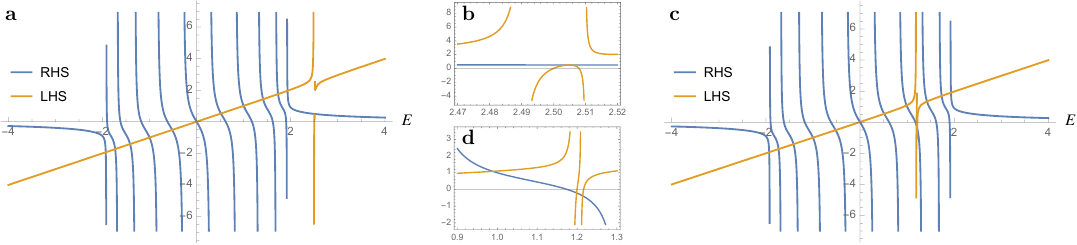}
  \caption{(Color on line.) Left-hand (LHS) and right-hand (RHS) sides of \eqref{E:consistency} for $L=11$, and $|t_{c,1}|=|t_{c,2}|=0.1$. Energy $E$ is dimensionless. The LHS curve crosses the RHS curve $L+2$ times. One solution has energy $\sim O(\lambda |t_{c,1}|^2)$, a consequence of weakly broken spectrum symmetry $E\leftrightarrow -E$ due to simultaneous nonzero $t_{c,1}$ and nonzero $\lambda$. \textbf{a} For $\lambda=\mu=2.5$, the crossings correspond to $L$ continuum states in the band, plus two bound states near $\lambda$. \textbf{b} Zoom-in of the neighborhood of $E = \lambda$ for $\lambda=\mu=2.5$. The nearly degenerate bound state energies $E=2.504980$ and $E=2.504987$ are not resolved yet, but we plot these states in Fig.~\ref{F:oddNgraph_bound_hybrid}.  \textbf{c} For $\lambda=\mu=1.2$ the LHS curve crosses the RHS curve $L+2$ times within the band. \textbf{d} Zoom-in of the neighborhood of $E = \lambda$ for $\lambda=\mu=1.2$, showing one continuous scattering mode (leftmost) and two hybridized scattering modes with energy separation $\sim 10^{-2}$.}\label{F:oddN_ab_maintext}
\end{figure*}

We analytically solve the Schr{\"o}dinger equation of two $\omega$-circulant $N$-level cores (with the same $\omega$, but possibly different core Hamiltonians $h_1$ and $h_2$) connected face-to-face by $N$ identical leads of $L$ sites. The Hamiltonian is
\beq
\begin{aligned}
H=&\frac{1}{2}\sum_{s=1}^2 \mathbf{a}_s^{\dagger}\cdot h_s \cdot \mathbf{a}_s + t_{c,1} \,\mathbf{a}_1^{\dagger}\cdot \mathbf{b}_{1}+ t_{c,2} \,\mathbf{a}_2^{\dagger}\cdot \mathbf{b}_{L}\\
 &+ \sum_{j=1}^{L-1} \mathbf{b}_j^{\dagger}\cdot\mathbf{b}_{j+1}^{\phantom{\dagger}}+\text{h.c.}\; ,
\end{aligned}
\eeq
where $\mathbf{a}_s=(a_{s,1} , \dots , a_{s,N})$ and $\mathbf{b}_j = (b_{1,j} , \dots , b_{N,j})$. Leads have hopping energies all equal, and set to unity (thus setting the scale for all energies). Eigenvalues of the unitary symmetry ($\omega$-rotation) correspond to superselection sectors. Without loss of generality, energy eigenstates may be chosen to have support in exactly one sector $k$. If $h_1, h_2$ are $\omega$-circulant with $\omega=e^{iq\pi/2}$ and $q=0,\dots ,3$, the change of basis Eqs.~\eqref{E:regularFourier},\eqref{E:Fourier_D}
\beq
\mathbf{a}_s\to \tilde{\mathbf{a}}_s=\mathcal{FD}^q\mathbf{a}_s \hspace{.5cm},\hspace{.5cm}\mathbf{b}_j \to  \tilde{\mathbf{b}}_j =\mathcal{FD}^q\mathbf{b}_j \noquad ,
\eeq
decouples the system into $N$ identical modes, each in the form of two dots of self-energies $\lambda_k , \mu_k$ connected by a finite lead of $L$ sites.
The single-particle Hamiltonian for one of these modes is
\beq
H_{1P}=\left[\begin{array}{c|ccc|c}
\lambda_k & t_{c,1} &  & & \\
\hline
t_{c,1}^* & 0 & 1 &  & \\
 & 1 & 0 & \ddots & \\
 &   & \ddots & \ddots & t_{c,2}\\
\hline
 &  &  & t_{c,2}^* & \mu_k
\end{array}\right].
\eeq
Let $|\psi\rangle = \alpha |0\rangle + \sum_{j=1}^L\beta_{j} |j\rangle + \gamma |L+1\rangle$ be a single-particle eigenstate in sector $k$, where $|j\rangle$ is the state with a single particle in the $j$th site. The Schr{\"o}dinger equation $E |\psi\rangle =H_{1P}|\psi\rangle$ yields the relations
\begin{eqnarray}
\beta_{1} &=& \Delta\beta_{0}\label{E:bdr1}\\
\beta_{j}+ \beta_{j+2} &=& E  \beta_{j+1}\qquad (0\leq j\leq L-1)\label{E:Fibonacci2}\\
\beta_{L} &=& \tilde\Delta\beta_{L+1}\label{E:bdr2}\\
\Delta &=&\frac{E -\lambda}{|t_{c,1}|^2}\\
\tilde\Delta &=&\frac{E -\mu}{|t_{c,2}|^2}\noquad .
\end{eqnarray}
where we have assumed that $t_{c,i}\neq 0$, and defined $\beta_0 =t_{c,1}^{*}\alpha$ and $\beta_{L+1}=t_{c,2}\gamma$.
The bulk equation~\eqref{E:Fibonacci2} is translation-invariant with general solution 
\beq
\label{eq:cat}
\beta_j = Ae^{ij\theta} +Be^{-ij\theta} \quad , \quad \theta=\cos^{-1}\frac{E}{2}.
\eeq
The boundary conditions~\eqref{E:bdr1},\eqref{E:bdr2} give the ratio
\beq
\label{eq:dog}
\frac{A}{B}=\frac{\Delta - e^{-i\theta}}{e^{i\theta}-\Delta}=e^{-2i\theta(L+1)}\left(\frac{\tilde\Delta - e^{i\theta}}{e^{-i\theta}-\tilde\Delta}\right),
\eeq
and the eigenenergies, $E=2\cos \theta$, in implicit form
\beq
(\Delta + \tilde\Delta)\sin \theta L - \Delta\tilde\Delta \sin \theta (L+1)-\sin \theta (L-1)=0.
\eeq
In terms of Chebyshev polynomials of the second kind
\beq
U_n(\cos \theta) = \frac{\sin (n+1)\theta}{\sin \theta},
\eeq
and recalling the recursion relation
\beq
U_{n}(x)=2xU_{n-1} (x)-U_{n-2}(x),
\eeq
we find 
\beq
\boxed{\frac{\Delta\tilde\Delta E - \Delta - \tilde\Delta}{\Delta\tilde\Delta - 1}=\frac{U_{L-2}(E/2)}{U_{L-1}(E/2)}} 
\eeq
as in Eq.\eqref{E:E_constraint} of the main text. In the simplest case where $\lambda=\mu$ and $|t_{c,1}|^2=|t_{c,2}|^2$ (identical dots, identical couplings), we have
\beq\label{E:consistency}
\frac{(E-\lambda)\left(E(E-\lambda)-2|t_{c,1}|^2\right)}{(E-\lambda+|t_{c,1}|^2)(E-\lambda-|t_{c,1}|^2)}=\frac{U_{L-2}(E/2)}{U_{L-1}(E/2)} .
\eeq
The system's $L+2$ eigenvalues coincide with the solutions of the above equation. Notice that for small $|t_{c,1}|$ the LHS is $\sim E+O(|t_{c,1}|^2)$ when $E$ is not in the vicinity of the singularities $\lambda\pm |t_{c,1}|^2$. This is illustrated in Fig.~\ref{F:evenN_ab_maintext} of the main text for $L$ even ($L=10$), and in Fig.~\ref{F:oddN_ab_maintext} for $L$ odd ($L=11$). In this symmetric case for which $\Delta=\tilde\Delta$, all eigenstates are either symmetric or antisymmetric. This is seen from~\eqref{eq:dog}, which becomes
$A/B=e^{-2i\theta (L+1)}(A/B)^{-1}$, implying
\beq
\left(\frac{A}{B}\right)=1\quad\text{ and }\quad e^{-i\theta (L+1)}=\pm \frac{A}{B}.
\eeq
Substituting in~\eqref{eq:cat} gives $\beta_{L+1-j}=\pm\beta_j$.We must consider two cases: when $E$ is outside the energy band of the leads, and when it is within this energy band.

\begin{figure}
  \centering
  	\includegraphics[]{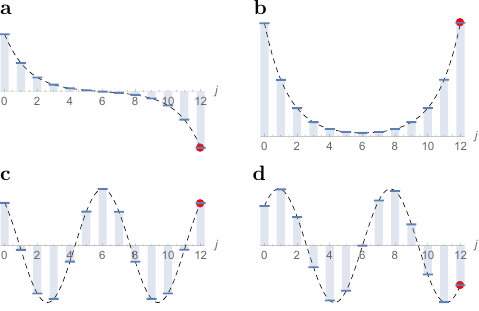}
\caption{(Color on line.) States from the case $L=11$, and $|t_{c,1}|=|t_{c,2}|=0.1$. \textbf{a,b} The two bound states for $\lambda=\mu=2.5$ \textbf{a}~Antisymmetric bound state, $E=2.504980$. \textbf{b}~Symmetric bound state, $E=2.504987$. The energy separation is $\sim 10^{-5}$. \textbf{c,d} The two hybridized states for $\lambda=\mu=1.2$. \textbf{c}~Symmetric hybridized state, $E=1.1990$. \textbf{d}~Antisymmetric hybridized state, $E=1.2142$. The energy separation is $\sim 10^{-2}$. In each graph, the red dot represents the value of $\beta_{L+1}$ from the consistency condition \eqref{E:bdr2}.}\label{F:oddNgraph_bound_hybrid}
\end{figure}

\subsection{Bound states ($E$ outside the band)}

When $E$ is outside the energy band of the leads, we have $\theta=i\xi$ if $E>2$, and $\theta=\pi +i\xi$ if $E<-2$, where $\xi=\cosh^{-1}(|E|/2)$ is a real parameter. Then
\beq
\beta_{j}=(\pm 1)^j \mathcal{N}\left[\tfrac{|E|\pm\lambda}{|t_{c,1}|^2}\sinh j\xi - \sinh (j-1)\xi\right]
\eeq
with $\pm = \text{sgn}(E)$, where $\mathcal{N}$ is a normalization factor, and where the eigenenergies satisfy the constraint
\beq
\Delta\tilde\Delta\sinh (L+1)\xi\mp (\Delta+\tilde\Delta)\sinh L\xi+\sinh (L-1)\xi=0.
\eeq
Alternatively, we can write the solution as
\beq
\beta_{L+1-j}=(\pm 1)^j \mathcal{N}'\left[\tfrac{|E|\pm\mu}{|t_{c,2}|^2}\sinh j\xi - \sinh (j-1)\xi\right].
\eeq
A generic feature of these bound states is their large amplitude at the dots. Because there are at least $L$ states from the continuous spectrum (see next section), there are at most two bound states. The bound states for $L=10$ and $\lambda=\mu=2.5$ are displayed in Fig.\ref{F:evenNgraph_boundstates} \textbf{a,b} of the main text. The bound states for $L=11$ are plotted in Fig.~\ref{F:oddNgraph_bound_hybrid} \textbf{a,b}. 

In the limit $L\to\infty$, the constraint is equivalent to
\beq
e^{2\xi}\mp\left(\frac{1}{\Delta}+\frac{1}{\tilde\Delta}\right)e^{\xi}+\frac{1}{\Delta\tilde\Delta}=\left(e^{\xi}\mp\frac{1}{\Delta}\right)\left(e^{\xi}\mp\frac{1}{\tilde\Delta}\right)=0,
\eeq
yielding $e^{-\xi}=\pm\Delta$ or $e^{-\xi}=\pm\tilde\Delta$. The resulting expressions for $\beta_{j}$ and $\beta_{L+1-j}$ describe bound states localized at either dot and decaying exponentially with distance over the characteristic length $\xi^{-1}$. Moreover, the constraint equations for $e^{-\xi}$ are equivalent to the fixed-point relations $E=\lambda \pm |t_{c,1}|^2 \Sigma(E)$ and $E=\mu \pm |t_{c,2}|^2 \Sigma(E)$, respectively. These relations can be obtained as normalizability conditions on the eigenstates of a dot connected to a \textit{semi-infinite} lead, by solving the corresponding Schr{\"o}dinger equation. Alternatively, the Green's function treatment of the core with semi-infinite leads, Appendix~\ref{S:effective_H_appendix}, gives the same fixed-point relations as effective self-energies of the core once the leads are traced out. The bound state condition on $\Delta$ for $E\geq 2$ amounts to
\beq
\lambda_k = \left(1-\frac{|t_c|^2}{2}\right)E+|t_c|^2\sqrt{\left(\frac{E}{2}\right)^2-1}\; ,
\eeq
or equivalently
\beq\label{E:fixed_point_solution_appendix}
E=\frac{1}{1-|t_c|^2}\left[\left(1-\frac{|t_c|^2}{2}\right)\lambda_k - \frac{|t_c|^2}{2}\sqrt{\lambda_k^2 - 4(1-|t_c|^2)}\right] ,
\eeq
with $\lambda_k\geq 2-|t_c|^2$, in agreement with~\eqref{E:fx_pts_E_1} from the main text. For $E\leq -2$, the condition on $\Delta$ gives instead
\beq\label{E:fixed_point_solution_appendix_negE}
E=\frac{1}{1+|t_c|^2}\left[\left(1+\frac{|t_c|^2}{2}\right)\lambda_k + \frac{|t_c|^2}{2}\sqrt{\lambda_k^2 - 4(1+|t_c|^2)}\right] ,
\eeq
with $\lambda_k\leq -2+|t_c|^2$. Similar relations hold for the bound state condition on $\tilde\Delta$.

\subsection{Scattering states ($E$ within the band)}

When $E\in[-2,2]$, $\theta$ is real and we find
\beq
\beta_{j}=\mathcal{N}\Big[\tfrac{E-\lambda}{|t_{c,1}|^2}\sin j\theta-\sin (j-1)\theta\Big],
\eeq
where $\mathcal{N}$ is a normalization factor. For any finite $L$, there are $L$ scattering states from the (perturbed) continuous spectrum of the lead. A generic feature of these $L$ states is their small amplitude at the quantum dots. Additionally, there can be two scattering hybridized states. A generic feature of these states is their relatively large amplitude on the dots. The hybridized states for $L=10$ and $\lambda=\mu=1.4$ are displayed in Fig.~\ref{F:evenNgraph_hybrid} \textbf{a,b}.
States from the (perturbed) continuous spectrum are displayed in Figs.~\ref{F:evenNgraph_boundstates} \textbf{c,d}, and~\ref{F:evenNgraph_hybrid} \textbf{c,d}.  The hybridized states for $L=11$ and $\lambda=\mu=1.2$ are displayed in Fig.~\ref{F:oddNgraph_bound_hybrid} \textbf{c,d}.

\bibliographystyle{apsrev4-1}
\bibliography{refs4dot}

\begin{thebibliography}{94}%
\makeatletter
\providecommand \@ifxundefined [1]{%
 \@ifx{#1\undefined}
}%
\providecommand \@ifnum [1]{%
 \ifnum #1\expandafter \@firstoftwo
 \else \expandafter \@secondoftwo
 \fi
}%
\providecommand \@ifx [1]{%
 \ifx #1\expandafter \@firstoftwo
 \else \expandafter \@secondoftwo
 \fi
}%
\providecommand \natexlab [1]{#1}%
\providecommand \enquote  [1]{``#1''}%
\providecommand \bibnamefont  [1]{#1}%
\providecommand \bibfnamefont [1]{#1}%
\providecommand \citenamefont [1]{#1}%
\providecommand \href@noop [0]{\@secondoftwo}%
\providecommand \href [0]{\begingroup \@sanitize@url \@href}%
\providecommand \@href[1]{\@@startlink{#1}\@@href}%
\providecommand \@@href[1]{\endgroup#1\@@endlink}%
\providecommand \@sanitize@url [0]{\catcode `\\12\catcode `\$12\catcode
  `\&12\catcode `\#12\catcode `\^12\catcode `\_12\catcode `\%12\relax}%
\providecommand \@@startlink[1]{}%
\providecommand \@@endlink[0]{}%
\providecommand \url  [0]{\begingroup\@sanitize@url \@url }%
\providecommand \@url [1]{\endgroup\@href {#1}{\urlprefix }}%
\providecommand \urlprefix  [0]{URL }%
\providecommand \Eprint [0]{\href }%
\providecommand \doibase [0]{http://dx.doi.org/}%
\providecommand \selectlanguage [0]{\@gobble}%
\providecommand \bibinfo  [0]{\@secondoftwo}%
\providecommand \bibfield  [0]{\@secondoftwo}%
\providecommand \translation [1]{[#1]}%
\providecommand \BibitemOpen [0]{}%
\providecommand \bibitemStop [0]{}%
\providecommand \bibitemNoStop [0]{.\EOS\space}%
\providecommand \EOS [0]{\spacefactor3000\relax}%
\providecommand \BibitemShut  [1]{\csname bibitem#1\endcsname}%
\let\auto@bib@innerbib\@empty
\bibitem [{\citenamefont {Nielsen}\ and\ \citenamefont
  {Chuang}(2010)}]{nielsen2010quantum}%
  \BibitemOpen
  \bibfield  {author} {\bibinfo {author} {\bibfnamefont {M.}~\bibnamefont
  {Nielsen}}\ and\ \bibinfo {author} {\bibfnamefont {I.}~\bibnamefont
  {Chuang}},\ }\href@noop {} {\emph {\bibinfo {title} {Quantum Computation and
  Quantum Information}}}\ (\bibinfo  {publisher} {Cambridge University Press},\
  \bibinfo {address} {Cambridge},\ \bibinfo {year} {2010})\ \bibinfo {note}
  {10th Anniversary Edition}\BibitemShut {NoStop}%
\bibitem [{\citenamefont {Altman}\ \emph {et~al.}(2021)\citenamefont {Altman},
  \citenamefont {Brown}, \citenamefont {Carleo}, \citenamefont {Carr},
  \citenamefont {Demler}, \citenamefont {Chin}, \citenamefont {DeMarco},
  \citenamefont {Economou}, \citenamefont {Eriksson}, \citenamefont {Fu},
  \citenamefont {Greiner}, \citenamefont {Hazzard}, \citenamefont {Hulet},
  \citenamefont {Koll\'ar}, \citenamefont {Lev}, \citenamefont {Lukin},
  \citenamefont {Ma}, \citenamefont {Mi}, \citenamefont {Misra}, \citenamefont
  {Monroe}, \citenamefont {Murch}, \citenamefont {Nazario}, \citenamefont {Ni},
  \citenamefont {Potter}, \citenamefont {Roushan}, \citenamefont {Saffman},
  \citenamefont {Schleier-Smith}, \citenamefont {Siddiqi}, \citenamefont
  {Simmonds}, \citenamefont {Singh}, \citenamefont {Spielman}, \citenamefont
  {Temme}, \citenamefont {Weiss}, \citenamefont {Vu\ifmmode \check{c}\else
  \v{c}\fi{}kovi\ifmmode~\acute{c}\else \'{c}\fi{}}, \citenamefont
  {Vuleti\ifmmode~\acute{c}\else \'{c}\fi{}}, \citenamefont {Ye},\ and\
  \citenamefont {Zwierlein}}]{altman2021quantum}%
  \BibitemOpen
  \bibfield  {author} {\bibinfo {author} {\bibfnamefont {E.}~\bibnamefont
  {Altman}}, \bibinfo {author} {\bibfnamefont {K.~R.}\ \bibnamefont {Brown}},
  \bibinfo {author} {\bibfnamefont {G.}~\bibnamefont {Carleo}}, \bibinfo
  {author} {\bibfnamefont {L.~D.}\ \bibnamefont {Carr}}, \bibinfo {author}
  {\bibfnamefont {E.}~\bibnamefont {Demler}}, \bibinfo {author} {\bibfnamefont
  {C.}~\bibnamefont {Chin}}, \bibinfo {author} {\bibfnamefont {B.}~\bibnamefont
  {DeMarco}}, \bibinfo {author} {\bibfnamefont {S.~E.}\ \bibnamefont
  {Economou}}, \bibinfo {author} {\bibfnamefont {M.~A.}\ \bibnamefont
  {Eriksson}}, \bibinfo {author} {\bibfnamefont {K.-M.~C.}\ \bibnamefont {Fu}},
  \bibinfo {author} {\bibfnamefont {M.}~\bibnamefont {Greiner}}, \bibinfo
  {author} {\bibfnamefont {K.~R.}\ \bibnamefont {Hazzard}}, \bibinfo {author}
  {\bibfnamefont {R.~G.}\ \bibnamefont {Hulet}}, \bibinfo {author}
  {\bibfnamefont {A.~J.}\ \bibnamefont {Koll\'ar}}, \bibinfo {author}
  {\bibfnamefont {B.~L.}\ \bibnamefont {Lev}}, \bibinfo {author} {\bibfnamefont
  {M.~D.}\ \bibnamefont {Lukin}}, \bibinfo {author} {\bibfnamefont
  {R.}~\bibnamefont {Ma}}, \bibinfo {author} {\bibfnamefont {X.}~\bibnamefont
  {Mi}}, \bibinfo {author} {\bibfnamefont {S.}~\bibnamefont {Misra}}, \bibinfo
  {author} {\bibfnamefont {C.}~\bibnamefont {Monroe}}, \bibinfo {author}
  {\bibfnamefont {K.}~\bibnamefont {Murch}}, \bibinfo {author} {\bibfnamefont
  {Z.}~\bibnamefont {Nazario}}, \bibinfo {author} {\bibfnamefont {K.-K.}\
  \bibnamefont {Ni}}, \bibinfo {author} {\bibfnamefont {A.~C.}\ \bibnamefont
  {Potter}}, \bibinfo {author} {\bibfnamefont {P.}~\bibnamefont {Roushan}},
  \bibinfo {author} {\bibfnamefont {M.}~\bibnamefont {Saffman}}, \bibinfo
  {author} {\bibfnamefont {M.}~\bibnamefont {Schleier-Smith}}, \bibinfo
  {author} {\bibfnamefont {I.}~\bibnamefont {Siddiqi}}, \bibinfo {author}
  {\bibfnamefont {R.}~\bibnamefont {Simmonds}}, \bibinfo {author}
  {\bibfnamefont {M.}~\bibnamefont {Singh}}, \bibinfo {author} {\bibfnamefont
  {I.}~\bibnamefont {Spielman}}, \bibinfo {author} {\bibfnamefont
  {K.}~\bibnamefont {Temme}}, \bibinfo {author} {\bibfnamefont {D.~S.}\
  \bibnamefont {Weiss}}, \bibinfo {author} {\bibfnamefont {J.}~\bibnamefont
  {Vu\ifmmode \check{c}\else \v{c}\fi{}kovi\ifmmode~\acute{c}\else
  \'{c}\fi{}}}, \bibinfo {author} {\bibfnamefont {V.}~\bibnamefont
  {Vuleti\ifmmode~\acute{c}\else \'{c}\fi{}}}, \bibinfo {author} {\bibfnamefont
  {J.}~\bibnamefont {Ye}}, \ and\ \bibinfo {author} {\bibfnamefont
  {M.}~\bibnamefont {Zwierlein}},\ }\href {\doibase
  10.1103/PRXQuantum.2.017003} {\bibfield  {journal} {\bibinfo  {journal} {PRX
  Quantum}\ }\textbf {\bibinfo {volume} {2}},\ \bibinfo {pages} {017003}
  (\bibinfo {year} {2021})}\BibitemShut {NoStop}%
\bibitem [{\citenamefont {Zeng}\ \emph {et~al.}(2015)\citenamefont {Zeng},
  \citenamefont {Chen}, \citenamefont {Zhou},\ and\ \citenamefont
  {Wen}}]{zeng2015quantum}%
  \BibitemOpen
  \bibfield  {author} {\bibinfo {author} {\bibfnamefont {B.}~\bibnamefont
  {Zeng}}, \bibinfo {author} {\bibfnamefont {X.}~\bibnamefont {Chen}}, \bibinfo
  {author} {\bibfnamefont {D.}~\bibnamefont {Zhou}}, \ and\ \bibinfo {author}
  {\bibfnamefont {X.-G.}\ \bibnamefont {Wen}},\ }\href@noop {} {\emph {\bibinfo
  {title} {Quantum Information Meets Quantum Matter -- From Quantum
  Entanglement to Topological Phase in Many-Body Systems}}}\ (\bibinfo
  {publisher} {Springer},\ \bibinfo {address} {New York},\ \bibinfo {year}
  {2015})\BibitemShut {NoStop}%
\bibitem [{\citenamefont {Giovannetti}\ \emph {et~al.}(2004)\citenamefont
  {Giovannetti}, \citenamefont {Lloyd},\ and\ \citenamefont
  {Maccone}}]{giovannetti2004quantum-enchanced}%
  \BibitemOpen
  \bibfield  {author} {\bibinfo {author} {\bibfnamefont {V.}~\bibnamefont
  {Giovannetti}}, \bibinfo {author} {\bibfnamefont {S.}~\bibnamefont {Lloyd}},
  \ and\ \bibinfo {author} {\bibfnamefont {L.}~\bibnamefont {Maccone}},\ }\href
  {\doibase 10.1126/science.1104149} {\bibfield  {journal} {\bibinfo  {journal}
  {Science (New York, N.Y.)}\ }\textbf {\bibinfo {volume} {306}},\ \bibinfo
  {pages} {1330} (\bibinfo {year} {2004})}\BibitemShut {NoStop}%
\bibitem [{\citenamefont {{Reck}}\ \emph {et~al.}(1994)\citenamefont {{Reck}},
  \citenamefont {{Zeilinger}}, \citenamefont {{Bernstein}},\ and\ \citenamefont
  {{Bertani}}}]{reck1994experimental}%
  \BibitemOpen
  \bibfield  {author} {\bibinfo {author} {\bibfnamefont {M.}~\bibnamefont
  {{Reck}}}, \bibinfo {author} {\bibfnamefont {A.}~\bibnamefont {{Zeilinger}}},
  \bibinfo {author} {\bibfnamefont {H.~J.}\ \bibnamefont {{Bernstein}}}, \ and\
  \bibinfo {author} {\bibfnamefont {P.}~\bibnamefont {{Bertani}}},\ }\href
  {\doibase 10.1103/PhysRevLett.73.58} {\bibfield  {journal} {\bibinfo
  {journal} {\prl}\ }\textbf {\bibinfo {volume} {73}},\ \bibinfo {pages} {58}
  (\bibinfo {year} {1994})}\BibitemShut {NoStop}%
\bibitem [{\citenamefont {DiVincenzo}(1995)}]{divincenzo1995twobit}%
  \BibitemOpen
  \bibfield  {author} {\bibinfo {author} {\bibfnamefont {D.~P.}\ \bibnamefont
  {DiVincenzo}},\ }\href {\doibase 10.1103/PhysRevA.51.1015} {\bibfield
  {journal} {\bibinfo  {journal} {Phys. Rev. A}\ }\textbf {\bibinfo {volume}
  {51}},\ \bibinfo {pages} {1015} (\bibinfo {year} {1995})}\BibitemShut
  {NoStop}%
\bibitem [{\citenamefont {Barenco}(1995)}]{Barenco1995universal}%
  \BibitemOpen
  \bibfield  {author} {\bibinfo {author} {\bibfnamefont {A.}~\bibnamefont
  {Barenco}},\ }\href@noop {} {\bibfield  {journal} {\bibinfo  {journal}
  {Proceedings of the Royal Society of London. Series A: Mathematical and
  Physical Sciences}\ }\textbf {\bibinfo {volume} {449}},\ \bibinfo {pages}
  {679 } (\bibinfo {year} {1995})}\BibitemShut {NoStop}%
\bibitem [{\citenamefont {Barenco}\ \emph {et~al.}(1995)\citenamefont
  {Barenco}, \citenamefont {Bennett}, \citenamefont {Cleve}, \citenamefont
  {DiVincenzo}, \citenamefont {Margolus}, \citenamefont {Shor}, \citenamefont
  {Sleator}, \citenamefont {Smolin},\ and\ \citenamefont
  {Weinfurter}}]{barenco1995elementary}%
  \BibitemOpen
  \bibfield  {author} {\bibinfo {author} {\bibfnamefont {A.}~\bibnamefont
  {Barenco}}, \bibinfo {author} {\bibfnamefont {C.}~\bibnamefont {Bennett}},
  \bibinfo {author} {\bibfnamefont {R.}~\bibnamefont {Cleve}}, \bibinfo
  {author} {\bibfnamefont {D.}~\bibnamefont {DiVincenzo}}, \bibinfo {author}
  {\bibfnamefont {N.}~\bibnamefont {Margolus}}, \bibinfo {author}
  {\bibfnamefont {P.}~\bibnamefont {Shor}}, \bibinfo {author} {\bibfnamefont
  {T.}~\bibnamefont {Sleator}}, \bibinfo {author} {\bibfnamefont
  {J.}~\bibnamefont {Smolin}}, \ and\ \bibinfo {author} {\bibfnamefont
  {H.}~\bibnamefont {Weinfurter}},\ }\href {\doibase 10.1103/PhysRevA.52.3457}
  {\bibfield  {journal} {\bibinfo  {journal} {Physical Review A}\ }\textbf
  {\bibinfo {volume} {52}} (\bibinfo {year} {1995}),\
  10.1103/PhysRevA.52.3457}\BibitemShut {NoStop}%
\bibitem [{\citenamefont {Kitaev}(1997)}]{kitaev1997quantum}%
  \BibitemOpen
  \bibfield  {author} {\bibinfo {author} {\bibfnamefont {A.~Y.}\ \bibnamefont
  {Kitaev}},\ }\href {\doibase 10.1070/rm1997v052n06abeh002155} {\bibfield
  {journal} {\bibinfo  {journal} {Russian Mathematical Surveys}\ }\textbf
  {\bibinfo {volume} {52}},\ \bibinfo {pages} {1191} (\bibinfo {year}
  {1997})}\BibitemShut {NoStop}%
\bibitem [{\citenamefont {Deutsch}\ \emph {et~al.}(1995)\citenamefont
  {Deutsch}, \citenamefont {Barenco},\ and\ \citenamefont
  {Ekert}}]{deutsch1995universality}%
  \BibitemOpen
  \bibfield  {author} {\bibinfo {author} {\bibfnamefont {D.}~\bibnamefont
  {Deutsch}}, \bibinfo {author} {\bibfnamefont {A.}~\bibnamefont {Barenco}}, \
  and\ \bibinfo {author} {\bibfnamefont {A.}~\bibnamefont {Ekert}},\ }\href
  {\doibase 10.1098/rspa.1995.0065} {\bibfield  {journal} {\bibinfo  {journal}
  {Proc. Roy. Soc. of London}\ }\textbf {\bibinfo {volume} {449}} (\bibinfo
  {year} {1995}),\ 10.1098/rspa.1995.0065}\BibitemShut {NoStop}%
\bibitem [{\citenamefont {Lloyd}(1995)}]{lloyd1995almost}%
  \BibitemOpen
  \bibfield  {author} {\bibinfo {author} {\bibfnamefont {S.}~\bibnamefont
  {Lloyd}},\ }\href {\doibase 10.1103/PhysRevLett.75.346} {\bibfield  {journal}
  {\bibinfo  {journal} {Phys. Rev. Lett.}\ }\textbf {\bibinfo {volume} {75}},\
  \bibinfo {pages} {346} (\bibinfo {year} {1995})}\BibitemShut {NoStop}%
\bibitem [{\citenamefont {Bacon}\ \emph {et~al.}(2001)\citenamefont {Bacon},
  \citenamefont {Kempe}, \citenamefont {DiVincenzo}, \citenamefont {Lidar},\
  and\ \citenamefont {Whaley}}]{bacon2001encoded}%
  \BibitemOpen
  \bibfield  {author} {\bibinfo {author} {\bibfnamefont {D.}~\bibnamefont
  {Bacon}}, \bibinfo {author} {\bibfnamefont {J.}~\bibnamefont {Kempe}},
  \bibinfo {author} {\bibfnamefont {D.~P.}\ \bibnamefont {DiVincenzo}},
  \bibinfo {author} {\bibfnamefont {D.~A.}\ \bibnamefont {Lidar}}, \ and\
  \bibinfo {author} {\bibfnamefont {K.~B.}\ \bibnamefont {Whaley}},\
  }\href@noop {} {\enquote {\bibinfo {title} {Encoded universality in physical
  implementations of a quantum computer},}\ } (\bibinfo {year} {2001}),\
  \Eprint {http://arxiv.org/abs/quant-ph/0102140} {arXiv:quant-ph/0102140
  [quant-ph]} \BibitemShut {NoStop}%
\bibitem [{\citenamefont {Kempe}\ \emph {et~al.}(2001)\citenamefont {Kempe},
  \citenamefont {Bacon}, \citenamefont {DiVincenzo},\ and\ \citenamefont
  {Whaley}}]{kempe2001encoded}%
  \BibitemOpen
  \bibfield  {author} {\bibinfo {author} {\bibfnamefont {J.}~\bibnamefont
  {Kempe}}, \bibinfo {author} {\bibfnamefont {D.}~\bibnamefont {Bacon}},
  \bibinfo {author} {\bibfnamefont {D.~P.}\ \bibnamefont {DiVincenzo}}, \ and\
  \bibinfo {author} {\bibfnamefont {K.~B.}\ \bibnamefont {Whaley}},\
  }\href@noop {} {\enquote {\bibinfo {title} {Encoded universality from a
  single physical interaction},}\ } (\bibinfo {year} {2001}),\ \Eprint
  {http://arxiv.org/abs/quant-ph/0112013} {arXiv:quant-ph/0112013 [quant-ph]}
  \BibitemShut {NoStop}%
\bibitem [{\citenamefont {Aharonov}(2003)}]{aharonov2003simple}%
  \BibitemOpen
  \bibfield  {author} {\bibinfo {author} {\bibfnamefont {D.}~\bibnamefont
  {Aharonov}},\ }\href@noop {} {\enquote {\bibinfo {title} {A simple proof that
  toffoli and hadamard are quantum universal},}\ } (\bibinfo {year} {2003}),\
  \Eprint {http://arxiv.org/abs/quant-ph/0301040} {arXiv:quant-ph/0301040
  [quant-ph]} \BibitemShut {NoStop}%
\bibitem [{\citenamefont {Shor}(1997)}]{shor1997polynomial}%
  \BibitemOpen
  \bibfield  {author} {\bibinfo {author} {\bibfnamefont {P.~W.}\ \bibnamefont
  {Shor}},\ }\href {\doibase 10.1137/s0097539795293172} {\bibfield  {journal}
  {\bibinfo  {journal} {SIAM Journal on Computing}\ }\textbf {\bibinfo {volume}
  {26}},\ \bibinfo {pages} {1484–1509} (\bibinfo {year} {1997})}\BibitemShut
  {NoStop}%
\bibitem [{\citenamefont {Lloyd}(1996)}]{lloyd1996universal}%
  \BibitemOpen
  \bibfield  {author} {\bibinfo {author} {\bibfnamefont {S.}~\bibnamefont
  {Lloyd}},\ }\href {\doibase 10.1126/science.273.5278.1073} {\bibfield
  {journal} {\bibinfo  {journal} {Science}\ }\textbf {\bibinfo {volume}
  {273}},\ \bibinfo {pages} {1073} (\bibinfo {year} {1996})},\ \Eprint
  {http://arxiv.org/abs/https://science.sciencemag.org/content/273/5278/\\
  1073.full.pdf} {https://science.sciencemag.org/content/273/5278/\\
  1073.full.pdf} \BibitemShut {NoStop}%
\bibitem [{\citenamefont {Raz}(1999)}]{raz1999exponential}%
  \BibitemOpen
  \bibfield  {author} {\bibinfo {author} {\bibfnamefont {R.}~\bibnamefont
  {Raz}},\ }in\ \href {\doibase 10.1145/301250.301343} {\emph {\bibinfo
  {booktitle} {Proceedings of the Thirty-First Annual ACM Symposium on Theory
  of Computing}}},\ \bibinfo {series and number} {STOC '99}\ (\bibinfo
  {publisher} {Association for Computing Machinery},\ \bibinfo {address} {New
  York, NY, USA},\ \bibinfo {year} {1999})\ p.\ \bibinfo {pages}
  {358–367}\BibitemShut {NoStop}%
\bibitem [{\citenamefont {Gavinsky}\ \emph {et~al.}(2007)\citenamefont
  {Gavinsky}, \citenamefont {Kempe}, \citenamefont {Kerenidis}, \citenamefont
  {Raz},\ and\ \citenamefont {de~Wolf}}]{gavinsky2007exponential}%
  \BibitemOpen
  \bibfield  {author} {\bibinfo {author} {\bibfnamefont {D.}~\bibnamefont
  {Gavinsky}}, \bibinfo {author} {\bibfnamefont {J.}~\bibnamefont {Kempe}},
  \bibinfo {author} {\bibfnamefont {I.}~\bibnamefont {Kerenidis}}, \bibinfo
  {author} {\bibfnamefont {R.}~\bibnamefont {Raz}}, \ and\ \bibinfo {author}
  {\bibfnamefont {R.}~\bibnamefont {de~Wolf}},\ }in\ \href {\doibase
  10.1145/1250790.1250866} {\emph {\bibinfo {booktitle} {Proceedings of the
  Thirty-Ninth Annual ACM Symposium on Theory of Computing}}},\ \bibinfo
  {series and number} {STOC '07}\ (\bibinfo  {publisher} {Association for
  Computing Machinery},\ \bibinfo {address} {New York, NY, USA},\ \bibinfo
  {year} {2007})\ p.\ \bibinfo {pages} {516–525}\BibitemShut {NoStop}%
\bibitem [{\citenamefont {Le~Gall}(2006)}]{legall2006exponential}%
  \BibitemOpen
  \bibfield  {author} {\bibinfo {author} {\bibfnamefont {F.}~\bibnamefont
  {Le~Gall}},\ }\href {\doibase 10.1007/s00224-007-9097-3} {\bibfield
  {journal} {\bibinfo  {journal} {Theory of Computing Systems}\ }\textbf
  {\bibinfo {volume} {45}} (\bibinfo {year} {2006}),\
  10.1007/s00224-007-9097-3}\BibitemShut {NoStop}%
\bibitem [{\citenamefont {Gottesman}(1997)}]{gottesman1997stabilizer}%
  \BibitemOpen
  \bibfield  {author} {\bibinfo {author} {\bibfnamefont {D.}~\bibnamefont
  {Gottesman}},\ }\href@noop {} {\enquote {\bibinfo {title} {Stabilizer codes
  and quantum error correction},}\ } (\bibinfo {year} {1997}),\ \bibinfo {note}
  {{Caltech} Ph.D. thesis},\ \Eprint {http://arxiv.org/abs/quant-ph/9705052}
  {arXiv:quant-ph/9705052 [quant-ph]} \BibitemShut {NoStop}%
\bibitem [{\citenamefont {Valiant}(2001)}]{valiant2001quantum}%
  \BibitemOpen
  \bibfield  {author} {\bibinfo {author} {\bibfnamefont {L.~G.}\ \bibnamefont
  {Valiant}},\ }in\ \href {\doibase 10.1145/380752.380785} {\emph {\bibinfo
  {booktitle} {Proceedings of the Thirty-Third Annual ACM Symposium on Theory
  of Computing}}},\ \bibinfo {series and number} {STOC '01}\ (\bibinfo
  {publisher} {Association for Computing Machinery},\ \bibinfo {address} {New
  York, NY, USA},\ \bibinfo {year} {2001})\ p.\ \bibinfo {pages}
  {114–123}\BibitemShut {NoStop}%
\bibitem [{\citenamefont {Terhal}\ and\ \citenamefont
  {DiVincenzo}(2001)}]{terhal2001classical}%
  \BibitemOpen
  \bibfield  {author} {\bibinfo {author} {\bibfnamefont {B.}~\bibnamefont
  {Terhal}}\ and\ \bibinfo {author} {\bibfnamefont {D.}~\bibnamefont
  {DiVincenzo}},\ }\href {\doibase 10.1103/PhysRevA.65.032325} {\bibfield
  {journal} {\bibinfo  {journal} {Physical Review A}\ }\textbf {\bibinfo
  {volume} {65}} (\bibinfo {year} {2001}),\
  10.1103/PhysRevA.65.032325}\BibitemShut {NoStop}%
\bibitem [{\citenamefont {Knill}(2001)}]{knill2001fermionic}%
  \BibitemOpen
  \bibfield  {author} {\bibinfo {author} {\bibfnamefont {E.}~\bibnamefont
  {Knill}},\ }\href@noop {} {\enquote {\bibinfo {title} {Fermionic linear
  optics and matchgates},}\ } (\bibinfo {year} {2001}),\ \Eprint
  {http://arxiv.org/abs/quant-ph/0108033} {arXiv:quant-ph/0108033 [quant-ph]}
  \BibitemShut {NoStop}%
\bibitem [{\citenamefont {Knill}\ \emph {et~al.}(2001)\citenamefont {Knill},
  \citenamefont {Laflamme},\ and\ \citenamefont {Milburn}}]{knill2001scheme}%
  \BibitemOpen
  \bibfield  {author} {\bibinfo {author} {\bibfnamefont {E.}~\bibnamefont
  {Knill}}, \bibinfo {author} {\bibfnamefont {R.}~\bibnamefont {Laflamme}}, \
  and\ \bibinfo {author} {\bibfnamefont {G.~J.}\ \bibnamefont {Milburn}},\
  }\href {\doibase 10.1038/35051009} {\bibfield  {journal} {\bibinfo  {journal}
  {Nature}\ }\textbf {\bibinfo {volume} {409}},\ \bibinfo {pages} {46}
  (\bibinfo {year} {2001})}\BibitemShut {NoStop}%
\bibitem [{\citenamefont {Bravyi}\ and\ \citenamefont
  {Kitaev}(2000)}]{bravyi2000fermionic}%
  \BibitemOpen
  \bibfield  {author} {\bibinfo {author} {\bibfnamefont {S.}~\bibnamefont
  {Bravyi}}\ and\ \bibinfo {author} {\bibfnamefont {A.}~\bibnamefont
  {Kitaev}},\ }\href {\doibase 10.1006/aphy.2002.6254} {\bibfield  {journal}
  {\bibinfo  {journal} {Annals of Physics}\ }\textbf {\bibinfo {volume} {298}}
  (\bibinfo {year} {2000}),\ 10.1006/aphy.2002.6254}\BibitemShut {NoStop}%
\bibitem [{\citenamefont {Valiant}(1979)}]{valiant1979complexity}%
  \BibitemOpen
  \bibfield  {author} {\bibinfo {author} {\bibfnamefont {L.}~\bibnamefont
  {Valiant}},\ }\href {\doibase https://doi.org/10.1016/0304-3975(79)90044-6}
  {\bibfield  {journal} {\bibinfo  {journal} {Theoretical Computer Science}\
  }\textbf {\bibinfo {volume} {8}},\ \bibinfo {pages} {189} (\bibinfo {year}
  {1979})}\BibitemShut {NoStop}%
\bibitem [{\citenamefont {Loss}\ and\ \citenamefont
  {DiVincenzo}(1998)}]{loss1998quantum}%
  \BibitemOpen
  \bibfield  {author} {\bibinfo {author} {\bibfnamefont {D.}~\bibnamefont
  {Loss}}\ and\ \bibinfo {author} {\bibfnamefont {D.~P.}\ \bibnamefont
  {DiVincenzo}},\ }\href {\doibase 10.1103/PhysRevA.57.120} {\bibfield
  {journal} {\bibinfo  {journal} {Phys. Rev. A}\ }\textbf {\bibinfo {volume}
  {57}},\ \bibinfo {pages} {120} (\bibinfo {year} {1998})}\BibitemShut
  {NoStop}%
\bibitem [{\citenamefont {Zwanenburg}\ \emph {et~al.}(2013)\citenamefont
  {Zwanenburg}, \citenamefont {Dzurak}, \citenamefont {Morello}, \citenamefont
  {Simmons}, \citenamefont {Hollenberg}, \citenamefont {Klimeck}, \citenamefont
  {Rogge}, \citenamefont {Coppersmith},\ and\ \citenamefont
  {Eriksson}}]{zwanenburg2013silicon}%
  \BibitemOpen
  \bibfield  {author} {\bibinfo {author} {\bibfnamefont {F.~A.}\ \bibnamefont
  {Zwanenburg}}, \bibinfo {author} {\bibfnamefont {A.~S.}\ \bibnamefont
  {Dzurak}}, \bibinfo {author} {\bibfnamefont {A.}~\bibnamefont {Morello}},
  \bibinfo {author} {\bibfnamefont {M.~Y.}\ \bibnamefont {Simmons}}, \bibinfo
  {author} {\bibfnamefont {L.~C.}\ \bibnamefont {Hollenberg}}, \bibinfo
  {author} {\bibfnamefont {G.}~\bibnamefont {Klimeck}}, \bibinfo {author}
  {\bibfnamefont {S.}~\bibnamefont {Rogge}}, \bibinfo {author} {\bibfnamefont
  {S.~N.}\ \bibnamefont {Coppersmith}}, \ and\ \bibinfo {author} {\bibfnamefont
  {M.~A.}\ \bibnamefont {Eriksson}},\ }\href@noop {} {\bibfield  {journal}
  {\bibinfo  {journal} {Reviews of modern physics}\ }\textbf {\bibinfo {volume}
  {85}},\ \bibinfo {pages} {961} (\bibinfo {year} {2013})}\BibitemShut
  {NoStop}%
\bibitem [{\citenamefont {Watson}\ \emph {et~al.}(2018)\citenamefont {Watson},
  \citenamefont {Philips}, \citenamefont {Kawakami}, \citenamefont {Ward},
  \citenamefont {Scarlino}, \citenamefont {Veldhorst}, \citenamefont {Savage},
  \citenamefont {Lagally}, \citenamefont {Friesen}, \citenamefont {Coppersmith}
  \emph {et~al.}}]{watson2018programmable}%
  \BibitemOpen
  \bibfield  {author} {\bibinfo {author} {\bibfnamefont {T.}~\bibnamefont
  {Watson}}, \bibinfo {author} {\bibfnamefont {S.}~\bibnamefont {Philips}},
  \bibinfo {author} {\bibfnamefont {E.}~\bibnamefont {Kawakami}}, \bibinfo
  {author} {\bibfnamefont {D.}~\bibnamefont {Ward}}, \bibinfo {author}
  {\bibfnamefont {P.}~\bibnamefont {Scarlino}}, \bibinfo {author}
  {\bibfnamefont {M.}~\bibnamefont {Veldhorst}}, \bibinfo {author}
  {\bibfnamefont {D.}~\bibnamefont {Savage}}, \bibinfo {author} {\bibfnamefont
  {M.}~\bibnamefont {Lagally}}, \bibinfo {author} {\bibfnamefont
  {M.}~\bibnamefont {Friesen}}, \bibinfo {author} {\bibfnamefont
  {S.}~\bibnamefont {Coppersmith}},  \emph {et~al.},\ }\href@noop {} {\bibfield
   {journal} {\bibinfo  {journal} {nature}\ }\textbf {\bibinfo {volume}
  {555}},\ \bibinfo {pages} {633} (\bibinfo {year} {2018})}\BibitemShut
  {NoStop}%
\bibitem [{\citenamefont {Li}\ \emph {et~al.}(2018)\citenamefont {Li},
  \citenamefont {Cao}, \citenamefont {Yu}, \citenamefont {Xiao}, \citenamefont
  {Guo}, \citenamefont {Jiang},\ and\ \citenamefont {Guo}}]{li2018controlled}%
  \BibitemOpen
  \bibfield  {author} {\bibinfo {author} {\bibfnamefont {H.-O.}\ \bibnamefont
  {Li}}, \bibinfo {author} {\bibfnamefont {G.}~\bibnamefont {Cao}}, \bibinfo
  {author} {\bibfnamefont {G.-D.}\ \bibnamefont {Yu}}, \bibinfo {author}
  {\bibfnamefont {M.}~\bibnamefont {Xiao}}, \bibinfo {author} {\bibfnamefont
  {G.-C.}\ \bibnamefont {Guo}}, \bibinfo {author} {\bibfnamefont {H.-W.}\
  \bibnamefont {Jiang}}, \ and\ \bibinfo {author} {\bibfnamefont {G.-P.}\
  \bibnamefont {Guo}},\ }\href@noop {} {\bibfield  {journal} {\bibinfo
  {journal} {Physical Review Applied}\ }\textbf {\bibinfo {volume} {9}},\
  \bibinfo {pages} {024015} (\bibinfo {year} {2018})}\BibitemShut {NoStop}%
\bibitem [{\citenamefont {Nakamura}\ \emph {et~al.}(1999)\citenamefont
  {Nakamura}, \citenamefont {Pashkin},\ and\ \citenamefont
  {Tsai}}]{nakamura1999coherent}%
  \BibitemOpen
  \bibfield  {author} {\bibinfo {author} {\bibfnamefont {Y.}~\bibnamefont
  {Nakamura}}, \bibinfo {author} {\bibfnamefont {Y.~A.}\ \bibnamefont
  {Pashkin}}, \ and\ \bibinfo {author} {\bibfnamefont {J.~S.}\ \bibnamefont
  {Tsai}},\ }\href@noop {} {\bibfield  {journal} {\bibinfo  {journal} {nature}\
  }\textbf {\bibinfo {volume} {398}},\ \bibinfo {pages} {786} (\bibinfo {year}
  {1999})}\BibitemShut {NoStop}%
\bibitem [{\citenamefont {Devoret}\ and\ \citenamefont
  {Schoelkopf}(2013)}]{devoret2013superconducting}%
  \BibitemOpen
  \bibfield  {author} {\bibinfo {author} {\bibfnamefont {M.~H.}\ \bibnamefont
  {Devoret}}\ and\ \bibinfo {author} {\bibfnamefont {R.~J.}\ \bibnamefont
  {Schoelkopf}},\ }\href@noop {} {\bibfield  {journal} {\bibinfo  {journal}
  {Science}\ }\textbf {\bibinfo {volume} {339}},\ \bibinfo {pages} {1169}
  (\bibinfo {year} {2013})}\BibitemShut {NoStop}%
\bibitem [{\citenamefont {O'Brien}\ \emph {et~al.}(2003)\citenamefont
  {O'Brien}, \citenamefont {Pryde}, \citenamefont {White}, \citenamefont
  {Ralph},\ and\ \citenamefont {Branning}}]{o2003demonstration}%
  \BibitemOpen
  \bibfield  {author} {\bibinfo {author} {\bibfnamefont {J.~L.}\ \bibnamefont
  {O'Brien}}, \bibinfo {author} {\bibfnamefont {G.~J.}\ \bibnamefont {Pryde}},
  \bibinfo {author} {\bibfnamefont {A.~G.}\ \bibnamefont {White}}, \bibinfo
  {author} {\bibfnamefont {T.~C.}\ \bibnamefont {Ralph}}, \ and\ \bibinfo
  {author} {\bibfnamefont {D.}~\bibnamefont {Branning}},\ }\href@noop {}
  {\bibfield  {journal} {\bibinfo  {journal} {Nature}\ }\textbf {\bibinfo
  {volume} {426}},\ \bibinfo {pages} {264} (\bibinfo {year}
  {2003})}\BibitemShut {NoStop}%
\bibitem [{\citenamefont {Blatt}\ and\ \citenamefont
  {Wineland}(2008)}]{blatt2008entangled}%
  \BibitemOpen
  \bibfield  {author} {\bibinfo {author} {\bibfnamefont {R.}~\bibnamefont
  {Blatt}}\ and\ \bibinfo {author} {\bibfnamefont {D.}~\bibnamefont
  {Wineland}},\ }\href@noop {} {\bibfield  {journal} {\bibinfo  {journal}
  {Nature}\ }\textbf {\bibinfo {volume} {453}},\ \bibinfo {pages} {1008}
  (\bibinfo {year} {2008})}\BibitemShut {NoStop}%
\bibitem [{\citenamefont {Yao}\ \emph {et~al.}(2012)\citenamefont {Yao},
  \citenamefont {Jiang}, \citenamefont {Gorshkov}, \citenamefont {Maurer},
  \citenamefont {Giedke}, \citenamefont {Cirac},\ and\ \citenamefont
  {Lukin}}]{yao2012scalable}%
  \BibitemOpen
  \bibfield  {author} {\bibinfo {author} {\bibfnamefont {N.~Y.}\ \bibnamefont
  {Yao}}, \bibinfo {author} {\bibfnamefont {L.}~\bibnamefont {Jiang}}, \bibinfo
  {author} {\bibfnamefont {A.~V.}\ \bibnamefont {Gorshkov}}, \bibinfo {author}
  {\bibfnamefont {P.~C.}\ \bibnamefont {Maurer}}, \bibinfo {author}
  {\bibfnamefont {G.}~\bibnamefont {Giedke}}, \bibinfo {author} {\bibfnamefont
  {J.~I.}\ \bibnamefont {Cirac}}, \ and\ \bibinfo {author} {\bibfnamefont
  {M.~D.}\ \bibnamefont {Lukin}},\ }\href@noop {} {\bibfield  {journal}
  {\bibinfo  {journal} {Nature communications}\ }\textbf {\bibinfo {volume}
  {3}},\ \bibinfo {pages} {1} (\bibinfo {year} {2012})}\BibitemShut {NoStop}%
\bibitem [{\citenamefont {Nayak}\ \emph {et~al.}(2008)\citenamefont {Nayak},
  \citenamefont {Simon}, \citenamefont {Stern}, \citenamefont {Freedman},\ and\
  \citenamefont {Sarma}}]{nayak2008non}%
  \BibitemOpen
  \bibfield  {author} {\bibinfo {author} {\bibfnamefont {C.}~\bibnamefont
  {Nayak}}, \bibinfo {author} {\bibfnamefont {S.~H.}\ \bibnamefont {Simon}},
  \bibinfo {author} {\bibfnamefont {A.}~\bibnamefont {Stern}}, \bibinfo
  {author} {\bibfnamefont {M.}~\bibnamefont {Freedman}}, \ and\ \bibinfo
  {author} {\bibfnamefont {S.~D.}\ \bibnamefont {Sarma}},\ }\href@noop {}
  {\bibfield  {journal} {\bibinfo  {journal} {Reviews of Modern Physics}\
  }\textbf {\bibinfo {volume} {80}},\ \bibinfo {pages} {1083} (\bibinfo {year}
  {2008})}\BibitemShut {NoStop}%
\bibitem [{\citenamefont {Kiktenko}\ \emph
  {et~al.}(2015{\natexlab{a}})\citenamefont {Kiktenko}, \citenamefont
  {Fedorov}, \citenamefont {Man'ko},\ and\ \citenamefont
  {Man'ko}}]{kiktenko2015multilevel}%
  \BibitemOpen
  \bibfield  {author} {\bibinfo {author} {\bibfnamefont {E.~O.}\ \bibnamefont
  {Kiktenko}}, \bibinfo {author} {\bibfnamefont {A.~K.}\ \bibnamefont
  {Fedorov}}, \bibinfo {author} {\bibfnamefont {O.~V.}\ \bibnamefont {Man'ko}},
  \ and\ \bibinfo {author} {\bibfnamefont {V.~I.}\ \bibnamefont {Man'ko}},\
  }\href {\doibase 10.1103/PhysRevA.91.042312} {\bibfield  {journal} {\bibinfo
  {journal} {Phys. Rev. A}\ }\textbf {\bibinfo {volume} {91}},\ \bibinfo
  {pages} {042312} (\bibinfo {year} {2015}{\natexlab{a}})}\BibitemShut
  {NoStop}%
\bibitem [{\citenamefont {Kiktenko}\ \emph
  {et~al.}(2015{\natexlab{b}})\citenamefont {Kiktenko}, \citenamefont
  {Fedorov}, \citenamefont {Strakhov},\ and\ \citenamefont
  {Man'ko}}]{kiktenko2015single}%
  \BibitemOpen
  \bibfield  {author} {\bibinfo {author} {\bibfnamefont {E.}~\bibnamefont
  {Kiktenko}}, \bibinfo {author} {\bibfnamefont {A.}~\bibnamefont {Fedorov}},
  \bibinfo {author} {\bibfnamefont {A.}~\bibnamefont {Strakhov}}, \ and\
  \bibinfo {author} {\bibfnamefont {V.}~\bibnamefont {Man'ko}},\ }\href
  {\doibase https://doi.org/10.1016/j.physleta.2015.03.023} {\bibfield
  {journal} {\bibinfo  {journal} {Physics Letters A}\ }\textbf {\bibinfo
  {volume} {379}},\ \bibinfo {pages} {1409 } (\bibinfo {year}
  {2015}{\natexlab{b}})}\BibitemShut {NoStop}%
\bibitem [{\citenamefont {Yan}\ \emph {et~al.}(2018)\citenamefont {Yan},
  \citenamefont {Krantz}, \citenamefont {Sung}, \citenamefont {Kjaergaard},
  \citenamefont {Campbell}, \citenamefont {Wang}, \citenamefont {Orlando},
  \citenamefont {Gustavsson},\ and\ \citenamefont {Oliver}}]{yan2018tunable}%
  \BibitemOpen
  \bibfield  {author} {\bibinfo {author} {\bibfnamefont {F.}~\bibnamefont
  {Yan}}, \bibinfo {author} {\bibfnamefont {P.}~\bibnamefont {Krantz}},
  \bibinfo {author} {\bibfnamefont {Y.}~\bibnamefont {Sung}}, \bibinfo {author}
  {\bibfnamefont {M.}~\bibnamefont {Kjaergaard}}, \bibinfo {author}
  {\bibfnamefont {D.}~\bibnamefont {Campbell}}, \bibinfo {author}
  {\bibfnamefont {J.}~\bibnamefont {Wang}}, \bibinfo {author} {\bibfnamefont
  {T.}~\bibnamefont {Orlando}}, \bibinfo {author} {\bibfnamefont
  {S.}~\bibnamefont {Gustavsson}}, \ and\ \bibinfo {author} {\bibfnamefont
  {W.}~\bibnamefont {Oliver}},\ }\href {\doibase
  10.1103/PhysRevApplied.10.054062} {\bibfield  {journal} {\bibinfo  {journal}
  {Physical Review Applied}\ }\textbf {\bibinfo {volume} {10}} (\bibinfo {year}
  {2018}),\ 10.1103/PhysRevApplied.10.054062}\BibitemShut {NoStop}%
\bibitem [{\citenamefont {Gottesman}(1998)}]{gottesman1998theory}%
  \BibitemOpen
  \bibfield  {author} {\bibinfo {author} {\bibfnamefont {D.}~\bibnamefont
  {Gottesman}},\ }\href@noop {} {\bibfield  {journal} {\bibinfo  {journal}
  {Physical Review A}\ }\textbf {\bibinfo {volume} {57}},\ \bibinfo {pages}
  {127} (\bibinfo {year} {1998})}\BibitemShut {NoStop}%
\bibitem [{\citenamefont {Sanders}\ \emph {et~al.}(2015)\citenamefont
  {Sanders}, \citenamefont {Wallman},\ and\ \citenamefont
  {Sanders}}]{sanders2015bounding}%
  \BibitemOpen
  \bibfield  {author} {\bibinfo {author} {\bibfnamefont {Y.~R.}\ \bibnamefont
  {Sanders}}, \bibinfo {author} {\bibfnamefont {J.~J.}\ \bibnamefont
  {Wallman}}, \ and\ \bibinfo {author} {\bibfnamefont {B.~C.}\ \bibnamefont
  {Sanders}},\ }\href@noop {} {\bibfield  {journal} {\bibinfo  {journal} {New
  Journal of Physics}\ }\textbf {\bibinfo {volume} {18}},\ \bibinfo {pages}
  {012002} (\bibinfo {year} {2015})}\BibitemShut {NoStop}%
\bibitem [{\citenamefont {Shor}(1995)}]{shor1995scheme}%
  \BibitemOpen
  \bibfield  {author} {\bibinfo {author} {\bibfnamefont {P.~W.}\ \bibnamefont
  {Shor}},\ }\href@noop {} {\bibfield  {journal} {\bibinfo  {journal} {Physical
  review A}\ }\textbf {\bibinfo {volume} {52}},\ \bibinfo {pages} {R2493}
  (\bibinfo {year} {1995})}\BibitemShut {NoStop}%
\bibitem [{\citenamefont {Palma}\ \emph {et~al.}(1996)\citenamefont {Palma},
  \citenamefont {Suominen},\ and\ \citenamefont {Ekert}}]{palma1996quantum}%
  \BibitemOpen
  \bibfield  {author} {\bibinfo {author} {\bibfnamefont {G.~M.}\ \bibnamefont
  {Palma}}, \bibinfo {author} {\bibfnamefont {K.-A.}\ \bibnamefont {Suominen}},
  \ and\ \bibinfo {author} {\bibfnamefont {A.}~\bibnamefont {Ekert}},\
  }\href@noop {} {\bibfield  {journal} {\bibinfo  {journal} {Proceedings of the
  Royal Society of London. Series A: Mathematical, Physical and Engineering
  Sciences}\ }\textbf {\bibinfo {volume} {452}},\ \bibinfo {pages} {567}
  (\bibinfo {year} {1996})}\BibitemShut {NoStop}%
\bibitem [{\citenamefont {Ischi}\ \emph {et~al.}(2005)\citenamefont {Ischi},
  \citenamefont {Hilke},\ and\ \citenamefont
  {Dub{\'e}}}]{ischi2005decoherence}%
  \BibitemOpen
  \bibfield  {author} {\bibinfo {author} {\bibfnamefont {B.}~\bibnamefont
  {Ischi}}, \bibinfo {author} {\bibfnamefont {M.}~\bibnamefont {Hilke}}, \ and\
  \bibinfo {author} {\bibfnamefont {M.}~\bibnamefont {Dub{\'e}}},\ }\href@noop
  {} {\bibfield  {journal} {\bibinfo  {journal} {Physical Review B}\ }\textbf
  {\bibinfo {volume} {71}},\ \bibinfo {pages} {195325} (\bibinfo {year}
  {2005})}\BibitemShut {NoStop}%
\bibitem [{\citenamefont {Martinis}\ \emph {et~al.}(2003)\citenamefont
  {Martinis}, \citenamefont {Nam}, \citenamefont {Aumentado}, \citenamefont
  {Lang},\ and\ \citenamefont {Urbina}}]{martinis2003decoherence}%
  \BibitemOpen
  \bibfield  {author} {\bibinfo {author} {\bibfnamefont {J.~M.}\ \bibnamefont
  {Martinis}}, \bibinfo {author} {\bibfnamefont {S.}~\bibnamefont {Nam}},
  \bibinfo {author} {\bibfnamefont {J.}~\bibnamefont {Aumentado}}, \bibinfo
  {author} {\bibfnamefont {K.}~\bibnamefont {Lang}}, \ and\ \bibinfo {author}
  {\bibfnamefont {C.}~\bibnamefont {Urbina}},\ }\href@noop {} {\bibfield
  {journal} {\bibinfo  {journal} {Physical Review B}\ }\textbf {\bibinfo
  {volume} {67}},\ \bibinfo {pages} {094510} (\bibinfo {year}
  {2003})}\BibitemShut {NoStop}%
\bibitem [{\citenamefont {Preskill}(2018)}]{Preskill2018quantum}%
  \BibitemOpen
  \bibfield  {author} {\bibinfo {author} {\bibfnamefont {J.}~\bibnamefont
  {Preskill}},\ }\href {\doibase 10.22331/q-2018-08-06-79} {\bibfield
  {journal} {\bibinfo  {journal} {{Quantum}}\ }\textbf {\bibinfo {volume}
  {2}},\ \bibinfo {pages} {79} (\bibinfo {year} {2018})}\BibitemShut {NoStop}%
\bibitem [{\citenamefont {Aidelsburger}\ \emph {et~al.}(2018)\citenamefont
  {Aidelsburger}, \citenamefont {Nascimbene},\ and\ \citenamefont
  {Goldman}}]{aidelsburger2018artificial}%
  \BibitemOpen
  \bibfield  {author} {\bibinfo {author} {\bibfnamefont {M.}~\bibnamefont
  {Aidelsburger}}, \bibinfo {author} {\bibfnamefont {S.}~\bibnamefont
  {Nascimbene}}, \ and\ \bibinfo {author} {\bibfnamefont {N.}~\bibnamefont
  {Goldman}},\ }\href {\doibase https://doi.org/10.1016/j.crhy.2018.03.002}
  {\bibfield  {journal} {\bibinfo  {journal} {Comptes Rendus Physique}\
  }\textbf {\bibinfo {volume} {19}},\ \bibinfo {pages} {394} (\bibinfo {year}
  {2018})},\ \bibinfo {note} {quantum simulation / Simulation
  quantique}\BibitemShut {NoStop}%
\bibitem [{\citenamefont {Vitanov}(2020)}]{Ivanov2020two-qubit}%
  \BibitemOpen
  \bibfield  {author} {\bibinfo {author} {\bibfnamefont {N.}~\bibnamefont
  {Vitanov}},\ }\href {\doibase 10.1038/s41598-020-61766-w} {\bibfield
  {journal} {\bibinfo  {journal} {Scientific Reports}\ }\textbf {\bibinfo
  {volume} {10}} (\bibinfo {year} {2020}),\
  10.1038/s41598-020-61766-w}\BibitemShut {NoStop}%
\bibitem [{\citenamefont {Frahm}\ \emph {et~al.}(2004)\citenamefont {Frahm},
  \citenamefont {Fleckinger},\ and\ \citenamefont
  {Shepelyansky}}]{Frahm2004quantum}%
  \BibitemOpen
  \bibfield  {author} {\bibinfo {author} {\bibfnamefont {K.~M.}\ \bibnamefont
  {Frahm}}, \bibinfo {author} {\bibfnamefont {R.}~\bibnamefont {Fleckinger}}, \
  and\ \bibinfo {author} {\bibfnamefont {D.~L.}\ \bibnamefont {Shepelyansky}},\
  }\href {\doibase 10.1140/epjd/e2004-00038-x} {\bibfield  {journal} {\bibinfo
  {journal} {The European Physical Journal D - Atomic, Molecular and Optical
  Physics}\ }\textbf {\bibinfo {volume} {29}},\ \bibinfo {pages} {139–155}
  (\bibinfo {year} {2004})}\BibitemShut {NoStop}%
\bibitem [{\citenamefont {Sleator}\ and\ \citenamefont
  {Weinfurter}(1995)}]{sleator1995realizable}%
  \BibitemOpen
  \bibfield  {author} {\bibinfo {author} {\bibfnamefont {T.}~\bibnamefont
  {Sleator}}\ and\ \bibinfo {author} {\bibfnamefont {H.}~\bibnamefont
  {Weinfurter}},\ }\href {\doibase 10.1103/PhysRevLett.74.4087} {\bibfield
  {journal} {\bibinfo  {journal} {Phys. Rev. Lett.}\ }\textbf {\bibinfo
  {volume} {74}},\ \bibinfo {pages} {4087} (\bibinfo {year}
  {1995})}\BibitemShut {NoStop}%
\bibitem [{\citenamefont {Wallman}\ and\ \citenamefont
  {Emerson}(2016)}]{Wallman2016noise}%
  \BibitemOpen
  \bibfield  {author} {\bibinfo {author} {\bibfnamefont {J.~J.}\ \bibnamefont
  {Wallman}}\ and\ \bibinfo {author} {\bibfnamefont {J.}~\bibnamefont
  {Emerson}},\ }\href {\doibase 10.1103/PhysRevA.94.052325} {\bibfield
  {journal} {\bibinfo  {journal} {Phys. Rev. A}\ }\textbf {\bibinfo {volume}
  {94}},\ \bibinfo {pages} {052325} (\bibinfo {year} {2016})}\BibitemShut
  {NoStop}%
\bibitem [{\citenamefont {Ware}\ \emph {et~al.}(2021)\citenamefont {Ware},
  \citenamefont {Ribeill}, \citenamefont {Rist\`e}, \citenamefont {Ryan},
  \citenamefont {Johnson},\ and\ \citenamefont
  {da~Silva}}]{Ware2021experimental}%
  \BibitemOpen
  \bibfield  {author} {\bibinfo {author} {\bibfnamefont {M.}~\bibnamefont
  {Ware}}, \bibinfo {author} {\bibfnamefont {G.}~\bibnamefont {Ribeill}},
  \bibinfo {author} {\bibfnamefont {D.}~\bibnamefont {Rist\`e}}, \bibinfo
  {author} {\bibfnamefont {C.~A.}\ \bibnamefont {Ryan}}, \bibinfo {author}
  {\bibfnamefont {B.}~\bibnamefont {Johnson}}, \ and\ \bibinfo {author}
  {\bibfnamefont {M.~P.}\ \bibnamefont {da~Silva}},\ }\href {\doibase
  10.1103/PhysRevA.103.042604} {\bibfield  {journal} {\bibinfo  {journal}
  {Phys. Rev. A}\ }\textbf {\bibinfo {volume} {103}},\ \bibinfo {pages}
  {042604} (\bibinfo {year} {2021})}\BibitemShut {NoStop}%
\bibitem [{\citenamefont {Hashim}\ \emph {et~al.}(2021)\citenamefont {Hashim},
  \citenamefont {Naik}, \citenamefont {Morvan}, \citenamefont {Ville},
  \citenamefont {Mitchell}, \citenamefont {Kreikebaum}, \citenamefont {Davis},
  \citenamefont {Smith}, \citenamefont {Iancu}, \citenamefont {O'Brien},
  \citenamefont {Hincks}, \citenamefont {Wallman}, \citenamefont {Emerson},\
  and\ \citenamefont {Siddiqi}}]{Hashim2021randomized}%
  \BibitemOpen
  \bibfield  {author} {\bibinfo {author} {\bibfnamefont {A.}~\bibnamefont
  {Hashim}}, \bibinfo {author} {\bibfnamefont {R.~K.}\ \bibnamefont {Naik}},
  \bibinfo {author} {\bibfnamefont {A.}~\bibnamefont {Morvan}}, \bibinfo
  {author} {\bibfnamefont {J.-L.}\ \bibnamefont {Ville}}, \bibinfo {author}
  {\bibfnamefont {B.}~\bibnamefont {Mitchell}}, \bibinfo {author}
  {\bibfnamefont {J.~M.}\ \bibnamefont {Kreikebaum}}, \bibinfo {author}
  {\bibfnamefont {M.}~\bibnamefont {Davis}}, \bibinfo {author} {\bibfnamefont
  {E.}~\bibnamefont {Smith}}, \bibinfo {author} {\bibfnamefont
  {C.}~\bibnamefont {Iancu}}, \bibinfo {author} {\bibfnamefont {K.~P.}\
  \bibnamefont {O'Brien}}, \bibinfo {author} {\bibfnamefont {I.}~\bibnamefont
  {Hincks}}, \bibinfo {author} {\bibfnamefont {J.~J.}\ \bibnamefont {Wallman}},
  \bibinfo {author} {\bibfnamefont {J.}~\bibnamefont {Emerson}}, \ and\
  \bibinfo {author} {\bibfnamefont {I.}~\bibnamefont {Siddiqi}},\ }\href
  {\doibase 10.1103/PhysRevX.11.041039} {\bibfield  {journal} {\bibinfo
  {journal} {Phys. Rev. X}\ }\textbf {\bibinfo {volume} {11}},\ \bibinfo
  {pages} {041039} (\bibinfo {year} {2021})}\BibitemShut {NoStop}%
\bibitem [{\citenamefont {Perez}\ \emph {et~al.}(2021)\citenamefont {Perez},
  \citenamefont {Varosy}, \citenamefont {Li}, \citenamefont {Roy},
  \citenamefont {Kapit},\ and\ \citenamefont
  {Schuster}}]{Perez2021error-divisible}%
  \BibitemOpen
  \bibfield  {author} {\bibinfo {author} {\bibfnamefont {D.~R.}\ \bibnamefont
  {Perez}}, \bibinfo {author} {\bibfnamefont {P.}~\bibnamefont {Varosy}},
  \bibinfo {author} {\bibfnamefont {Z.}~\bibnamefont {Li}}, \bibinfo {author}
  {\bibfnamefont {T.}~\bibnamefont {Roy}}, \bibinfo {author} {\bibfnamefont
  {E.}~\bibnamefont {Kapit}}, \ and\ \bibinfo {author} {\bibfnamefont
  {D.}~\bibnamefont {Schuster}},\ }\href@noop {} {\  (\bibinfo {year}
  {2021})},\ \Eprint {http://arxiv.org/abs/2110.11537} {arXiv:2110.11537
  [quant-ph]} \BibitemShut {NoStop}%
\bibitem [{\citenamefont {Peres}(1984)}]{peres1984stability}%
  \BibitemOpen
  \bibfield  {author} {\bibinfo {author} {\bibfnamefont {A.}~\bibnamefont
  {Peres}},\ }\href {\doibase 10.1103/PhysRevA.30.1610} {\bibfield  {journal}
  {\bibinfo  {journal} {Phys. Rev. A}\ }\textbf {\bibinfo {volume} {30}},\
  \bibinfo {pages} {1610} (\bibinfo {year} {1984})}\BibitemShut {NoStop}%
\bibitem [{\citenamefont {Cucchietti}\ \emph {et~al.}(2005)\citenamefont
  {Cucchietti}, \citenamefont {Paz},\ and\ \citenamefont
  {Zurek}}]{Cucchietti2005decoherence}%
  \BibitemOpen
  \bibfield  {author} {\bibinfo {author} {\bibfnamefont {F.~M.}\ \bibnamefont
  {Cucchietti}}, \bibinfo {author} {\bibfnamefont {J.~P.}\ \bibnamefont {Paz}},
  \ and\ \bibinfo {author} {\bibfnamefont {W.~H.}\ \bibnamefont {Zurek}},\
  }\href {\doibase 10.1103/physreva.72.052113} {\bibfield  {journal} {\bibinfo
  {journal} {Physical Review A}\ }\textbf {\bibinfo {volume} {72}} (\bibinfo
  {year} {2005}),\ 10.1103/physreva.72.052113}\BibitemShut {NoStop}%
\bibitem [{\citenamefont {Goussev}\ \emph {et~al.}(2012)\citenamefont
  {Goussev}, \citenamefont {Jalabert}, \citenamefont {Pastawski},\ and\
  \citenamefont {Wisniacki}}]{Wisniacki2012loschmidt}%
  \BibitemOpen
  \bibfield  {author} {\bibinfo {author} {\bibfnamefont {A.}~\bibnamefont
  {Goussev}}, \bibinfo {author} {\bibfnamefont {R.~A.}\ \bibnamefont
  {Jalabert}}, \bibinfo {author} {\bibfnamefont {H.~M.}\ \bibnamefont
  {Pastawski}}, \ and\ \bibinfo {author} {\bibfnamefont {A.}~\bibnamefont
  {Wisniacki}},\ }\href {\doibase 10.4249/scholarpedia.11687} {\bibfield
  {journal} {\bibinfo  {journal} {Scholarpedia}\ }\textbf {\bibinfo {volume}
  {7}},\ \bibinfo {pages} {11687} (\bibinfo {year} {2012})}\BibitemShut
  {NoStop}%
\bibitem [{\citenamefont {Berry}(1984)}]{berry1984quantal}%
  \BibitemOpen
  \bibfield  {author} {\bibinfo {author} {\bibfnamefont {M.~V.}\ \bibnamefont
  {Berry}},\ }\href {http://www.jstor.org/stable/2397741} {\bibfield  {journal}
  {\bibinfo  {journal} {Proceedings of the Royal Society of London. Series A,
  Mathematical and Physical Sciences}\ }\textbf {\bibinfo {volume} {392}},\
  \bibinfo {pages} {45} (\bibinfo {year} {1984})}\BibitemShut {NoStop}%
\bibitem [{\citenamefont {Simon}(1983)}]{simon1983holonomy}%
  \BibitemOpen
  \bibfield  {author} {\bibinfo {author} {\bibfnamefont {B.}~\bibnamefont
  {Simon}},\ }\href {\doibase 10.1103/PhysRevLett.51.2167} {\bibfield
  {journal} {\bibinfo  {journal} {Phys. Rev. Lett.}\ }\textbf {\bibinfo
  {volume} {51}},\ \bibinfo {pages} {2167} (\bibinfo {year}
  {1983})}\BibitemShut {NoStop}%
\bibitem [{\citenamefont {S\o{}rensen}\ \emph {et~al.}(2005)\citenamefont
  {S\o{}rensen}, \citenamefont {Demler},\ and\ \citenamefont
  {Lukin}}]{sorensen2005fractional}%
  \BibitemOpen
  \bibfield  {author} {\bibinfo {author} {\bibfnamefont {A.~S.}\ \bibnamefont
  {S\o{}rensen}}, \bibinfo {author} {\bibfnamefont {E.}~\bibnamefont {Demler}},
  \ and\ \bibinfo {author} {\bibfnamefont {M.~D.}\ \bibnamefont {Lukin}},\
  }\href {\doibase 10.1103/PhysRevLett.94.086803} {\bibfield  {journal}
  {\bibinfo  {journal} {Phys. Rev. Lett.}\ }\textbf {\bibinfo {volume} {94}},\
  \bibinfo {pages} {086803} (\bibinfo {year} {2005})}\BibitemShut {NoStop}%
\bibitem [{\citenamefont {Wilczek}\ and\ \citenamefont
  {Zee}(1984)}]{wilczek1984appearance}%
  \BibitemOpen
  \bibfield  {author} {\bibinfo {author} {\bibfnamefont {F.}~\bibnamefont
  {Wilczek}}\ and\ \bibinfo {author} {\bibfnamefont {A.}~\bibnamefont {Zee}},\
  }\href {\doibase 10.1103/PhysRevLett.52.2111} {\bibfield  {journal} {\bibinfo
   {journal} {Phys. Rev. Lett.}\ }\textbf {\bibinfo {volume} {52}},\ \bibinfo
  {pages} {2111} (\bibinfo {year} {1984})}\BibitemShut {NoStop}%
\bibitem [{\citenamefont {Mead}(1992)}]{mead1992geometric}%
  \BibitemOpen
  \bibfield  {author} {\bibinfo {author} {\bibfnamefont {C.~A.}\ \bibnamefont
  {Mead}},\ }\href {\doibase 10.1103/RevModPhys.64.51} {\bibfield  {journal}
  {\bibinfo  {journal} {Rev. Mod. Phys.}\ }\textbf {\bibinfo {volume} {64}},\
  \bibinfo {pages} {51} (\bibinfo {year} {1992})}\BibitemShut {NoStop}%
\bibitem [{\citenamefont {Xiao}\ \emph {et~al.}(2010)\citenamefont {Xiao},
  \citenamefont {Chang},\ and\ \citenamefont {Niu}}]{xiao2010berry}%
  \BibitemOpen
  \bibfield  {author} {\bibinfo {author} {\bibfnamefont {D.}~\bibnamefont
  {Xiao}}, \bibinfo {author} {\bibfnamefont {M.-C.}\ \bibnamefont {Chang}}, \
  and\ \bibinfo {author} {\bibfnamefont {Q.}~\bibnamefont {Niu}},\ }\href
  {\doibase 10.1103/RevModPhys.82.1959} {\bibfield  {journal} {\bibinfo
  {journal} {Rev. Mod. Phys.}\ }\textbf {\bibinfo {volume} {82}},\ \bibinfo
  {pages} {1959} (\bibinfo {year} {2010})}\BibitemShut {NoStop}%
\bibitem [{\citenamefont {Grifoni}\ and\ \citenamefont
  {H{\"a}nggi}(1998)}]{grifoni1998driven}%
  \BibitemOpen
  \bibfield  {author} {\bibinfo {author} {\bibfnamefont {M.}~\bibnamefont
  {Grifoni}}\ and\ \bibinfo {author} {\bibfnamefont {P.}~\bibnamefont
  {H{\"a}nggi}},\ }\href {\doibase
  https://doi.org/10.1016/S0370-1573(98)00022-2} {\bibfield  {journal}
  {\bibinfo  {journal} {Physics Reports}\ }\textbf {\bibinfo {volume} {304}},\
  \bibinfo {pages} {229 } (\bibinfo {year} {1998})}\BibitemShut {NoStop}%
\bibitem [{\citenamefont {Dunlap}\ and\ \citenamefont
  {Kenkre}(1986)}]{dunlap1986dynamic}%
  \BibitemOpen
  \bibfield  {author} {\bibinfo {author} {\bibfnamefont {D.~H.}\ \bibnamefont
  {Dunlap}}\ and\ \bibinfo {author} {\bibfnamefont {V.~M.}\ \bibnamefont
  {Kenkre}},\ }\href {\doibase 10.1103/PhysRevB.34.3625} {\bibfield  {journal}
  {\bibinfo  {journal} {Phys. Rev. B}\ }\textbf {\bibinfo {volume} {34}},\
  \bibinfo {pages} {3625} (\bibinfo {year} {1986})}\BibitemShut {NoStop}%
\bibitem [{\citenamefont {Castro~Neto}\ \emph {et~al.}(2007)\citenamefont
  {Castro~Neto}, \citenamefont {Guinea}, \citenamefont {Peres}, \citenamefont
  {Novoselov},\ and\ \citenamefont {Geim}}]{castro-neto2007electronic}%
  \BibitemOpen
  \bibfield  {author} {\bibinfo {author} {\bibfnamefont {A.}~\bibnamefont
  {Castro~Neto}}, \bibinfo {author} {\bibfnamefont {F.}~\bibnamefont {Guinea}},
  \bibinfo {author} {\bibfnamefont {N.}~\bibnamefont {Peres}}, \bibinfo
  {author} {\bibfnamefont {K.}~\bibnamefont {Novoselov}}, \ and\ \bibinfo
  {author} {\bibfnamefont {A.}~\bibnamefont {Geim}},\ }\href {\doibase
  10.1103/RevModPhys.81.109} {\bibfield  {journal} {\bibinfo  {journal} {Review
  of Modern Physics}\ }\textbf {\bibinfo {volume} {81}} (\bibinfo {year}
  {2007}),\ 10.1103/RevModPhys.81.109}\BibitemShut {NoStop}%
\bibitem [{\citenamefont {Hafezi}(2013)}]{hafezi2013synthetic}%
  \BibitemOpen
  \bibfield  {author} {\bibinfo {author} {\bibfnamefont {M.}~\bibnamefont
  {Hafezi}},\ }\href {\doibase 10.1142/S0217979214410021} {\bibfield  {journal}
  {\bibinfo  {journal} {International Journal of Modern Physics B}\ }\textbf
  {\bibinfo {volume} {28}} (\bibinfo {year} {2013}),\
  10.1142/S0217979214410021}\BibitemShut {NoStop}%
\bibitem [{\citenamefont {Ozawa}\ \emph {et~al.}(2019)\citenamefont {Ozawa},
  \citenamefont {Price}, \citenamefont {Amo}, \citenamefont {Goldman},
  \citenamefont {Hafezi}, \citenamefont {Lu}, \citenamefont {Rechtsman},
  \citenamefont {Schuster}, \citenamefont {Simon}, \citenamefont {Zilberberg},\
  and\ \citenamefont {Carusotto}}]{ozawa2019topological}%
  \BibitemOpen
  \bibfield  {author} {\bibinfo {author} {\bibfnamefont {T.}~\bibnamefont
  {Ozawa}}, \bibinfo {author} {\bibfnamefont {H.~M.}\ \bibnamefont {Price}},
  \bibinfo {author} {\bibfnamefont {A.}~\bibnamefont {Amo}}, \bibinfo {author}
  {\bibfnamefont {N.}~\bibnamefont {Goldman}}, \bibinfo {author} {\bibfnamefont
  {M.}~\bibnamefont {Hafezi}}, \bibinfo {author} {\bibfnamefont
  {L.}~\bibnamefont {Lu}}, \bibinfo {author} {\bibfnamefont {M.~C.}\
  \bibnamefont {Rechtsman}}, \bibinfo {author} {\bibfnamefont {D.}~\bibnamefont
  {Schuster}}, \bibinfo {author} {\bibfnamefont {J.}~\bibnamefont {Simon}},
  \bibinfo {author} {\bibfnamefont {O.}~\bibnamefont {Zilberberg}}, \ and\
  \bibinfo {author} {\bibfnamefont {I.}~\bibnamefont {Carusotto}},\ }\href
  {\doibase 10.1103/RevModPhys.91.015006} {\bibfield  {journal} {\bibinfo
  {journal} {Rev. Mod. Phys.}\ }\textbf {\bibinfo {volume} {91}},\ \bibinfo
  {pages} {015006} (\bibinfo {year} {2019})}\BibitemShut {NoStop}%
\bibitem [{\citenamefont {Anderson}\ \emph {et~al.}(2011)\citenamefont
  {Anderson}, \citenamefont {Juzeliūnas}, \citenamefont {Spielman},\ and\
  \citenamefont {Galitski}}]{anderson2011synthetic}%
  \BibitemOpen
  \bibfield  {author} {\bibinfo {author} {\bibfnamefont {B.}~\bibnamefont
  {Anderson}}, \bibinfo {author} {\bibfnamefont {G.}~\bibnamefont
  {Juzeliūnas}}, \bibinfo {author} {\bibfnamefont {I.}~\bibnamefont
  {Spielman}}, \ and\ \bibinfo {author} {\bibfnamefont {V.}~\bibnamefont
  {Galitski}},\ }\href {\doibase 10.1103/PhysRevLett.108.235301} {\bibfield
  {journal} {\bibinfo  {journal} {Physical Review Letters}\ }\textbf {\bibinfo
  {volume} {108}} (\bibinfo {year} {2011}),\
  10.1103/PhysRevLett.108.235301}\BibitemShut {NoStop}%
\bibitem [{\citenamefont {Struck}\ \emph {et~al.}(2012)\citenamefont {Struck},
  \citenamefont {\"Olschl\"ager}, \citenamefont {Weinberg}, \citenamefont
  {Hauke}, \citenamefont {Simonet}, \citenamefont {Eckardt}, \citenamefont
  {Lewenstein}, \citenamefont {Sengstock},\ and\ \citenamefont
  {Windpassinger}}]{struck2012tunable}%
  \BibitemOpen
  \bibfield  {author} {\bibinfo {author} {\bibfnamefont {J.}~\bibnamefont
  {Struck}}, \bibinfo {author} {\bibfnamefont {C.}~\bibnamefont
  {\"Olschl\"ager}}, \bibinfo {author} {\bibfnamefont {M.}~\bibnamefont
  {Weinberg}}, \bibinfo {author} {\bibfnamefont {P.}~\bibnamefont {Hauke}},
  \bibinfo {author} {\bibfnamefont {J.}~\bibnamefont {Simonet}}, \bibinfo
  {author} {\bibfnamefont {A.}~\bibnamefont {Eckardt}}, \bibinfo {author}
  {\bibfnamefont {M.}~\bibnamefont {Lewenstein}}, \bibinfo {author}
  {\bibfnamefont {K.}~\bibnamefont {Sengstock}}, \ and\ \bibinfo {author}
  {\bibfnamefont {P.}~\bibnamefont {Windpassinger}},\ }\href {\doibase
  10.1103/PhysRevLett.108.225304} {\bibfield  {journal} {\bibinfo  {journal}
  {Phys. Rev. Lett.}\ }\textbf {\bibinfo {volume} {108}},\ \bibinfo {pages}
  {225304} (\bibinfo {year} {2012})}\BibitemShut {NoStop}%
\bibitem [{\citenamefont {Hauke}\ \emph {et~al.}(2012)\citenamefont {Hauke},
  \citenamefont {Tieleman}, \citenamefont {Celi}, \citenamefont
  {\"Olschl\"ager}, \citenamefont {Simonet}, \citenamefont {Struck},
  \citenamefont {Weinberg}, \citenamefont {Windpassinger}, \citenamefont
  {Sengstock}, \citenamefont {Lewenstein},\ and\ \citenamefont
  {Eckardt}}]{hauke2012non-abelian}%
  \BibitemOpen
  \bibfield  {author} {\bibinfo {author} {\bibfnamefont {P.}~\bibnamefont
  {Hauke}}, \bibinfo {author} {\bibfnamefont {O.}~\bibnamefont {Tieleman}},
  \bibinfo {author} {\bibfnamefont {A.}~\bibnamefont {Celi}}, \bibinfo {author}
  {\bibfnamefont {C.}~\bibnamefont {\"Olschl\"ager}}, \bibinfo {author}
  {\bibfnamefont {J.}~\bibnamefont {Simonet}}, \bibinfo {author} {\bibfnamefont
  {J.}~\bibnamefont {Struck}}, \bibinfo {author} {\bibfnamefont
  {M.}~\bibnamefont {Weinberg}}, \bibinfo {author} {\bibfnamefont
  {P.}~\bibnamefont {Windpassinger}}, \bibinfo {author} {\bibfnamefont
  {K.}~\bibnamefont {Sengstock}}, \bibinfo {author} {\bibfnamefont
  {M.}~\bibnamefont {Lewenstein}}, \ and\ \bibinfo {author} {\bibfnamefont
  {A.}~\bibnamefont {Eckardt}},\ }\href {\doibase
  10.1103/PhysRevLett.109.145301} {\bibfield  {journal} {\bibinfo  {journal}
  {Phys. Rev. Lett.}\ }\textbf {\bibinfo {volume} {109}},\ \bibinfo {pages}
  {145301} (\bibinfo {year} {2012})}\BibitemShut {NoStop}%
\bibitem [{\citenamefont {Aidelsburger}\ \emph {et~al.}(2013)\citenamefont
  {Aidelsburger}, \citenamefont {Atala}, \citenamefont {Lohse}, \citenamefont
  {Barreiro}, \citenamefont {Paredes},\ and\ \citenamefont
  {Bloch}}]{aidelsburger2013realization}%
  \BibitemOpen
  \bibfield  {author} {\bibinfo {author} {\bibfnamefont {M.}~\bibnamefont
  {Aidelsburger}}, \bibinfo {author} {\bibfnamefont {M.}~\bibnamefont {Atala}},
  \bibinfo {author} {\bibfnamefont {M.}~\bibnamefont {Lohse}}, \bibinfo
  {author} {\bibfnamefont {J.~T.}\ \bibnamefont {Barreiro}}, \bibinfo {author}
  {\bibfnamefont {B.}~\bibnamefont {Paredes}}, \ and\ \bibinfo {author}
  {\bibfnamefont {I.}~\bibnamefont {Bloch}},\ }\href {\doibase
  10.1103/PhysRevLett.111.185301} {\bibfield  {journal} {\bibinfo  {journal}
  {Phys. Rev. Lett.}\ }\textbf {\bibinfo {volume} {111}},\ \bibinfo {pages}
  {185301} (\bibinfo {year} {2013})}\BibitemShut {NoStop}%
\bibitem [{\citenamefont {Galitski}\ and\ \citenamefont
  {Spielman}(2013)}]{galitski2013spin-orbit}%
  \BibitemOpen
  \bibfield  {author} {\bibinfo {author} {\bibfnamefont {V.}~\bibnamefont
  {Galitski}}\ and\ \bibinfo {author} {\bibfnamefont {I.}~\bibnamefont
  {Spielman}},\ }\href {\doibase 10.1038/nature11841} {\bibfield  {journal}
  {\bibinfo  {journal} {Nature}\ }\textbf {\bibinfo {volume} {494}},\ \bibinfo
  {pages} {49} (\bibinfo {year} {2013})}\BibitemShut {NoStop}%
\bibitem [{\citenamefont {{Galitski}}\ \emph {et~al.}(2019)\citenamefont
  {{Galitski}}, \citenamefont {{Juzeli{\={u}}nas}},\ and\ \citenamefont
  {{Spielman}}}]{galitski2019artificial}%
  \BibitemOpen
  \bibfield  {author} {\bibinfo {author} {\bibfnamefont {V.}~\bibnamefont
  {{Galitski}}}, \bibinfo {author} {\bibfnamefont {G.}~\bibnamefont
  {{Juzeli{\={u}}nas}}}, \ and\ \bibinfo {author} {\bibfnamefont {I.~B.}\
  \bibnamefont {{Spielman}}},\ }\href {\doibase 10.1063/PT.3.4111} {\bibfield
  {journal} {\bibinfo  {journal} {Physics Today}\ }\textbf {\bibinfo {volume}
  {72}},\ \bibinfo {pages} {38} (\bibinfo {year} {2019})},\ \Eprint
  {http://arxiv.org/abs/1901.03705} {arXiv:1901.03705 [cond-mat.quant-gas]}
  \BibitemShut {NoStop}%
\bibitem [{\citenamefont {Bermudez}\ \emph {et~al.}(2011)\citenamefont
  {Bermudez}, \citenamefont {Schaetz},\ and\ \citenamefont
  {Porras}}]{bermudez2011synthetic}%
  \BibitemOpen
  \bibfield  {author} {\bibinfo {author} {\bibfnamefont {A.}~\bibnamefont
  {Bermudez}}, \bibinfo {author} {\bibfnamefont {T.}~\bibnamefont {Schaetz}}, \
  and\ \bibinfo {author} {\bibfnamefont {D.}~\bibnamefont {Porras}},\ }\href
  {\doibase 10.1103/PhysRevLett.107.150501} {\bibfield  {journal} {\bibinfo
  {journal} {Phys. Rev. Lett.}\ }\textbf {\bibinfo {volume} {107}},\ \bibinfo
  {pages} {150501} (\bibinfo {year} {2011})}\BibitemShut {NoStop}%
\bibitem [{\citenamefont {Bermudez}\ \emph {et~al.}(2012)\citenamefont
  {Bermudez}, \citenamefont {Schaetz},\ and\ \citenamefont
  {Porras}}]{bermudez2012photon-assisted}%
  \BibitemOpen
  \bibfield  {author} {\bibinfo {author} {\bibfnamefont {A.}~\bibnamefont
  {Bermudez}}, \bibinfo {author} {\bibfnamefont {T.}~\bibnamefont {Schaetz}}, \
  and\ \bibinfo {author} {\bibfnamefont {D.}~\bibnamefont {Porras}},\ }\href
  {\doibase 10.1088/1367-2630/14/5/053049} {\bibfield  {journal} {\bibinfo
  {journal} {New Journal of Physics}\ }\textbf {\bibinfo {volume} {14}},\
  \bibinfo {pages} {053049} (\bibinfo {year} {2012})}\BibitemShut {NoStop}%
\bibitem [{\citenamefont {Aharony}\ \emph {et~al.}(2012)\citenamefont
  {Aharony}, \citenamefont {Gurvitz}, \citenamefont {Tokura}, \citenamefont
  {Entin-Wohlman},\ and\ \citenamefont {Dattagupta}}]{aharony2012partial}%
  \BibitemOpen
  \bibfield  {author} {\bibinfo {author} {\bibfnamefont {A.}~\bibnamefont
  {Aharony}}, \bibinfo {author} {\bibfnamefont {S.}~\bibnamefont {Gurvitz}},
  \bibinfo {author} {\bibfnamefont {Y.}~\bibnamefont {Tokura}}, \bibinfo
  {author} {\bibfnamefont {O.}~\bibnamefont {Entin-Wohlman}}, \ and\ \bibinfo
  {author} {\bibfnamefont {S.}~\bibnamefont {Dattagupta}},\ }\href {\doibase
  10.1088/0031-8949/2012/t151/014018} {\bibfield  {journal} {\bibinfo
  {journal} {Physica Scripta}\ }\textbf {\bibinfo {volume} {T151}},\ \bibinfo
  {pages} {014018} (\bibinfo {year} {2012})}\BibitemShut {NoStop}%
\bibitem [{\citenamefont {Nilsson}\ \emph {et~al.}(2006)\citenamefont
  {Nilsson}, \citenamefont {Thelander}, \citenamefont {Fröberg}, \citenamefont
  {Wagner},\ and\ \citenamefont {Samuelson}}]{nilsson2006nanowire}%
  \BibitemOpen
  \bibfield  {author} {\bibinfo {author} {\bibfnamefont {H.}~\bibnamefont
  {Nilsson}}, \bibinfo {author} {\bibfnamefont {C.}~\bibnamefont {Thelander}},
  \bibinfo {author} {\bibfnamefont {L.}~\bibnamefont {Fröberg}}, \bibinfo
  {author} {\bibfnamefont {J.}~\bibnamefont {Wagner}}, \ and\ \bibinfo {author}
  {\bibfnamefont {L.}~\bibnamefont {Samuelson}},\ }\href {\doibase
  10.1063/1.2362594} {\bibfield  {journal} {\bibinfo  {journal} {Applied
  Physics Letters}\ }\textbf {\bibinfo {volume} {89}},\ \bibinfo {pages}
  {163101} (\bibinfo {year} {2006})}\BibitemShut {NoStop}%
\bibitem [{\citenamefont {Aharon}\ \emph {et~al.}(2016)\citenamefont {Aharon},
  \citenamefont {Pozner}, \citenamefont {Lifshitz},\ and\ \citenamefont
  {Peskin}}]{aharon2016multi-bit}%
  \BibitemOpen
  \bibfield  {author} {\bibinfo {author} {\bibfnamefont {E.}~\bibnamefont
  {Aharon}}, \bibinfo {author} {\bibfnamefont {R.}~\bibnamefont {Pozner}},
  \bibinfo {author} {\bibfnamefont {E.}~\bibnamefont {Lifshitz}}, \ and\
  \bibinfo {author} {\bibfnamefont {U.}~\bibnamefont {Peskin}},\ }\href
  {\doibase 10.1063/1.4972340} {\bibfield  {journal} {\bibinfo  {journal}
  {Journal of Applied Physics}\ }\textbf {\bibinfo {volume} {120}},\ \bibinfo
  {pages} {244301} (\bibinfo {year} {2016})}\BibitemShut {NoStop}%
\bibitem [{\citenamefont {Heshami}\ \emph {et~al.}(2015)\citenamefont
  {Heshami}, \citenamefont {England}, \citenamefont {Humphreys}, \citenamefont
  {Bustard}, \citenamefont {Acosta}, \citenamefont {Nunn},\ and\ \citenamefont
  {Sussman}}]{heshami2015quantum}%
  \BibitemOpen
  \bibfield  {author} {\bibinfo {author} {\bibfnamefont {K.}~\bibnamefont
  {Heshami}}, \bibinfo {author} {\bibfnamefont {D.}~\bibnamefont {England}},
  \bibinfo {author} {\bibfnamefont {P.}~\bibnamefont {Humphreys}}, \bibinfo
  {author} {\bibfnamefont {P.}~\bibnamefont {Bustard}}, \bibinfo {author}
  {\bibfnamefont {V.}~\bibnamefont {Acosta}}, \bibinfo {author} {\bibfnamefont
  {J.}~\bibnamefont {Nunn}}, \ and\ \bibinfo {author} {\bibfnamefont
  {B.}~\bibnamefont {Sussman}},\ }\href {\doibase
  10.1080/09500340.2016.1148212} {\bibfield  {journal} {\bibinfo  {journal}
  {Journal of Modern Optics}\ }\textbf {\bibinfo {volume} {63}} (\bibinfo
  {year} {2015}),\ 10.1080/09500340.2016.1148212}\BibitemShut {NoStop}%
\bibitem [{\citenamefont {Saglamyurek}\ \emph {et~al.}(2018)\citenamefont
  {Saglamyurek}, \citenamefont {Hrushevskyi}, \citenamefont {Rastogi},
  \citenamefont {Heshami},\ and\ \citenamefont
  {LeBlanc}}]{saglamyurek2018coherent}%
  \BibitemOpen
  \bibfield  {author} {\bibinfo {author} {\bibfnamefont {E.}~\bibnamefont
  {Saglamyurek}}, \bibinfo {author} {\bibfnamefont {T.}~\bibnamefont
  {Hrushevskyi}}, \bibinfo {author} {\bibfnamefont {A.}~\bibnamefont
  {Rastogi}}, \bibinfo {author} {\bibfnamefont {K.}~\bibnamefont {Heshami}}, \
  and\ \bibinfo {author} {\bibfnamefont {L.}~\bibnamefont {LeBlanc}},\ }\href
  {\doibase 10.1038/s41566-018-0279-0} {\bibfield  {journal} {\bibinfo
  {journal} {Nature Photonics}\ }\textbf {\bibinfo {volume} {12}} (\bibinfo
  {year} {2018}),\ 10.1038/s41566-018-0279-0}\BibitemShut {NoStop}%
\bibitem [{\citenamefont {Koong}\ \emph {et~al.}(2020)\citenamefont {Koong},
  \citenamefont {Ballesteros-Garcia}, \citenamefont {Proux}, \citenamefont
  {Dalacu}, \citenamefont {Poole},\ and\ \citenamefont
  {Gerardot}}]{koong2020multiplexed}%
  \BibitemOpen
  \bibfield  {author} {\bibinfo {author} {\bibfnamefont {Z.-X.}\ \bibnamefont
  {Koong}}, \bibinfo {author} {\bibfnamefont {G.}~\bibnamefont
  {Ballesteros-Garcia}}, \bibinfo {author} {\bibfnamefont {R.}~\bibnamefont
  {Proux}}, \bibinfo {author} {\bibfnamefont {D.}~\bibnamefont {Dalacu}},
  \bibinfo {author} {\bibfnamefont {P.~J.}\ \bibnamefont {Poole}}, \ and\
  \bibinfo {author} {\bibfnamefont {B.~D.}\ \bibnamefont {Gerardot}},\ }\href
  {\doibase 10.1103/PhysRevApplied.14.034011} {\bibfield  {journal} {\bibinfo
  {journal} {Phys. Rev. Applied}\ }\textbf {\bibinfo {volume} {14}},\ \bibinfo
  {pages} {034011} (\bibinfo {year} {2020})}\BibitemShut {NoStop}%
\bibitem [{\citenamefont {Blais}\ \emph {et~al.}(2020)\citenamefont {Blais},
  \citenamefont {Grimsmo}, \citenamefont {Girvin},\ and\ \citenamefont
  {Wallraff}}]{blais2020circuit}%
  \BibitemOpen
  \bibfield  {author} {\bibinfo {author} {\bibfnamefont {A.}~\bibnamefont
  {Blais}}, \bibinfo {author} {\bibfnamefont {A.~L.}\ \bibnamefont {Grimsmo}},
  \bibinfo {author} {\bibfnamefont {S.~M.}\ \bibnamefont {Girvin}}, \ and\
  \bibinfo {author} {\bibfnamefont {A.}~\bibnamefont {Wallraff}},\ }\href@noop
  {} {\enquote {\bibinfo {title} {Circuit quantum electrodynamics},}\ }
  (\bibinfo {year} {2020}),\ \Eprint {http://arxiv.org/abs/2005.12667}
  {arXiv:2005.12667 [quant-ph]} \BibitemShut {NoStop}%
\bibitem [{\citenamefont {Steane}(1996)}]{steane1996multiple}%
  \BibitemOpen
  \bibfield  {author} {\bibinfo {author} {\bibfnamefont {A.}~\bibnamefont
  {Steane}},\ }\href {\doibase 10.1098/rspa.1996.0136} {\bibfield  {journal}
  {\bibinfo  {journal} {Proceedings of the Royal Society A: Mathematical,
  Physical and Engineering Sciences}\ }\textbf {\bibinfo {volume} {452}}
  (\bibinfo {year} {1996}),\ 10.1098/rspa.1996.0136}\BibitemShut {NoStop}%
\bibitem [{\citenamefont {Bechmann-Pasquinucci}\ and\ \citenamefont
  {Tittel}(2000)}]{bechmann-pasquinucci1999quantum}%
  \BibitemOpen
  \bibfield  {author} {\bibinfo {author} {\bibfnamefont {H.}~\bibnamefont
  {Bechmann-Pasquinucci}}\ and\ \bibinfo {author} {\bibfnamefont
  {W.}~\bibnamefont {Tittel}},\ }\href {\doibase 10.1103/PhysRevA.61.062308}
  {\bibfield  {journal} {\bibinfo  {journal} {Phys. Rev. A}\ }\textbf {\bibinfo
  {volume} {61}},\ \bibinfo {pages} {062308} (\bibinfo {year}
  {2000})}\BibitemShut {NoStop}%
\bibitem [{\citenamefont {Gottesman}(1999)}]{gottesman1999fault}%
  \BibitemOpen
  \bibfield  {author} {\bibinfo {author} {\bibfnamefont {D.}~\bibnamefont
  {Gottesman}},\ }\href {\doibase 10.1016/S0960-0779(98)00218-5} {\bibfield
  {journal} {\bibinfo  {journal} {Chaos Solitons Fractals}\ }\textbf {\bibinfo
  {volume} {10}},\ \bibinfo {pages} {1749} (\bibinfo {year} {1999})},\ \Eprint
  {http://arxiv.org/abs/quant-ph/9802007} {arXiv:quant-ph/9802007} \BibitemShut
  {NoStop}%
\bibitem [{\citenamefont {Gottesman}\ \emph {et~al.}(2001)\citenamefont
  {Gottesman}, \citenamefont {Kitaev},\ and\ \citenamefont
  {Preskill}}]{gottesman2001encoding}%
  \BibitemOpen
  \bibfield  {author} {\bibinfo {author} {\bibfnamefont {D.}~\bibnamefont
  {Gottesman}}, \bibinfo {author} {\bibfnamefont {A.}~\bibnamefont {Kitaev}}, \
  and\ \bibinfo {author} {\bibfnamefont {J.}~\bibnamefont {Preskill}},\ }\href
  {\doibase 10.1103/PhysRevA.64.012310} {\bibfield  {journal} {\bibinfo
  {journal} {Phys. Rev. A}\ }\textbf {\bibinfo {volume} {64}},\ \bibinfo
  {pages} {012310} (\bibinfo {year} {2001})}\BibitemShut {NoStop}%
\bibitem [{\citenamefont {Rosen}(2002)}]{rosen2002discrete}%
  \BibitemOpen
  \bibfield  {author} {\bibinfo {author} {\bibfnamefont {K.~H.}\ \bibnamefont
  {Rosen}},\ }\href@noop {} {\emph {\bibinfo {title} {Discrete Mathematics and
  Its Applications}}},\ \bibinfo {edition} {5th}\ ed.\ (\bibinfo  {publisher}
  {McGraw-Hill Higher Education},\ \bibinfo {year} {2002})\BibitemShut
  {NoStop}%
\bibitem [{\citenamefont {Bandyopadhyay}\ \emph {et~al.}(2001)\citenamefont
  {Bandyopadhyay}, \citenamefont {Boykin}, \citenamefont {Roychowdhury},\ and\
  \citenamefont {Vatan}}]{bandyopadhyay2001new}%
  \BibitemOpen
  \bibfield  {author} {\bibinfo {author} {\bibfnamefont {S.}~\bibnamefont
  {Bandyopadhyay}}, \bibinfo {author} {\bibfnamefont {P.}~\bibnamefont
  {Boykin}}, \bibinfo {author} {\bibfnamefont {V.}~\bibnamefont
  {Roychowdhury}}, \ and\ \bibinfo {author} {\bibfnamefont {F.}~\bibnamefont
  {Vatan}},\ }\href@noop {} {\bibfield  {journal} {\bibinfo  {journal}
  {Algorithmica (New York)}\ }\textbf {\bibinfo {volume} {34}} (\bibinfo {year}
  {2001})}\BibitemShut {NoStop}%
\bibitem [{\citenamefont {Lawrence}\ \emph {et~al.}(2001)\citenamefont
  {Lawrence}, \citenamefont {Brukner},\ and\ \citenamefont
  {Zeilinger}}]{lawrence2001mutually}%
  \BibitemOpen
  \bibfield  {author} {\bibinfo {author} {\bibfnamefont {J.}~\bibnamefont
  {Lawrence}}, \bibinfo {author} {\bibfnamefont {C.}~\bibnamefont {Brukner}}, \
  and\ \bibinfo {author} {\bibfnamefont {A.}~\bibnamefont {Zeilinger}},\ }\href
  {\doibase 10.1103/PhysRevA.65.032320} {\bibfield  {journal} {\bibinfo
  {journal} {Physical Review A}\ }\textbf {\bibinfo {volume} {65}} (\bibinfo
  {year} {2001}),\ 10.1103/PhysRevA.65.032320}\BibitemShut {NoStop}%
\bibitem [{\citenamefont {Durt}\ \emph {et~al.}(2010)\citenamefont {Durt},
  \citenamefont {Englert}, \citenamefont {Bengtsson},\ and\ \citenamefont
  {Życzkowski}}]{durt2010mutually}%
  \BibitemOpen
  \bibfield  {author} {\bibinfo {author} {\bibfnamefont {T.}~\bibnamefont
  {Durt}}, \bibinfo {author} {\bibfnamefont {B.-G.}\ \bibnamefont {Englert}},
  \bibinfo {author} {\bibfnamefont {I.}~\bibnamefont {Bengtsson}}, \ and\
  \bibinfo {author} {\bibfnamefont {K.}~\bibnamefont {Życzkowski}},\ }\href
  {\doibase 10.1142/S0219749910006502} {\bibfield  {journal} {\bibinfo
  {journal} {International Journal of Quantum Information}\ }\textbf {\bibinfo
  {volume} {08}},\ \bibinfo {pages} {535} (\bibinfo {year} {2010})},\ \Eprint
  {http://arxiv.org/abs/https://doi.org/10.1142/S0219749910006502}
  {https://doi.org/10.1142/S0219749910006502} \BibitemShut {NoStop}%
\bibitem [{\citenamefont {Benedetti}\ \emph {et~al.}(2019)\citenamefont
  {Benedetti}, \citenamefont {Lloyd}, \citenamefont {Sack},\ and\ \citenamefont
  {Fiorentini}}]{Benedetti2019parameterized}%
  \BibitemOpen
  \bibfield  {author} {\bibinfo {author} {\bibfnamefont {M.}~\bibnamefont
  {Benedetti}}, \bibinfo {author} {\bibfnamefont {E.}~\bibnamefont {Lloyd}},
  \bibinfo {author} {\bibfnamefont {S.}~\bibnamefont {Sack}}, \ and\ \bibinfo
  {author} {\bibfnamefont {M.}~\bibnamefont {Fiorentini}},\ }\href {\doibase
  10.1088/2058-9565/ab4eb5} {\bibfield  {journal} {\bibinfo  {journal} {Quantum
  Science and Technology}\ }\textbf {\bibinfo {volume} {4}},\ \bibinfo {pages}
  {043001} (\bibinfo {year} {2019})}\BibitemShut {NoStop}%
\bibitem [{\citenamefont {Sim}\ \emph {et~al.}(2019)\citenamefont {Sim},
  \citenamefont {Johnson},\ and\ \citenamefont
  {Aspuru-Guzik}}]{sim2019expressibility}%
  \BibitemOpen
  \bibfield  {author} {\bibinfo {author} {\bibfnamefont {S.}~\bibnamefont
  {Sim}}, \bibinfo {author} {\bibfnamefont {P.}~\bibnamefont {Johnson}}, \ and\
  \bibinfo {author} {\bibfnamefont {A.}~\bibnamefont {Aspuru-Guzik}},\ }\href
  {\doibase 10.1002/qute.201900070} {\bibfield  {journal} {\bibinfo  {journal}
  {Advanced Quantum Technologies}\ }\textbf {\bibinfo {volume} {2}} (\bibinfo
  {year} {2019}),\ 10.1002/qute.201900070}\BibitemShut {NoStop}%
\bibitem [{\citenamefont {Holmes}\ \emph {et~al.}(2022)\citenamefont {Holmes},
  \citenamefont {Sharma}, \citenamefont {Cerezo},\ and\ \citenamefont
  {Coles}}]{holmes2022connecting}%
  \BibitemOpen
  \bibfield  {author} {\bibinfo {author} {\bibfnamefont {Z.}~\bibnamefont
  {Holmes}}, \bibinfo {author} {\bibfnamefont {K.}~\bibnamefont {Sharma}},
  \bibinfo {author} {\bibfnamefont {M.}~\bibnamefont {Cerezo}}, \ and\ \bibinfo
  {author} {\bibfnamefont {P.~J.}\ \bibnamefont {Coles}},\ }\href {\doibase
  10.1103/PRXQuantum.3.010313} {\bibfield  {journal} {\bibinfo  {journal} {PRX
  Quantum}\ }\textbf {\bibinfo {volume} {3}},\ \bibinfo {pages} {010313}
  (\bibinfo {year} {2022})}\BibitemShut {NoStop}%
\end{thebibliography}%


\end{document}